\documentclass{article}
\usepackage{arxiv}
\usepackage[utf8]{inputenc} 
\usepackage[T1]{fontenc}    
\usepackage{hyperref}       
\usepackage{url}            
\usepackage{booktabs}       
\usepackage{amsfonts}       
\usepackage{nicefrac}       
\usepackage{microtype}      
\usepackage{lipsum}
\usepackage{graphicx}
\usepackage{bm}
\usepackage{amsmath}
\usepackage{booktabs}
\usepackage{amsfonts}
\usepackage{amssymb}
\usepackage{url}
\usepackage[version=3]{mhchem}
\usepackage{enumitem}
\usepackage{caption}
\usepackage{subcaption}
\usepackage[labelformat=simple]{subcaption}

\usepackage{color}
\usepackage[disable]{todonotes}
\usepackage{float}

\usepackage{authblk}
\usepackage{bbold}

\usepackage{tikz}
\usepackage{xcolor}

\newcommand{\AQA}{\textrm{AAQAA}_3}

\title{Interpretable embeddings from molecular simulations using Gaussian mixture variational autoencoders}
\author[1, 2, *]{Yasemin Bozkurt Varolg{{\"u}}ne{{\c{s}}}}
\author[1]{Tristan Bereau}
\author[1]{Joseph F. Rudzinski}
\affil[1]{Max Planck Institute for Polymer Research, Mainz 55128, Germany}
\affil[2]{Department of Electrical \& Electronics Engineering, Koc University, Sariyer, Istanbul 34450, Turkey}
\affil[*]{\textit {bozkurty@mpip-mainz.mpg.de}}

\begin{document}
\maketitle

\begin{abstract}
Extracting insight from the enormous quantity of data generated from molecular simulations requires the identification of a small number of collective variables whose corresponding low-dimensional free-energy landscape retains the essential features of the underlying system.
Data-driven techniques provide a systematic route to constructing this landscape, without the need for extensive a priori intuition into the relevant driving forces.
In particular, autoencoders are powerful tools for dimensionality reduction, as they naturally force an information bottleneck and, thereby, a low-dimensional embedding of the essential features.
While variational autoencoders ensure continuity of the embedding by assuming a unimodal Gaussian prior, this is at odds with the multi-basin free-energy landscapes that typically arise from the identification of meaningful collective variables.
In this work, we incorporate this physical intuition into the prior by employing a Gaussian mixture variational autoencoder (GMVAE), which encourages the separation of metastable states within the embedding.
The GMVAE performs dimensionality reduction and clustering within a single unified framework, and is capable of identifying the inherent dimensionality of the input data, in terms of the number of Gaussians required to categorize the data. 
We illustrate our approach on two toy models, alanine dipeptide, and a challenging disordered peptide ensemble, demonstrating the enhanced clustering effect of the GMVAE prior compared to standard VAEs.
The resulting embeddings appear to be promising representations for constructing Markov state models, highlighting the transferability of the dimensionality reduction from static equilibrium properties to dynamics.

\end{abstract} 
\section{Introduction}  \label{sec:intro}

Particle-based computer simulations can provide unprecedented
mechanistic insight into the driving forces of complex molecular
systems, in contexts ranging from biochemistry to materials
science~\cite{binder1995monte, karplus2002molecular,
bottaro2018biophysical}. These simulations rely on numerical
integration of the relevant equations of motion as a means to navigate
the system's conformational space. Due to the high dimensionality of
this space, which prevents the exhaustive enumeration of all
microstates, exploration is typically achieved through importance
sampling~\cite{Allen:1987}. Conformational sampling leads to an
estimate of the potential energy landscape (PEL), which follows a
Boltzmann distribution at equilibrium.
%
Unfortunately, characterization of the PEL suffers from the so-called \emph{curse of dimensionality}~\cite{bellman2015adaptive}---organization of the data in the high-dimensional space is challenging due to low population density. 
This problem is often remedied by projecting the PEL onto a lower-dimensional manifold, i.e., by performing a dimensionality reduction.
By averaging over presumably unimportant degrees of freedom, the resulting low-dimensional surface represents a \emph{free}-energy landscape (FEL).
The ideal FEL distinguishes between microstates that are separated by large barriers on the PEL, yielding a partitioning of configuration space into collections of microstates, i.e., metastable basins.
If all the largest barriers are accounted for, intra-basin diffusion will occur much faster than inter-barrier crossing events, allowing an accurate, albeit coarse-grained, description of both the static and dynamical properties of the system.

The essential degrees of freedom that define the low-dimensional
representation, commonly referred to as collective variables (CVs), are
traditionally identified through expert physical/chemical intuition
that is often rather specific for the particular system or process of
interest~\cite{kevrekidis2003equation, dobson2003protein,
onuchic2004theory, weinan2007heterogeneous}.
Beyond the characterization of the FEL, these CVs can also be used for enhanced sampling~\cite{valsson2016enhancing}, or for the construction of low-dimensional configuration-space discretizations, for instance when building Markov state models (MSMs)~\cite{Husic:2018}.
Although the manual selection of CVs can be extremely effective for practitioners with insight into the system, the approach is difficult to extend systematically and is susceptible to missing unanticipated or subtle features of the FEL that may nonetheless play an important role in the relevant phenomena.
Data-driven techniques provide an alternative route by inferring the important features directly from the data.
There is a long history of methods for finding an optimal low-dimensional representation from a given set of data, employing both linear (e.g., principal component analysis~\cite{pca}, time-lagged independent component analysis~\cite{molgedey1994separation}) and nonlinear (e.g., Isomap~\cite{tenenbaum2000global}, Sketchmap~\cite{sketchmap}) transformations.

In the last couple years, there has been a growing interest in
applying (deep) neural networks to automate the discovery of
CVs~\cite{chen2018molecular, ribeiro2018reweighted, bonati2019neural, ma2005automatic, geiger2013neural}.
One architecture that stands out as conceptually appealing is the
autoencoder~\cite{autoencoder}. An autoencoder is a bow-tie-shaped
network that forces an information compression in the bottleneck
region. While the first half of the network (the encoder) reduces the
input to a predefined lower dimension, the second half (the decoder)
aims at transforming from the low-dimensional to the original
representation.
The weights of the neural network are tuned to minimize an objective or \emph{loss} function, which typically penalizes deviations between input and output data.
As such, the autoencoder aims at discovering a \emph{latent space} (embedding) that faithfully describes the essential features of the high-dimensional input data. 
This makes autoencoders well suited for constructing low-dimensional FELs from molecular simulation data~\cite{doerr2017dimensionality,Chen:2018b,encodermap}.

Traditional autoencoders lack continuity in the latent space, preventing interpolation between training points and, thus, its generative ability.
Variational autoencoders (VAEs) remedy this limitation by modeling the input probability distribution using Bayesian inference~\cite{kingma2013auto}. 
VAEs enable sampling new data from the learned distribution (i.e., VAEs are generative models), and are also well-suited to provide interpretable and disentangled data representations in the low-dimensional space~\cite{adel2018discovering}.
%
Within the VAE framework, the latent distribution is forced to resemble a predefined probability distribution, called the \emph{prior}. 
Although the VAE framework does not impose any particular prior distribution, it is often chosen as a normal distribution for computational convenience.
This prior induces an ``anti-clustering'' effect in the latent space, which can prohibit the identification of meaningful clusters and impede the construction of optimal FELs from molecular simulations.
%
The autoencoder-based approaches were recently extended to explicitly
incorporate the temporal nature of the data via a time-lag in the
network architecture~\cite{wehmeyer2018time,hernandez2018variational}.
These time-lagged autoencoders aim to retain information about the
slowest dynamical modes sampled in the underlying simulation
trajectory and, as a consequence, may encourage metastable clustering
in the latent space. However, they are also limited in terms of
characterizing the hierarchy of long timescale
processes~\cite{Chen:2019}, and only indirectly address the
anti-clustering issue. 

In this work, we propose to directly acknowledge the multi-basin
structure of an ideal FEL by employing a Gaussian mixture
model~\cite{dilokthanakul2016deep} as the prior distribution for the
VAE latent space.
The resulting \emph{Gaussian mixture variational autoencoder} (GMVAE) retains the computational ease and reconstruction fidelity of traditional VAEs, while enforcing a more faithful description of the underlying physics: the resulting FEL clearly distinguishes between metastable basins separated by large free-energy barriers.
We demonstrate the benefits of the GMVAE approach through explicit comparisons with the traditional VAE for two widely-studied toy models and for the standard benchmark system for conformational dynamics, alanine dipeptide, as well as a more challenging disordered peptide ensemble.
%
To ensure the presence of distinct distributions in the latent space, the GMVAE introduces a categorical variable that (probabilistically) assigns each input configuration to the set of clusters. 
Thus, the GMVAE simultaneously performs dimensionality reduction and unsupervised clustering.
Remarkably, the GMVAE clustering is capable of identifying the inherent dimensionality of the input data, in terms of the number of Gaussians required to categorize the data.
In the case of hierarchical input data (i.e., data with distinct dimensionality depending on the level of resolution), we show that the GMVAE makes a reasonable prediction for the number of clusters, independent of the given hyperparameter, based on the dimensionality of the latent space and characteristics of the data.
%
%
Beyond the representation of static equilibrium properties, by
constructing MSMs from the GMVAE embedding, we show that our approach
is also a promising avenue for accurately describing the long
timescale \emph{dynamical} properties of the data. In contrast to
recent deep neural-network approaches that aim to directly model the
\emph{propagator} of the system's
dynamics~\cite{mardt2018vampnets,lusch2018deep}, the construction of
MSMs from the learned FEL offers a different strategy: explicitly
testing to what extent a representation appropriate for the statics is
directly amenable for the dynamics. 

\section{Theory and Methods} \label{sec:theory}

\subsection{Autoencoder}
\begin{figure}[htbp]
  \centering
  \includegraphics[width=0.8\linewidth]{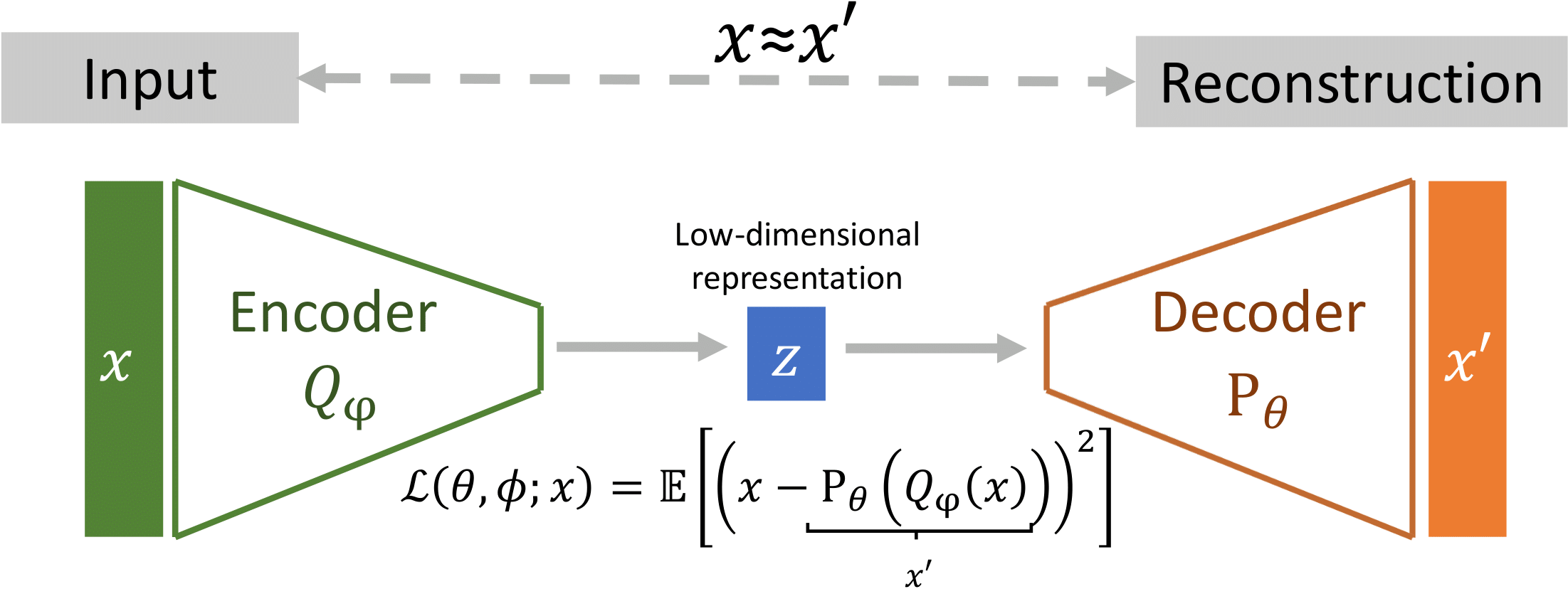}
  \caption{Schematic of an autoencoder architecture with mean-squared error reconstruction loss.}
  \label{fig:autoencoder}
\end{figure}

Autoencoders are special types of neural networks that are used for the task of representation learning in an unsupervised manner. 
They are composed of two connected parts: the encoder compresses the input signal to a low-dimensional representation, whereas, 
the decoder aims to reconstruct the input at full dimensionality from the reduced-space representation. 
The \emph{reconstruction loss}, usually defined as either the mean-squared error or cross-entropy between the input, $x$, and the output, $x'$, is minimized via backpropagation. 
Since the bottleneck dimension is typically much less than the original dimension, autoencoders learn the most compact representation of the input. 
Furthermore, because neural networks are universal function approximators, the learned data projections can generally preserve much more of the relevant information than with PCA or other basic linear projection techniques. 
Figure~\ref{fig:autoencoder}  shows the schematic structure of autoencoder with mean-squared error loss. 
There are different types of autoencoders which are tailored for special tasks. For instance, 
sparse autoencoders impose sparsity constraints during optimization, whereas convolutional autoencoders utilize convolutional layers 
instead of fully-connected layers, in which case they learn the optimal filters. 
Variational autoencoders, which model the latent space probabilistically,
 are used for generative purposes, i.e., they can create new samples that look like the ones 
in the training dataset without simple data replication. 

\subsection{Variational Autoencoder (VAE)}
Variational autoencoders were introduced in~\cite{kingma2013auto}. 
In general, the theory of VAEs is approached from two different perspectives: variational inference and neural networks. 
This section starts with the former interpretation and then illustrates the connection between them. 
We mostly follow the notation and reasoning used in~\cite{vae}.
The input data and the latent variable are denoted by $x$ and $z$, respectively. 

The objective of the VAE is to find the posterior distribution $P(z|x)$, which can be written in terms of the likelihood $P(x|z)$, the prior $P(z)$, and the marginal probability density of $x$, $P(x)$, using Bayes law as 
\begin{equation} \label{eq:bayes}
\begin{split}
P(z|x) &= \dfrac{P(x|z) P(z)}{P(x)} \,\, . 
\end{split}
\end{equation}
%
The denominator $P(x)$ is called the evidence and it could, in principle, be calculated using 
%
\begin{equation}
	P(x) = \int \textrm{d}z \, P(x|z) P(z) \,\, , 	\label{eq:vae_joint_b}
\end{equation}
once the prior is selected. However, the calculation is typically intractable, as it needs to be evaluated over all configurations of the latent variable $z$. Therefore, the posterior is approximated using \emph{variational inference} with a chosen easy-to-evaluate family of distributions $Q_{\phi}(z|x)$, e.g., Gaussian functions, where ${\phi}$ is the variational parameter of the distribution. 
In particular, $P(z|x)$ is inferred using $Q_{\phi}(z|x)$ by reformulating the problem within an optimization framework, such that the Kullback-Leibler divergence between $Q_{\phi}(z|x)$ and $P(z|x)$ is minimized. 
The KL divergence between $Q$ and $P$ is defined as
\begin{equation} \label{eq:eqa}
\begin{split}
D_\textrm{KL}[Q_{\phi}(z | x) || P(z | x)] &= \sum_z Q_{\phi}(z | x) \, \log \frac{Q_{\phi}(z | x)}{P(z | x)} \\
                            &= \mathbb{E} \left[ \log \frac{Q_{\phi}(z | x)}{P(z | x)} \right] \\
                            &= \mathbb{E}[\log Q_{\phi}(z | x) - \log P(z | x)] \,\, .
\end{split}
\end{equation}
Equation~\ref{eq:bayes} is then inserted into the posterior definition.
\begin{equation} \label{eq:eqb}
\begin{split}
D_\textrm{KL}[Q_{\phi}(z | x) || P(z | x)] &= \mathbb{E} \left[ \log Q_{\phi}(z | x) - \log \frac{P(x | z) P(z)}{P(x)} \right] \\
                                        &= \mathbb{E}[\log Q_{\phi}(z | x) - \log P(x | z) - \log P(z) + \log P(x)] \,\, . 
\end{split}
\end{equation}
Since the expectation is taken over $z$, $P(x)$ can be moved out of the expectation. 
%
\begin{equation} 
D_\textrm{KL}[Q_{\phi}(z | x) || P(z | x)] - \log P(x) = - \underbrace{\mathbb{E}[\log P(x, z) - \log Q_{\phi}(z | x)] \,\, .}_{\textrm{ELBO}(\phi)} 
\label{eq:elbo}
\end{equation}
The initial objective of minimizing the KL divergence between the exact and the approximate posterior is equivalent to maximizing the ELBO (Evidence Lower BOund), defined in Equation~\ref{eq:elbo}. 

Equation~\ref{eq:elbo} can also be rewritten in terms of a different KL divergence:
\begin{equation} \label{eq:vae_obj}
D_\textrm{KL}[Q_{\phi}(z | x) || P(z | x)] - \log P(x) = D_\textrm{KL}[Q_{\phi}(z | x) || P(z)] - \mathbb{E}[\log P(x | z)] \,\, .
\end{equation}
%
%
%
Here the neural network perspective comes into play, as depicted schematically in Figure~\ref{fig:encoding-decoding}(a). 
$Q_{\phi}(z | x)$ acts like an encoder (\emph{inference}), and transforms the data into the latent variable $z$. 
On the other hand, $P(z | x)$ (which can also be parametrized with the network parameter $\theta$ as $P_{\theta}(z | x)$\footnote{Both of the notations are used interchangeably.})
generates the data from the latent representation, analogous to a decoder (\emph{generator}). 
The parameters correspond to the weights and biases of the neural networks.
Note that the initial aim is to minimize $D_\textrm{KL}[Q_{\phi}(z | x) || P(z | x)]$, which is equivalent to 
minimizing the RHS of Equation~\ref{eq:vae_obj}. 
The first term enforces the encoder to be similar to the chosen prior $P(z)$, which acts as a regularization, whereas the second term on the RHS deals with how well the reconstructions match the original input. 


\subsubsection{Standard Selections for the Family of Inference Distributions and for the Prior Distribution}

In order to use Equation~\ref{eq:vae_obj} in an optimization procedure, both the family of distributions for inference, $Q_{\phi}(z | x)$, as well as the prior distribution, $P(z)$, must be specified. 
The most common assumption is that $Q_{\phi}(z | x)$ ($P(z)$) is a unimodal Gaussian distribution with mean $\mu(x)$ (0) and diagonal covariance $\Sigma(x)$ ($\mathbb{1}$).
Then, $D_\textrm{KL}[Q_{\phi}(z | x) || P(z)]$ has a closed form solution:
%
\begin{equation} \label{eq:KL_G}
\begin{split}
D_\textrm{KL}[Q_{\phi}(z | x) || P(z)] &= D_\textrm{KL}[\mathcal{N}(\mu(x), \Sigma(x)) || \mathcal{N}(0, \bm{1})] \\ 
																&= \frac{1}{2} \, \left( \textrm{tr}(\Sigma(x)) + 
																	 \mu(x)^T\mu(x) - d - \log \, \det(\Sigma(x)) \right) \,\, ,
\end{split}
\end{equation}
where $d$ is the dimension of the Gaussian and $\textrm{tr}$ denotes the trace.
Although the unimodal Gaussian assumption simplifies the calculations, it also restricts the possible latent space representations, and may hinder the performance of the variational autoencoder by pushing the latent space to be described by highly-overlapping clusters.
%
\begin{figure}[htbp]
	\begin{subfigure}{\textwidth}
  \centering
  \includegraphics[width=0.8\linewidth]{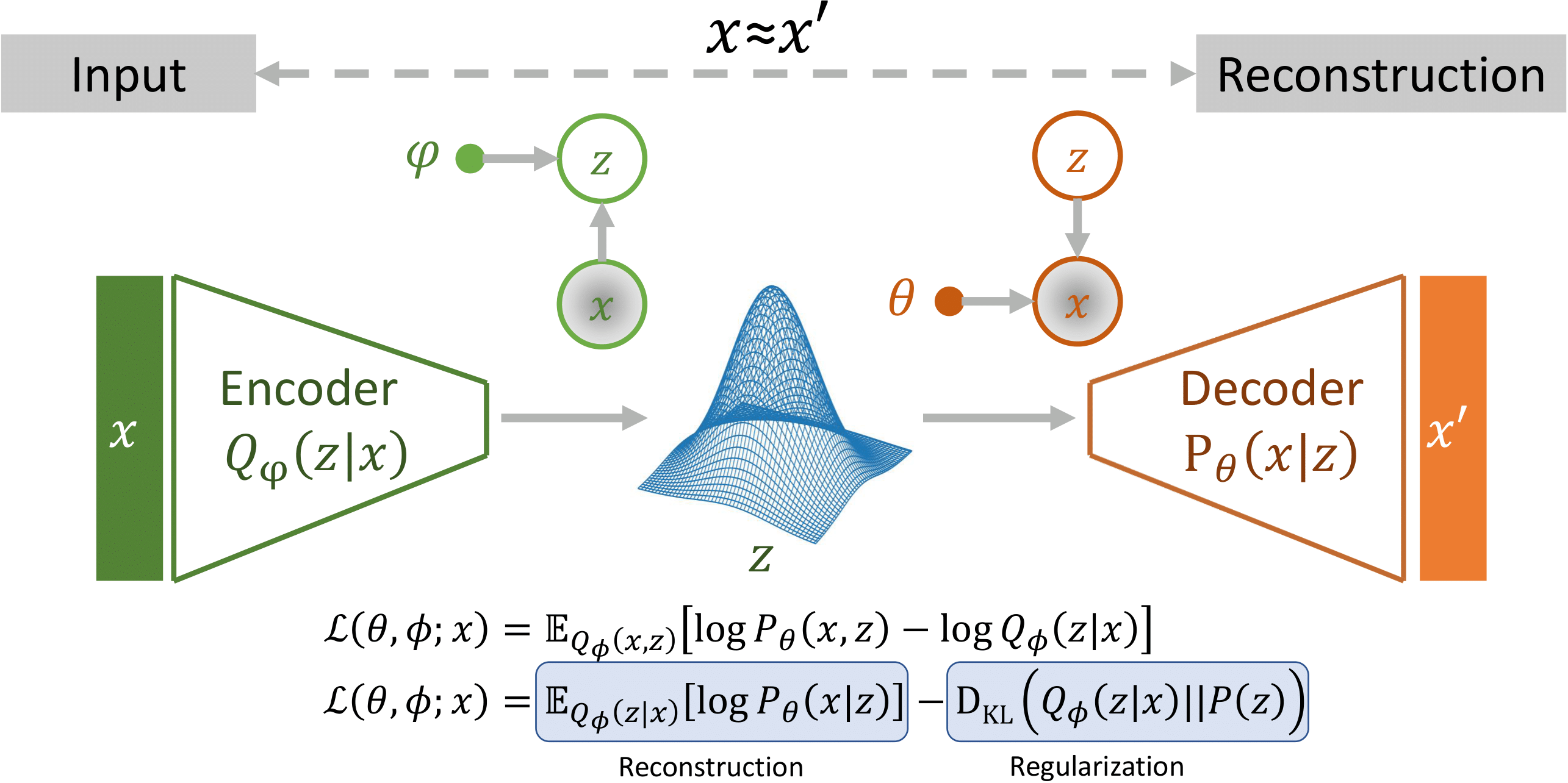}
  \caption{Schematic of a variational autoencoder with unimodal Gaussian prior.}
  \label{fig:sub1}
\end{subfigure} \\
\begin{subfigure}{\textwidth}
  \centering
  \includegraphics[width=0.8\linewidth]{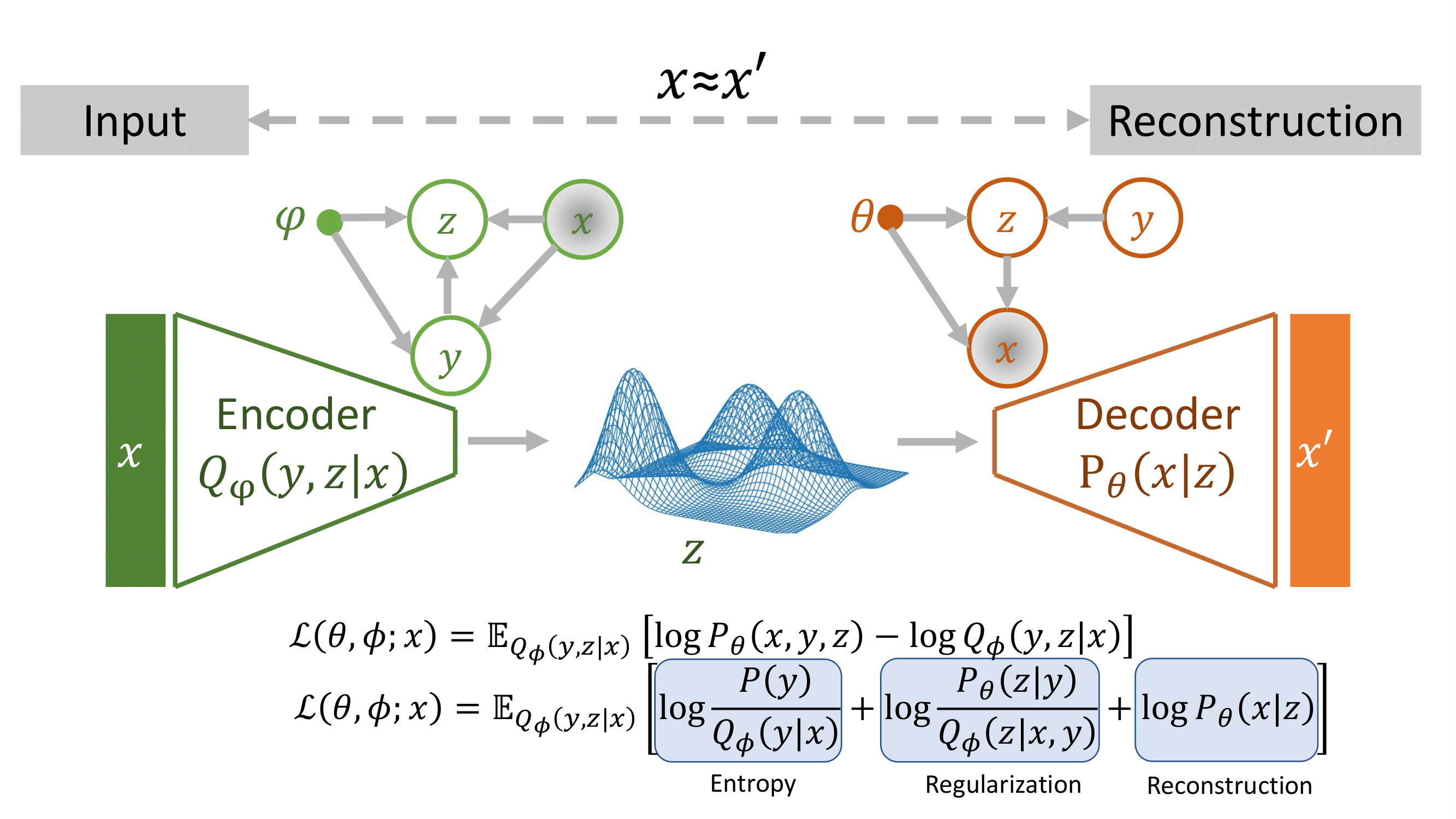}
  \caption{Schematic of a Gaussian mixture variational autoencoder.}
  \label{fig:sub2}
\end{subfigure}%
	\caption{(a) The VAE and (b) GMVAE architectures. In the probabilistic graph representation, circle nodes represent the random variables, and directed edges represent statistical dependencies between the variables in the two ends. Dot nodes are used to indicate the parameters of the model, while some of the nodes are intentionally filled to differentiate the observed random variables from the non-observed ones which are left empty. }
	\label{fig:VAEAndGMVAEArchitectures}
	\end{figure}
\subsection{Gaussian Mixture Variational Autoencoder} \label{sec:GMVAE_theory}

This section is largely distilled from the discussion and insights presented in~\cite{ruishu}. 
The term Gaussian mixture variational autoencoder is open to misinterpretations. 
There exist several distinct architectures given this name, with variations in the choice of generative or inference models~\cite{dilokthanakul2016deep, zhao2019variational, nalisnick2016approximate, shi2019fixing}.
In the present work, we take both the approximate posterior, (i.e., the family of distribution functions for inference), $Q_{\phi}(y,z|x)$, and the latent space distribution (i.e., the prior), $P(z)$, to be Gaussian mixtures.
Note that we have introduced a categorical variable, $y$, which identifies which Gaussian each particular data point belongs to.
The inference model can be written as
\begin{equation} \label{eq:gmvae_inf}
\begin{split}
        Q_{\phi}(y,z|x) &= Q_{\phi}(y|x)Q_{\phi}(z|x,y) \,\, .
\end{split}
\end{equation}
The latent space is composed of $k$ distinct Gaussians, i.e.,  $Q_{\phi}(z|x, y_i)$ is assumed to be Gaussian, where $i\in{0, 1, \dots, k-1}$. Thus, the approximate posterior becomes a Gaussian mixture. 
%

\begin{figure}[htbp]
	\centering
	\includegraphics[width = 1\textwidth]{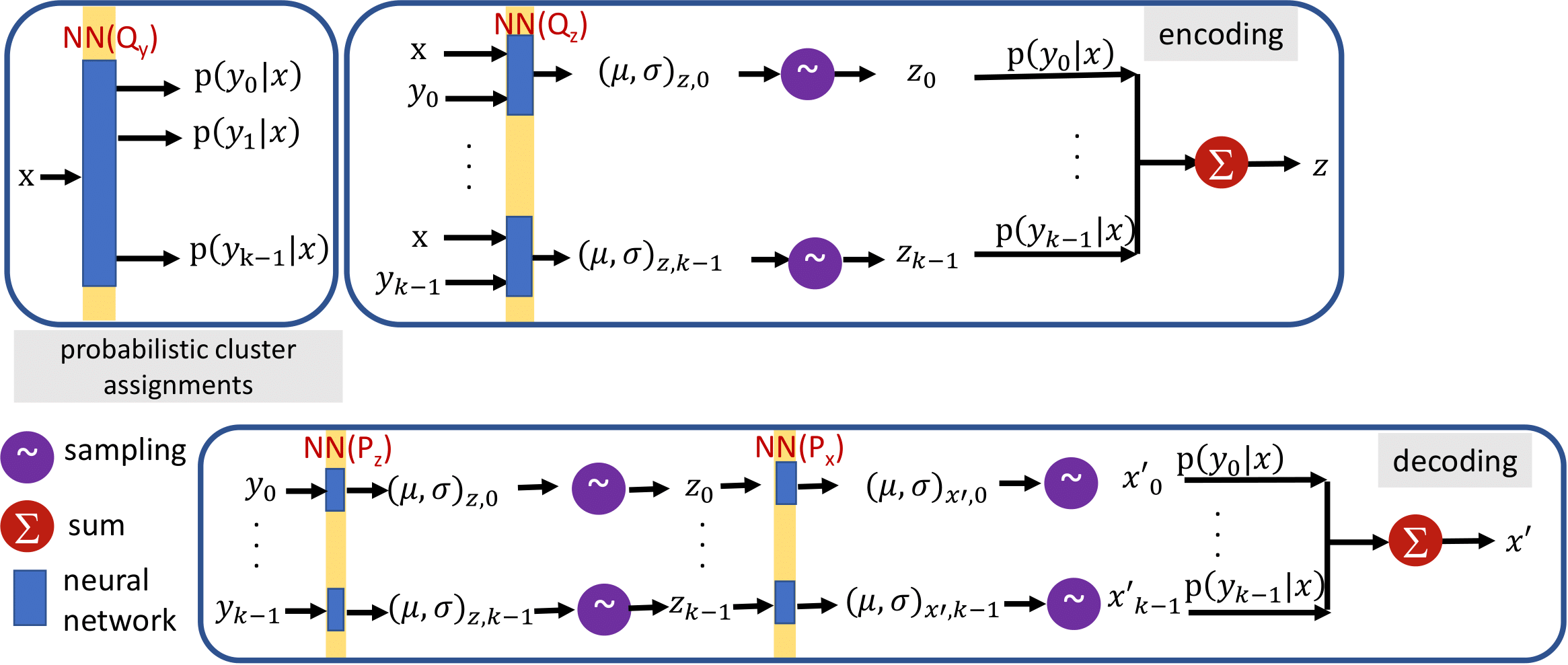}
	\caption{Schematic of the GMVAE workflow.}
	\label{fig:encoding-decoding}
\end{figure}

Similar to Equation~\ref{eq:elbo}, the ELBO can be written as 
%
\begin{equation} \label{eq:elbo_m}
\begin{split}
\textrm{ELBO}_\textrm{m} = \mathbb{E}_{Q{_\phi}(y,z|x)}[\log P_{\theta}(x,y,z) - \log Q_{\phi}(y,z|x)] \,\, ,
\end{split}
\end{equation}
%
where the number of Gaussians, $k$, is a hyperparameter, and the subscript $\textrm{m}$ is used to distinguish $\textrm{ELBO}_\textrm{m}$
from the VAE $\textrm{ELBO}$.
%
%
$P_{\theta}(x,y,z)$ can be written as $P_{\theta}(x,y,z) = P_{\theta}(x| y, z) P_{\theta}(z|y) P(y)$ using conditioning without any assumptions.
Then, by assuming that $x$ is conditionally independent of $y$, i.e.,  $P_{\theta}(x| y, z) = P_{\theta}(x|z)$ (see the graph representation in Figure~\ref{fig:sub2}), the joint probability can be expressed as 
\begin{equation} \label{eq:gmvae_gen}
\begin{split}
        P_{\theta}(x,y,z) &= P_{\theta}(x|z) P_{\theta}(z|y)  P(y) \,\, .
\end{split}
\end{equation}
%
%
%
By inserting Equations~\ref{eq:gmvae_inf} and~\ref{eq:gmvae_gen} into Equation~\ref{eq:elbo_m}, $\textrm{ELBO}_m$ becomes
\begin{equation} \label{eq:elbo_m2}
\begin{split}
\textrm{ELBO}_\textrm{m} 
			 &= \mathbb{E}_{Q(y,z|x)}[\log P(y)P_{\theta}(z|y)P_{\theta}(x|z) - \log Q_{\phi}(y|x)Q_{\phi}(z|x,y)] \\
			&= \mathbb{E}_{Q(y,z|x)}\Big[\log P(y) - \log {Q_{\phi}(y|x)} + \log \frac{P_{\theta}(z|y)}{Q_{\phi}(z|x,y)} + 
			\log P_{\theta}(x|z) \Big] \,\, .
\end{split}
\end{equation}
Similar to the VAE, the third and fourth terms represent regularization and reconstruction contributions to the loss, respectively. The initial prior on $y$ is selected as a uniform multinomial distribution, while $\mathbb{E}_{Q(y,z|x)} [\log {Q_{\phi}(y|x)}]$ can be interpreted as a conditional entropy, reflecting how informative $x$ is on $y$. 
To directly control the impact of the clustering relative to the other loss terms during training, we introduced a weighting factor, $\alpha$, on the mutual information between $x$ and $y$:
%
\begin{equation} \label{eq:elbo_m2_alpha}
\begin{split}
\textrm{ELBO}_\textrm{m} &= \mathbb{E}_{Q(y,z|x)}\Big[\log P(y) - \alpha \log {Q_{\phi}(y|x)} + \log \frac{P_{\theta}(z|y)}{Q_{\phi}(z|x,y)} + 
			\log P_{\theta}(x|z) \Big] \,\, .
\end{split}
\end{equation}
Figure~\ref{fig:encoding-decoding} presents a more detailed schematic of the GMVAE architecture, while Table~\ref{tab:dist} presents a summary of the probability distributions utilized in the model. 
%
First, data points are probabilistically assigned to $k$ clusters (NN(Q$_y$)). $Q(y|x)$ represents these cluster assignment probabilities, and has multinomial distribution. Since each cluster is assumed to have Gaussian distribution in the latent space, the mean and variance of each of these Gaussians ($Q(z|x,y)$) are learned via the encoder part of the neural network (NN(Q$_z$)). The low-dimensional representation, $z$, is then obtained by first sampling and then taking the expected value of these samples, i.e., $z = \sum_{i=0}^{k-1} p(y_{i}|x) z_i$. As the first step in decoding, the moments of the corresponding low-dimensional representation $z$ is learned by NN(P$_z$) from each Gaussian-distributed individual cluster $y_i$, which is then followed by a sampling operation. $P(y)$ in the decoder is assumed to be uniformly distributed among the $k$ clusters. Next, using the encodings, $z_i$'s, the associated $x$ reconstructions are obtained again by sampling from the $x'$ by the NN(P$_x$). Similar to the encoder, the decoder obtains a fixed reconstruction by taking the expected value of $x'_i$'s.
\begin{table}[!h]
\large
\begin{center}
 \begin{tabular}{ l l l | l l } 
	 $Q(z|x,y)$ & $= \mathcal{N}(\mu_z(x, y), \sigma_{z}^2(x, y))$ && 	$P(y)$     & $= \mathrm{Uniform}(\frac{1}{k})$  \\
	 $Q(y|x)$   & $= \mathrm{Multinomial}(f(x))$ && $P(z|y)$   & $= \mathcal{N}(\mu_z(y), \sigma_{z}^2(y))$ \\
	&&& $P(x|z)$   & $= \mathcal{N}(\mu_x(z), \sigma_{x}^2(z))$  
  \end{tabular}
\caption{Distributions in the GMVAE model. Left (right) column corresponds to the distributions in the encoder (decoder) part.} 
\label{tab:dist}
\end{center}
\end{table}
%

%

\subsubsection{Determination of Cluster Labels and Thresholding Scheme} \label{sec:threshold}

%
The clustering within the GMVAE is probabilistic, i.e., each data point is assigned membership probabilities (between $0$ and $1$) to each of the clusters.
Since most configurations are assigned predominantly to a single cluster, we perform a hard cluster assignment by assigning each data point to the cluster with highest membership probability.
However, in cases where a configuration has similar membership probabilities for multiple clusters, this simple assignment may introduce errors when determining properties (e.g., transition probabilities) of the clusters.
Thus, we also considered a different approach by enforcing a thresholding value for cluster assignment. 
More specifically, each configuration is only assigned to a cluster if the largest membership probability is above a chosen cut-off value. 
A naive coring scheme followed the thresholding operation such that the points that had been identified as noise were assigned back to their previous cluster index for all other dynamical analyses.


\subsubsection{GMVAE Architecture and Training Hyperparameters}
The GMVAE algorithm was implemented in Tensorflow~\cite{tensorflow}, and is available at \url{https://github.com/yabozkurt/gmvae}.
Training was performed in all cases with fully connected layers, using the Adam optimization algorithm~\cite{kingma2014adam}. 
The Softmax activation function was used for probabilistic cluster assignments, while ReLu activation functions were employed in all hidden layers.
The means were obtained without any activation, whereas Softplus activation was employed to obtain the variances. 
Table~\ref{tab:hyper} shows the values of the hyperparameters for each example system. 
Default values were employed wherever the parameters are not specified. 
Number of nodes (NN($\cdot$)) columns correspond to the neural networks labeled in Figure~\ref{fig:encoding-decoding}. 
NN(Q$_y$) performs probabilistic cluster assignments, NN(Q$_z$) is for learning the moments of each Gaussian distribution in the encoding, whereas  NN(P$_z$) and NN(P$_x$) are for the decoding of the $z$ and $x$, respectively. 
The lengths of the ``Number of nodes'' entries correspond to the number of hidden layers. 
Hyperparameter optimization was carried out as follows.
The number of nodes was initialized as [16, 16].
The number of nodes in the decoder (NN(P$_{x}$)) was then increased whenever a large and non-decreasing reconstruction loss was observed.
Our overall observation for the considered examples is that the learning rate and batch size should be kept relatively low to promote the formation of distinct cluster. The VAE results (with unimodal Gaussian prior) that are provided as comparison are obtained using $k = 1$, while keeping the remaining parameters equal to the values in the corresponding GMVAE model. 
\begin{table}[!h]
\begin{center}
 \begin{tabular}{||l | l | l | l | l | l ||} 
 \hline
            							& 1D 4-well 	& M{{\"u}}ller-Brown 	& Dipeptide 			& $\AQA$ - I & $\AQA$ - II \\ [0.5ex]	
\hline
Number of clusters (k)		& 4 		 			& 5							& 8				& 10			& 6       \\	
Input dimension (n) 	 	& 1  					& 2							& 25			& 60			& 126     \\	
Latent dimension (d)   		& 1  					& 1							& 2				& 2				& 2	      \\
Number of nodes (NN(Q$_y$))	& [16, 16] 				& [32] 					& [32]		& [16, 16] 		& [128]	\\		
Number of nodes (NN(Q$_z$))	& [16, 16] 				& [16] 					& [16]		& [16, 16] 		& [16]	\\		
Number of nodes (NN(P$_z$))	& [16, 16] 				& [16] 					& [16]		& [16, 16] 		& [16]	\\		
Number of nodes (NN(P$_x$))	& [16, 16] 				& [128] 				& [128]		& [16, 16] 		& [256]	\\		
$\alpha$ 					& 0.5 					& 0.05					& 0.05		& 0.3			& 0.95  \\
Batch size 					& 32000 				& 5000					& 5000		& 10000 		& 3000    \\
Learning rate 				& 0.00005 				& 0.0001				& 0.00015	& 0.001 		& 0.00005 \\
Number of epochs   			& 50 					& 400					& 100		& 300			& 2000    \\
Probability cut-off			& None					& None					& None		& 0.95 			& 0.98    \\
 \hline
\end{tabular}
\caption{Architecture specification and training hyperparameters.}
\label{tab:hyper}
\end{center}
\end{table}

\subsection{Markov State Models} \label{sec:msm}
Markov state models (MSMs) represent the dynamics generated by a molecular simulation trajectory as a series of memoryless jumps between a discrete set of states~\cite{Bowman:2014}.
Given a configuration-space discretization, a transition probability matrix, $\textbf{P}(\tau)$, is obtained by counting the transitions between pairs of states within a given lag time, $\tau$, and then performing a maximum likelihood optimization~\cite{prinz2011markov}. 
The eigenvalues of $\textbf{P}(\tau)$, $\{ \lambda_i({\tau}) \}$, are related to characteristic timescales of the system's dynamics:
\begin{equation} \label{eq:its}
\begin{split}
t_i(\tau) = - \frac{\tau}{\ln |{\lambda_i({\tau})}|} \,\, ,
\end{split}
\end{equation}
where $t_i(\tau)$ is the timescale corresponding to the $i^{th}$ eigenvalue, $\lambda_i({\tau})$.
The time lag parameter $\tau$ is typically chosen by performing the ``implied timescale test'', which assesses the Markovianity of $\textbf{P}(\tau)$ through the convergence of its timescales with increasing $\tau$.
In other words, $\{ t_i(\tau) \}$ is plotted as a function of $\tau$, and $\tau$ is then chosen as small as possible such that the largest timescales are \emph{sufficiently} converged.
%
%
Once $\tau$ is chosen, the accuracy of $\textbf{P}(\tau)$ is determined via the Chapman-Kolmogorov (CK) test, which compares the estimated and predicted probability decay out of a given state.
The predicted values are obtained using the CK equation, i.e., using the Markovian property of the model:
\begin{equation} \label{eq:ck_test}
p_{ij}({m \tau}) = p_{ij}^m({\tau}) \,\, ,
\end{equation} 
where $p_{ij}({\tau})$ is the probability of transitioning from state $i$ to state $j$ within time $\tau$, and $m$ is a positive integer. The CK test is often performed on metastables states of the system---collections of quickly interconverting microstates. 

Within the standard Markov-state-modeling workflow, microstates are typically defined on low-dimensional projections of the full-dimensional configuration space. 
Therefore, obtaining a relevant transformation of the molecular simulation data is the key. 
To this end, time-lagged independent component analysis (TICA)~\cite{molgedey1994separation,Perez-Hernandez:2013} is one of the most commonly used dimensionality reduction methods, as its objective is to maximize the autocorrelation of the data at the given lag time, making it especially well suited for kinetic modeling purposes. 
Metastable states are typically obtained via a dynamical coarse-graining procedure, e.g., PCCA+~\cite{roblitz2013fuzzy} whose objective is to retain an accurate description of the dominant eigenvectors of the transition probability matrix.
The resulting metastable states are then used as representative collections of microstates for performing the CK test.
In many cases, a coarse-grained MSM at the resolution of the metastable states is constructed, providing an easily interpretable, albeit often qualitative, picture of the long timescale processes.
In this study, the GMVAE performs the dimensionality reduction and clustering simultaneously, yielding a coarse-grained description of configuration space directly, without the need for further dynamical clustering.
The (coarse-grained) MSMs are constructed from the discretized trajectories obtained using the simple cluster assignment based on the GMVAE membership probabilities as described in Section~\ref{sec:threshold}. 
MSM construction and analysis was performed using the PyEMMA package~\cite{scherer2015pyemma}. 
%
%

\subsection{Peptide Analysis} \label{sec:peptide_analysis}


The helical propensity of the peptide was determined using the Lifson-Roig perspective, which assigns each residue to either a helical (h) or coil (c) state, according to the dihedral angles along the peptide backbone (i.e., the Ramachandran plot)~\cite{lifson1961theory, doig2008helix}. 
Therefore, the number of different conformations of the peptide is limited to $2^N$, where $N$ is the number of residues; $N =$ 15 for $\AQA$.
The propensity of residue $i$ to be part of a ``helical segment'', $\langle h_i \rangle$, is then defined as the probability that residue $i$ as well as its two neighboring residues are simultaneously found in a helical state.
The average fraction of helical segments, $\langle f_h \rangle$, is obtained by averaging $\langle h_i \rangle$ over all residue positions: $\sum_{i=0}^{N-1} \frac{1}{N} \langle h_i \rangle$.
%
To distinguish between partial helical structures occuring at the N- and C-terminus ends of the peptide backbone, we define $ \langle h_N \rangle = \sum_{i=1}^{6} \frac{1}{6} \langle h_i \rangle$ and $ \langle h_C \rangle = \sum_{i=8}^{13} \frac{1}{6} \langle h_i \rangle$.
Note that the terminus residue from each end is not taken into consideration. 

The dRMSD measures the average deviation of internal distances from the corresponding distances in a reference structure, and is calculated as
\begin{equation} \label{eq:drmsd}
\textrm{dRMSD}(\textbf{X}(t), \textbf{X}^r) = \sqrt{ \sum_{i \neq j} ({||{\textbf{X}_{i}(t)} - {\textbf{X}_{j}(t)}||}
                                                                                                                                                                                                         - ||{\textbf{X}^r_{i}} - {\textbf{X}^r_{j}}||)^2 } \,\, ,
\end{equation}
where $\textbf{X}(t)$ represents the conformation at time t, $\textbf{X}^r$ is the conformation for the reference structure, and
$||\cdot||$ denotes the Euclidean norm.
Note that, unlike other RMSD metrics, no pre-alignment of structures is required.
In this study, due to the large fluctuations of the end residues, two residues from each end of the peptide were excluded in the dRMSD calculations.
dRMSD was calculated using the positions of the $C_{\alpha}$ atoms only.
Helix, hairpin-like, and extended (coil) structures were separately considered as reference structures as illustrated in Figure~\ref{fig:aaqaa-I-drmsds}.


\section{Results}  \label{sec:results}

Variational autoencoders (VAEs) have been previously applied for dimensionality reduction of molecular simulation data~\cite{hernandez2018variational, ribeiro2018reweighted, bhowmik2018deep}. VAEs typically employ a normal distribution to represent both the prior distribution in the latent space and the family of distributions for variational inference.
In this work, we extend traditional VAEs by representing these distributions with Gaussian mixture models.
The resulting Gaussian mixture VAE (GMVAE) adopts the physics-based viewpoint that an optimal embedding of the simulation data should give rise to a free-energy landscape (FEL) with well-separated clusters of configurations, which correspond to metastable states that are separated by large barriers along the high-dimensional potential energy landscape.
The GMVAE introduces a categorical variable, $y$, which represents the various underlying Gaussian distributions to which each configuration will be (probabilistically) assigned.
As a consequence, the approach simultaneously performs a dimensionality reduction and clustering, while enabling direct control over the organization of configurations in the latent space.
We demonstrate the properties of this architecture by considering two model systems and molecular simulations of alanine dipeptide as well as a more challenging disordered peptide ensemble.
In the following, X $\in \mathbb{R}^{n}$ represents the $n$ dimensional input. The latent variable in the bottleneck is represented by $z \in \mathbb{R}^d, d \le n$.

\subsection{One-dimensional 4-well Potential}

\begin{figure}[htbp] 
\centering
	\begin{subfigure}[b]{0.45\textwidth}
	\centering
	\caption{Potential}	
	\label{fig:1d4w-potential}
	\includegraphics[height = 0.225\textheight]{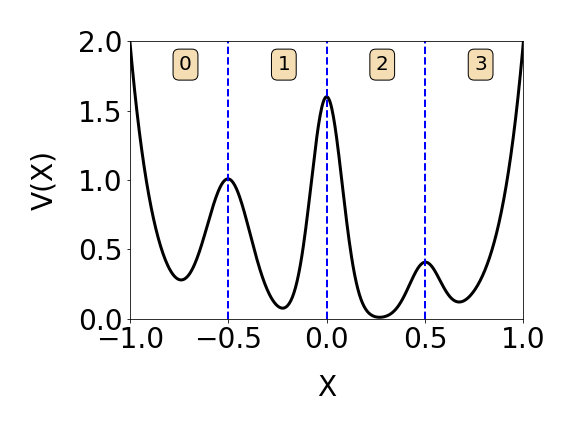}
	\end{subfigure}%
	~ 
	\begin{subfigure}[b]{0.45\textwidth}
	\centering
	\caption{Confusion matrix}
	\label{fig:1d4w-conf}
	\includegraphics[height = 0.225\textheight]{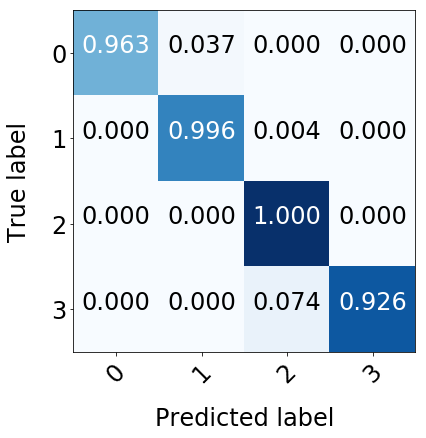}
	\end{subfigure}%
	\\
	\begin{subfigure}[b]{0.45\textwidth}
	\centering
	\caption{$z$ via the GMVAE} 
	\label{fig:1d4w-zhist}
	\includegraphics[height = 0.225\textheight]{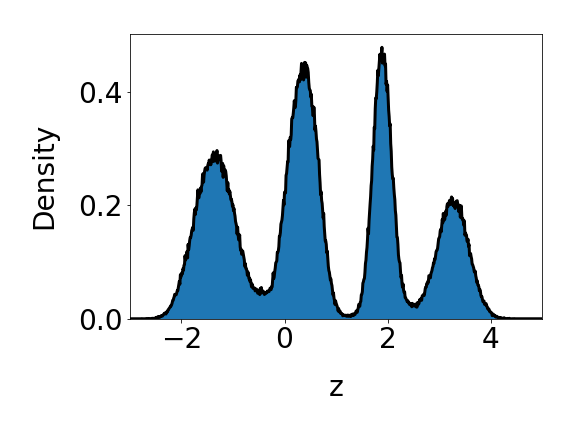}
	\end{subfigure}
        ~
    \begin{subfigure}[b]{0.45\textwidth}
    \caption{$z$ via the VAE}
    \label{fig:1d4w_vae}
    \includegraphics[height = 0.225\textheight]{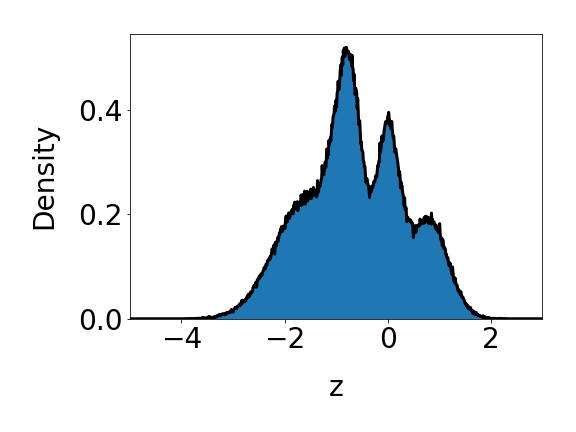}
    \end{subfigure}
\caption{(a) 1D 4-well potential with the true labels. (b) Confusion matrix constructed with the true labels shown in (a) and the predicted labels obtained via the GMVAE. Population size increases from light to dark blue. Normalized histograms of the 1D latent variable via the (c) GMVAE (d) VAE.}
\label{fig:1d4w_potential}
\end{figure}

We first consider a single particle in one-dimension interacting with a 4-well external potential, which has been previously employed for testing methods associated with constructing MSMs~\cite{schwantes2015modeling, chen2019nonlinear}.
Figure~\ref{fig:1d4w-potential} presents the potential, whose functional form and simulation details are given in Section~\ref{sec:app-1d4w}.
We employ a GMVAE with a latent space dimension of 1, which assesses the clustering performance of the architecture in the absence of any dimensionality reduction.
The GMVAE was trained with $k =$ 4 according to the parameters in Table~\ref{tab:hyper}.
Figure~\ref{fig:1d4w-conf} presents the confusion matrix of the resulting model, which quantifies the probability that the model assigns a predicted label (x-axis) given the true label (y-axis).
The true labels were determined using a coarse-grained representation of the system, where four metastable states are defined based on simple dividing surfaces, chosen as the maxima of the barriers between each potential well (dashed vertical lines in Figure~\ref{fig:1d4w-potential}).
The GMVAE assigns the state labels with 97$\%$ overall accuracy.
%

Figure~\ref{fig:1d4w-zhist} shows a normalized histogram of $z$ values.
Without dimensionality reduction, the GMVAE largely retains the description of the input space within the latent dimension. 
As a consequence, the decoder is able to quite accurately reconstruct the input from the latent variable (See Figure~\ref{fig:app-1d4w-reconst}).
This behavior is in stark contrast to traditional VAEs, which employ a Gaussian prior to represent the latent space distribution.
As a result, anti-clustering effects can arise, leading to highly overlapping clusters of data in the reduced space.
To demonstrate this effect, we constructed a traditional VAE for the present example.
Figure~\ref{fig:1d4w_vae} presents the corresponding normalized histogram of $z$ values.
In this case, even without a reduction in dimension, significant information is lost due to the constraint of the assumed prior distribution.

\begin{figure}[htbp] 	
\centering
	\begin{subfigure}[b]{0.45\textwidth}
	\centering
	\caption{Implied timescales}
	\label{fig:1d4w-its}
	\includegraphics[height = 0.225\textheight]{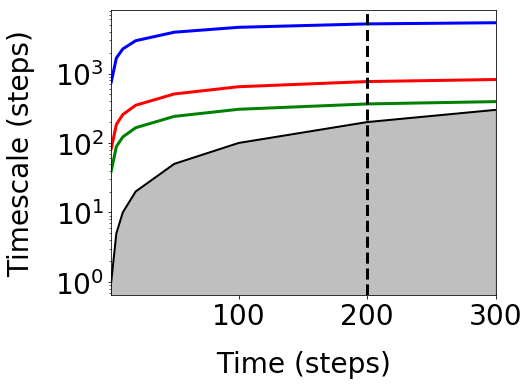}
	\end{subfigure}%
	~
	\begin{subfigure}[b]{0.45\textwidth}
	\centering
	\caption{Chapman-Kolmogorov test}
	\label{fig:1d4w-ck}
	\includegraphics[height = 0.225\textheight]{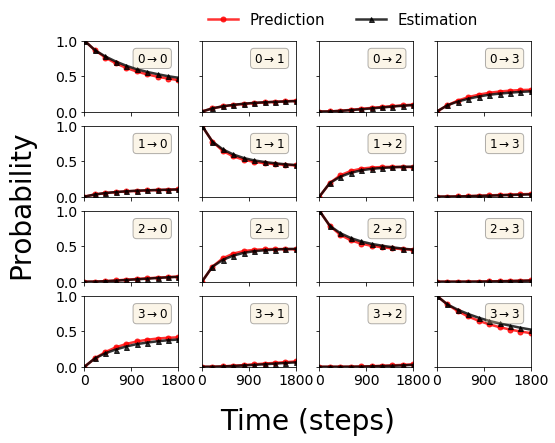}
	\end{subfigure}
\caption{Markovianity check of the kinetic model built for 1D 4-well potential system. The MSM was constructed directly using the cluster labels obtained from the GMVAE. (a) Implied timescale test. (b) Chapman-Kolmogorov test (at lag = 200 steps).}
\label{fig:1d4w_potential_msm}
\end{figure}

To further characterize the quality of the GMVAE clustering, we constructed an MSM from the trajectories of the predicted cluster IDs.
Figure~\ref{fig:1d4w-its} presents the standard implied timescale test, which assesses the convergence of the characteristic timescales with increasing lag time parameter $\tau$.
Convergence indicates that the simulation dynamics, within the discrete-state representation, can be described within a Markovian approximation.
The grey area indicates timescales that cannot be resolved by the model, since they are faster than the chosen lag time.
From the test, the MSM with $\tau = 200$ was chosen for further analysis. The accuracy of this model was assessed with the Chapman-Kolmogorov test, which compares the simulated and predicted decay of probability from a chosen set of metastable states.
Figure~\ref{fig:1d4w-ck} demonstrates that the predicted ``cluster dynamics'' accurately represent the long timescale kinetic properties of the underlying simulation trajectory.

\subsection{M{{\"u}}ller-Brown Potential}
\begin{figure}[htbp] 	
	\begin{subfigure}[b]{0.32\textwidth}
	\centering
	\caption{Free-energy landscape}
	\label{fig:mb-fe}
	\includegraphics[height = 40mm]{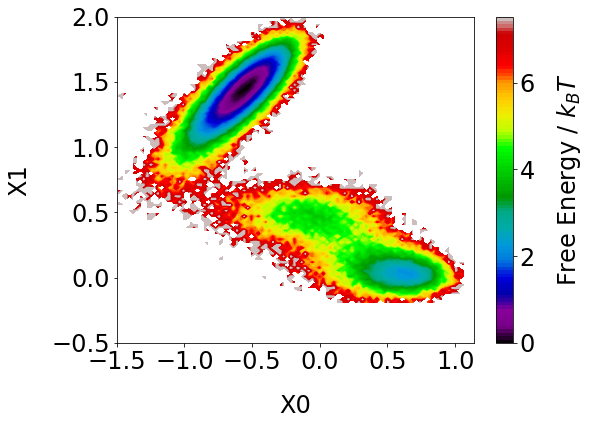}
	\end{subfigure}%
	~
	\begin{subfigure}[b]{0.32\textwidth}
	\centering
	\caption{GMVAE clusters}
	\label{fig:mb-clusters}
	\includegraphics[height = 40mm]{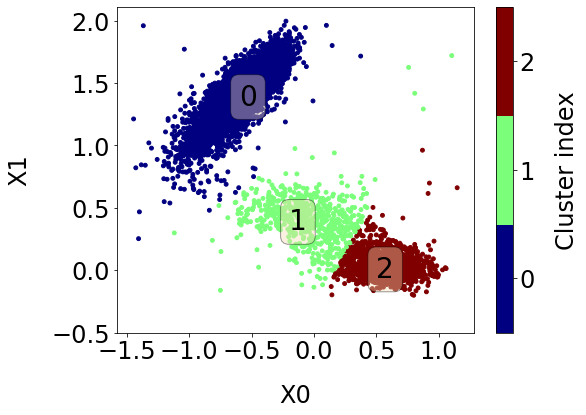}
	\end{subfigure}%
	~
	\begin{subfigure}[b]{0.32\textwidth}
	\centering
	\caption{Confusion matrix}
	\label{fig:mb-conf}
	\includegraphics[height = 40mm]{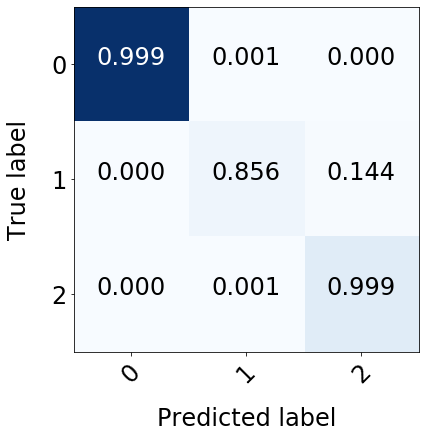}
	\end{subfigure}%
	\\
    \begin{subfigure}[b]{0.45\textwidth}
    \centering
    \caption{$z$ via the GMVAE}
    \label{fig:mb-zhist}
    \includegraphics[height = 0.225\textheight]{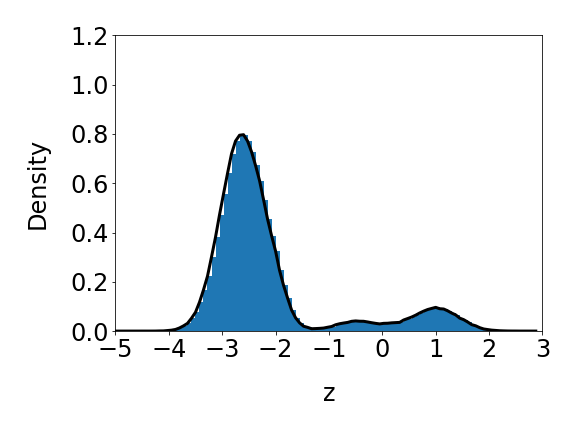}
    \end{subfigure}
    ~
	\begin{subfigure}[b]{0.45\textwidth}
	\caption{z via the VAE}
	\label{fig:mb-vae}
	\includegraphics[height = 0.225\textheight]{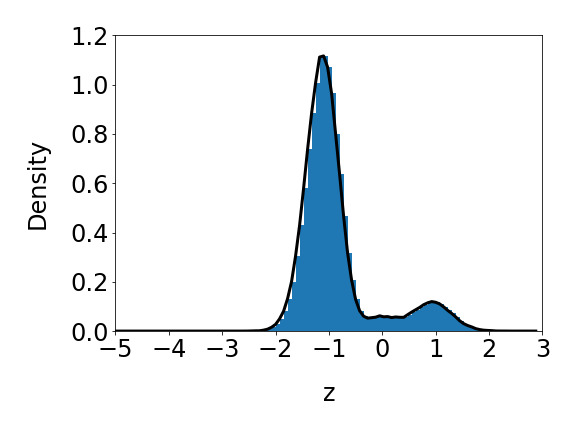}
	\end{subfigure}
\caption{2D M{{\"u}}ller-Brown potential. (a) Free-energy landscape. (b) Clusters obtained from the GMVAE. (c) Confusion matrix with the true labels determined with linear dividing surfaces (Figure~\ref{fig:mb-labels}) and predicted labels obtained via the GMVAE. Population size increases from light to dark blue. Normalized histograms of the 1D latent variable via the (d) GMVAE (e) VAE. }
\label{mb-projections}
\end{figure}

To assess both the dimensionality reduction and clustering performance of the GMVAE approach, we next consider a single Brownian particle in two dimensions interacting with an external M{{\"u}}ller-Brown potential. 
The trajectory data was generated 
as the procedure suggested in~\cite{hernandez2018variational} with the standard parameters~\cite{muller1979location} (see Section~\ref{sec:app-mb} for more details). 
As depicted in Figure~\ref{fig:mb-fe}, the resulting FEL contains two deep minima along with a less stable intermediate state.
We employ a GMVAE that is trained with a latent space dimension of 1 and with $k =$ 5, according to the parameters in Table~\ref{tab:hyper}.

Despite employing $k =$ 5, the resulting GMVAE model identified only 3 states with non-zero membership probabilities.
Thus, somewhat remarkably, the GMVAE architecture was able to identify the inherent organization of the input data in the high-dimensional space, independent of the hyperparameter $k$. Figure~\ref{fig:mb-clusters} shows the identified clusters. 
We define the true cluster labels in this case using linear dividing surfaces, as shown in Figure~\ref{fig:mb-labels}. Figure~\ref{fig:mb-conf} presents the confusion matrix from the GMVAE model with respect to these defined labels.
Although it appears that there are errors in assigning state 1, this error is sensitively dependent on the precise definition of the true label dividing surfaces.
Moreover, the overall classification accuracy is actually $99\%$, since state 1 corresponds to a very rarely sampled intermediate state.
The model also demonstrates relatively high reconstruction accuracy (See Figures~\ref{fig:mb-reconstX1} and ~\ref{fig:mb-reconstX2}).
Figures~\ref{fig:mb-zhist} and \ref{fig:mb-vae} present normalized histograms of $z$ values obtained from the GMVAE model and a traditional VAE model trained on the same data, respectively.
The low-dimensional representations obtained from the GMVAE clearly demonstrate a better separation of metastable states.
Additionally, the ability of the GMVAE to learn a nonlinear manifold is demonstrated in Figure~\ref{fig:mb-manifold}, with respect to the linear embedding obtained using time-lagged independent component analysis (TICA).

\begin{figure}[htbp] 	
\centering
	\begin{subfigure}[b]{0.45\textwidth}
	\centering
	\caption{Implied timescales}
	\label{fig:mb-its}
	\includegraphics[height = 0.225\textheight]{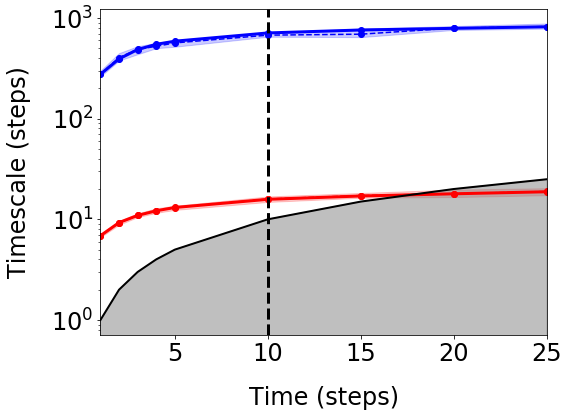}
	\end{subfigure}%
	~
	\begin{subfigure}[b]{0.45\textwidth}
	\centering
	\caption{Chapman-Kolmogorov test}
	\label{fig:mb-ck}
	\includegraphics[height = 0.225\textheight]{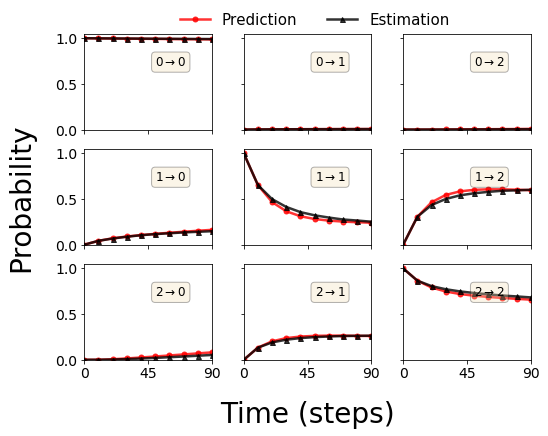}
	\end{subfigure}
\caption{Markovianity check of the MSM built for 2D M{{\"u}}ller-Brown potential via the GMVAE. (a) Implied timescales. (b) Chapman-Kolmogorov test (at lag=10 steps)}
\label{fig:mb_msm}
\end{figure}

To further characterize the quality of the GMVAE clustering, we again constructed an MSM from the trajectories of the predicted cluster IDs.
The implied timescale test (Figure~\ref{fig:mb-its}) shows two dominant processes.
The MSM with $\tau = 10$ was chosen for further analysis. Figure~\ref{fig:mb-ck}) presents the Chapman-Kolmogorov test, which further verifies the accuracy of the GMVAE embedding.

\subsection{Alanine Dipeptide}
\begin{figure}[htbp] 	
\centering
	\begin{subfigure}[b]{0.45\textwidth}
	\centering
	\caption{Free-energy landscape along the $(\phi$, $\psi)$ angles.}
	\label{fig:ala2-fe-rama}
	\includegraphics[height = 0.225\textheight]{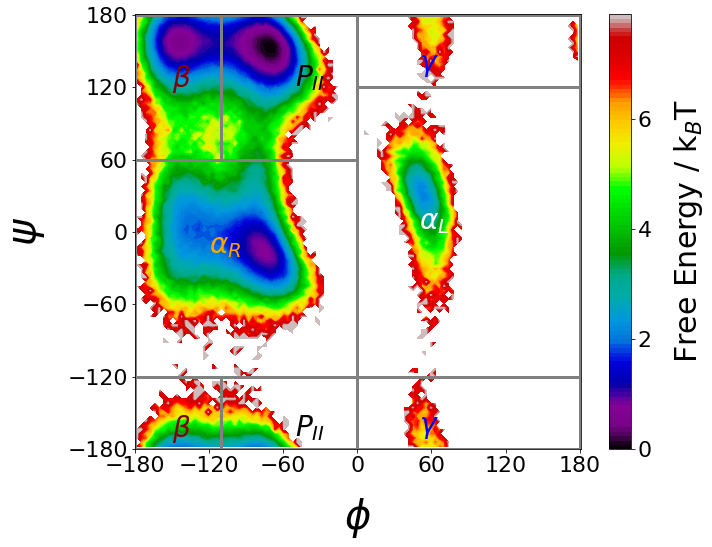}
	\end{subfigure}%
	~
	\begin{subfigure}[b]{0.45\textwidth}
	\centering
	\caption{GMVAE clusters}
	\label{fig:ala2-gmvae-rama}
	\includegraphics[height = 0.225\textheight]{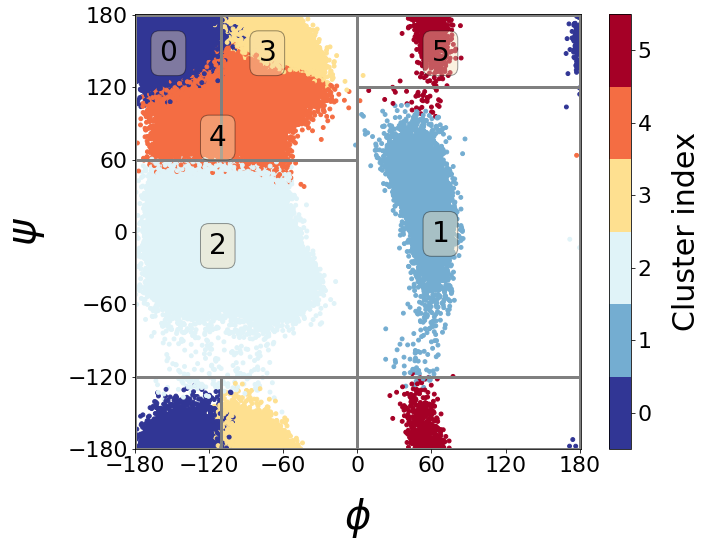}
	\end{subfigure}%
\caption{(a) Free energy landscape of alanine dipeptide, (b) GMVAE clusters on the Ramachandran plot.}
\label{fig:ala2_general}
\end{figure}

Alanine dipeptide is a representative model system for the characterization of conformational dynamics. 
Previous work~\cite{nuske2017markov, mardt2018vampnets, zhang2019deep, wehmeyer2018time, chen2019nonlinear} has shown that the $(\phi$, $\psi)$ backbone dihedral angles act as ideal collective variables for describing the metastable configurational basins and associated transition kinetics, making it an excellent system for testing the GMVAE framework within a more realistic molecular simulation context.
Since in general the optimal set of input features is unknown a priori, we use this example to test the ability of the GMVAE to identify the proper collective variables from a larger set of input features.
More specifically, we consider as input features both the normalized pairwise distances between heavy atoms as well as the $(\phi$, $\psi)$ dihedral angles (obtained from~\cite{Markovmodel}).
The pairwise distances were pre-processed using a kurtosis filter (with the threshold value of 0.03, see Figure~\ref{fig:ala2-features} for more detail), to reduce the input dimension by removing the low-variance features. 
The dihedral angles were pre-processed by applying $\sin$ and $\cos$ transformations in order to account for periodicity~\cite{Altis:2008}. 
Figure~\ref{fig:ala2-fe-rama} shows the FEL in the backbone dihedral angle space, with four labeled metastable basins corresponding to $\alpha_\textrm{R}$, $\alpha_\textrm{L}$, $\beta$, P$_{\textrm{II}}$, and $\gamma$ conformations~\cite{Chodera:2007}.
The gray lines are drawn for reference and do not represent any sort of optimal dividing surface.

\begin{figure}[htbp]
\centering
        \begin{subfigure}[b]{0.45\textwidth}
        \centering
        \caption{FEL via the GMVAE}
        \label{fig:ala2-gmvae-fe}
        \includegraphics[height = 0.225\textheight]{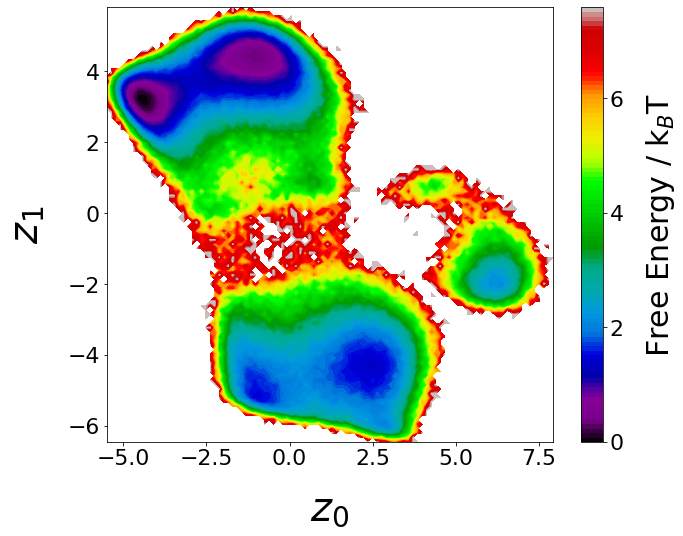}
        \end{subfigure}%
        ~
        \begin{subfigure}[b]{0.45\textwidth}
        \centering
        \caption{Clusters}
        \label{fig:ala2-gmvae-clusters}
        \includegraphics[height = 0.225\textheight]{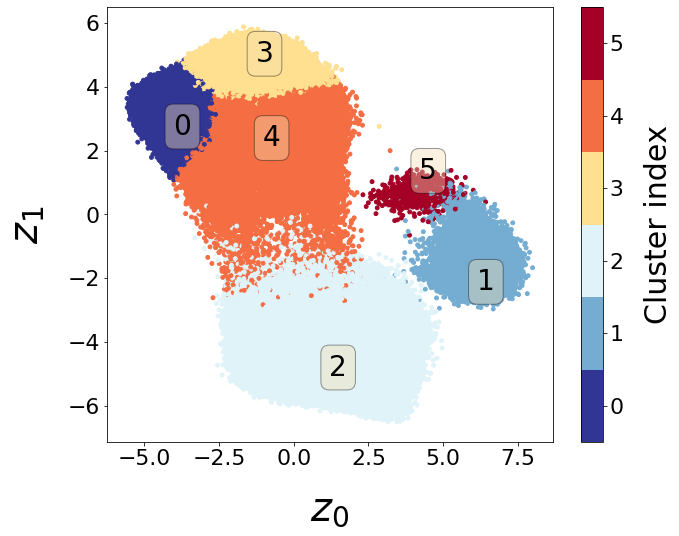}
        \end{subfigure}
\caption{(a) FEL obtained for the alanine dipeptide by the GMVAE. The GMVAE clusters on the (b) GMVAE landscape.}
\label{fig:ala2_gmvae}
\end{figure}

\begin{figure}[htbp] 	
\centering
	\begin{subfigure}[b]{0.45\textwidth}
	\centering
	\caption{Implied timescales}
	\label{fig:ala2-its}
	\includegraphics[height = 0.225\textheight]{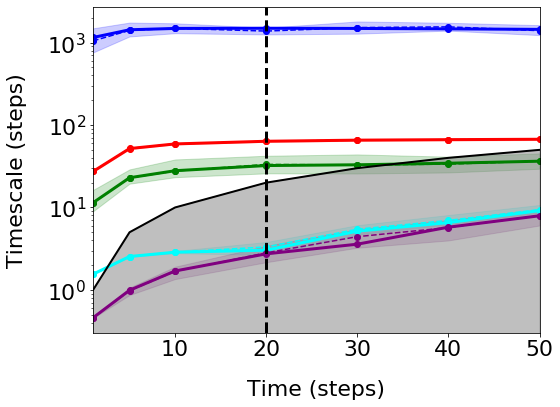}
	\end{subfigure}%
	~
	\begin{subfigure}[b]{0.45\textwidth}
	\centering
	\caption{Chapman-Kolmogorov test}
	\label{fig:ala2-ck}
	\includegraphics[height = 0.225\textheight]{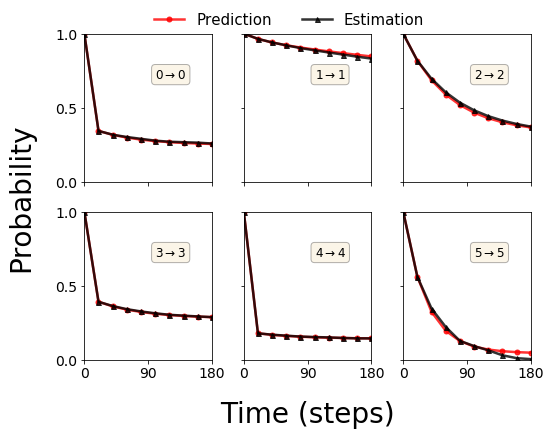}
	\end{subfigure}
\caption{Markovianity check of the MSM built for alanine dipeptide via the GMVAE. (a) Implied timescales. (b) Chapman-Kolmogorov test (at lag=20 steps).}
\label{fig:ala2_MSM}
\end{figure}

Figure~\ref{fig:ala2-gmvae-fe} presents the two-dimensional embedding found using the GMVAE, and Figure~\ref{fig:ala2-gmvae-clusters} shows the simultaneously-obtained 6 clusters (indexed from $0$ to $5$) as a part of the GMVAE algorithm.
The GMVAE again obtains a FEL that better separates clusters of conformations, relative to a standard VAE (Figure~\ref{fig:ala2-vae}).
The distribution of these clusters on the Ramachandran plot (Figure~\ref{fig:ala2-gmvae-rama}) already strongly indicates their suitability for a kinetic analysis.
The GMVAE clustering distinguishes all 5 of the metastable states, as well as a transition region between the $\alpha_\textrm{R}$ and $\beta$ states (cluster 4).
An MSM was again constructed from the coarse GMVAE cluster assignments.
The implied timescale and Chapman-Komolgorov tests are presented in Figure~\ref{fig:ala2_MSM}, demonstrating the accuracy of this kinetic model.

We found in this example that, unlike the toy systems, the clustering obtained using the GMVAE did not appear to be completely robust.
In particular, the precise clustering probabilities depend on the random effects of the training procedure (e.g., random weight initialization and the random shuffling of the input data).
This issue was most pronounced for the lowest populated state, whose probability differs from the other states by two orders of magnitude (Figure~\ref{fig:ala2-meta-hist}).
As a consequence, the $\gamma$ state was not always sufficiently separated from the $\alpha_\textrm{L}$ state, resulting in a loss of one of the resolved kinetic processes (although the accuracy of the MSM remained intact, see Figure~\ref{fig:ala2_MSM_np}).
Despite this issue, the obtained FEL appeared rather robust with respect to changes in the random factors during training.
We observed a much more robust clustering for all other applications considered.

\subsection{$\AQA$ Peptide - I} \label{sec:aaqaa-I}

As a more challenging test, we consider simulation trajectories of the capped helix forming peptide AC-($\textrm{AAQAA}$)$_3$-NH2, which is a representative system for investigating helix-coil transitions. 
We employ a coarse-grained model~\cite{Rudzinski:2018b}, which describes the dominant attractive interactions, e.g., hydrogen bonding and effective hydrophobic interactions between side chains, with simple potentials between the $C_\alpha$ and $C_\beta$ atoms.
These interactions are the minimum required to sample the proper range of structures, (i.e., helix, coil, and hairpin-like).
This model also represents excluded volume effects in near-atomic detail, which was demonstrated to be important for accurately characterizing the helix-coil kinetics.
Here we employ a parametrization of the model that most accurately reproduces the experimental cooperativity of the helix-coil transition for $\AQA$.
As a result, hairpin-like structures appear to have relatively low metastability (similar to the intermediate state in the M{{\"u}}ller-Brown example, and the $\gamma$ state in alanine dipeptide), as we discuss further below.
The model and simulation protocol are discussed further in the Supporting Information, and also in~\cite{Rudzinski:2018b,rudzinski2018role}.
The considered simulation trajectories correspond to a disordered ensemble of peptide configurations, representing a stringent test for dimensionality and clustering methods~\cite{Kukharenko:2016}.

%
%
%

Similar to alanine dipeptide, the set of $\sin$ and $\cos$ augmented $(\phi, \psi)$ dihedral angles along the peptide backbone were used as conformational descriptors. 
Thus, the input dimension is $60$ for the 15-residue $\AQA$ peptide. 
We chose to consider only a latent space dimension of 2, given that the ultimate goal of dimensionality reduction is often to reduce the high-dimensional description to something that is easily visualizable.
Unlike the simple model systems above, the number of clusters, $k$, is completely unclear a priori.
In fact, we expect that this ensemble to have a hierarchical structure, such that differing number of clusters may be appropriate depending on the chosen level of resolution.
While we initially considered the GMVAE with varying number of clusters, we found that the number of ``non-zero clusters'' (i.e., clusters with a significant probability of configuration assignment) was extremely insensitive to this choice, as discussed below. The GMVAE was trained according to the parameters in Table~\ref{tab:hyper}.
Also in contrast to the previous examples, there is no definitive reference kinetic model with corresponding known metastable states.
Instead, the analysis below assesses the GMVAE embedding and clustering (in terms of both statics and kinetics) with respect to the landscapes obtained using a standard VAE and also following the standard MSM workflow (i.e., TICA~\cite{molgedey1994separation,Perez-Hernandez:2013}, see Section~\ref{sec:msm} for more details).

\begin{figure}[htbp]
\centering
	\begin{subfigure}[b]{0.45\textwidth}
	\centering
    \caption{FEL via the GMVAE}
    \label{fig:aaqaa-I-FEL_gmvae}
    \includegraphics[height = 0.225\textheight]{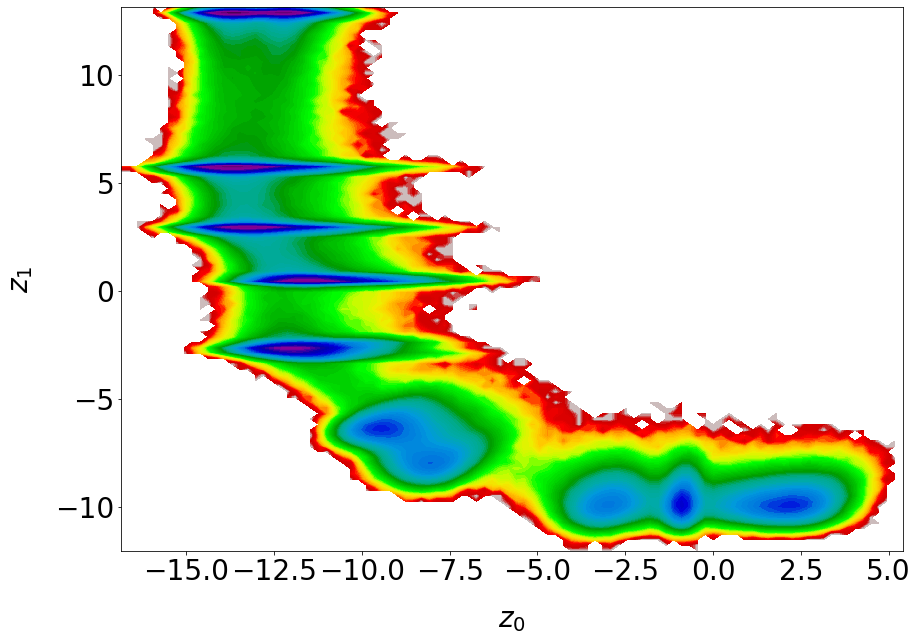}
    \end{subfigure}%
    ~
    \begin{subfigure}[b]{0.45\textwidth}
    \centering
    \caption{FEL via the VAE}
    \label{fig:aaqaa-I-FEL_vae}
    \includegraphics[height = 0.225\textheight]{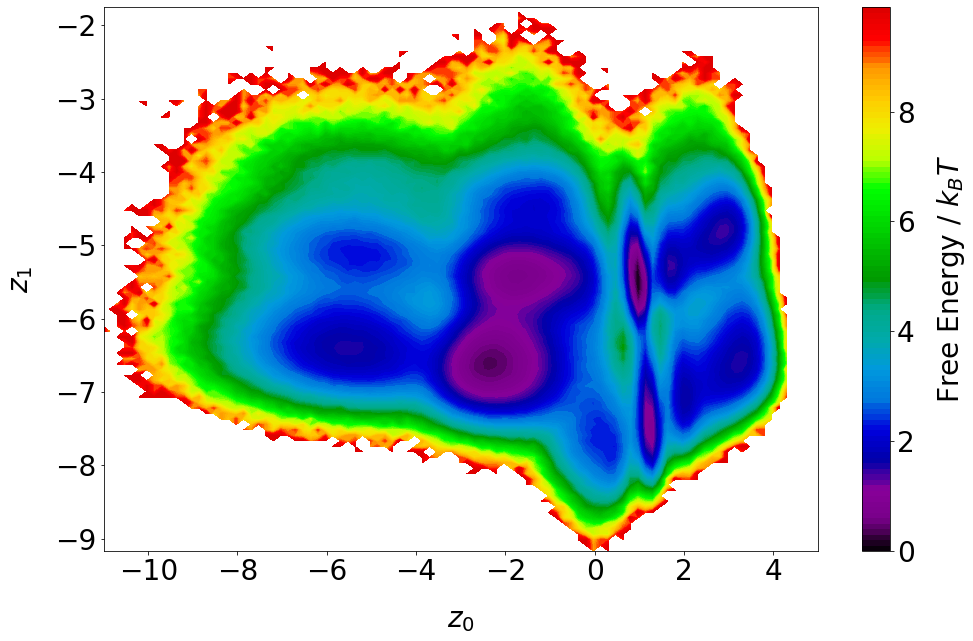}
    \end{subfigure}%
\caption{Free-energy landscapes of $\AQA$ - I peptide obtained by (a) the GMVAE, and (b) the VAE.}
\label{fig:aaqaa-I-FELs}
\end{figure}

\begin{figure}[htbp] 	
\centering
	\begin{subfigure}[b]{1\textwidth}
	\centering
	\caption{Clusters}
	\label{fig:aaqaa-I-clusters}	
	\includegraphics[width=0.5\textwidth]{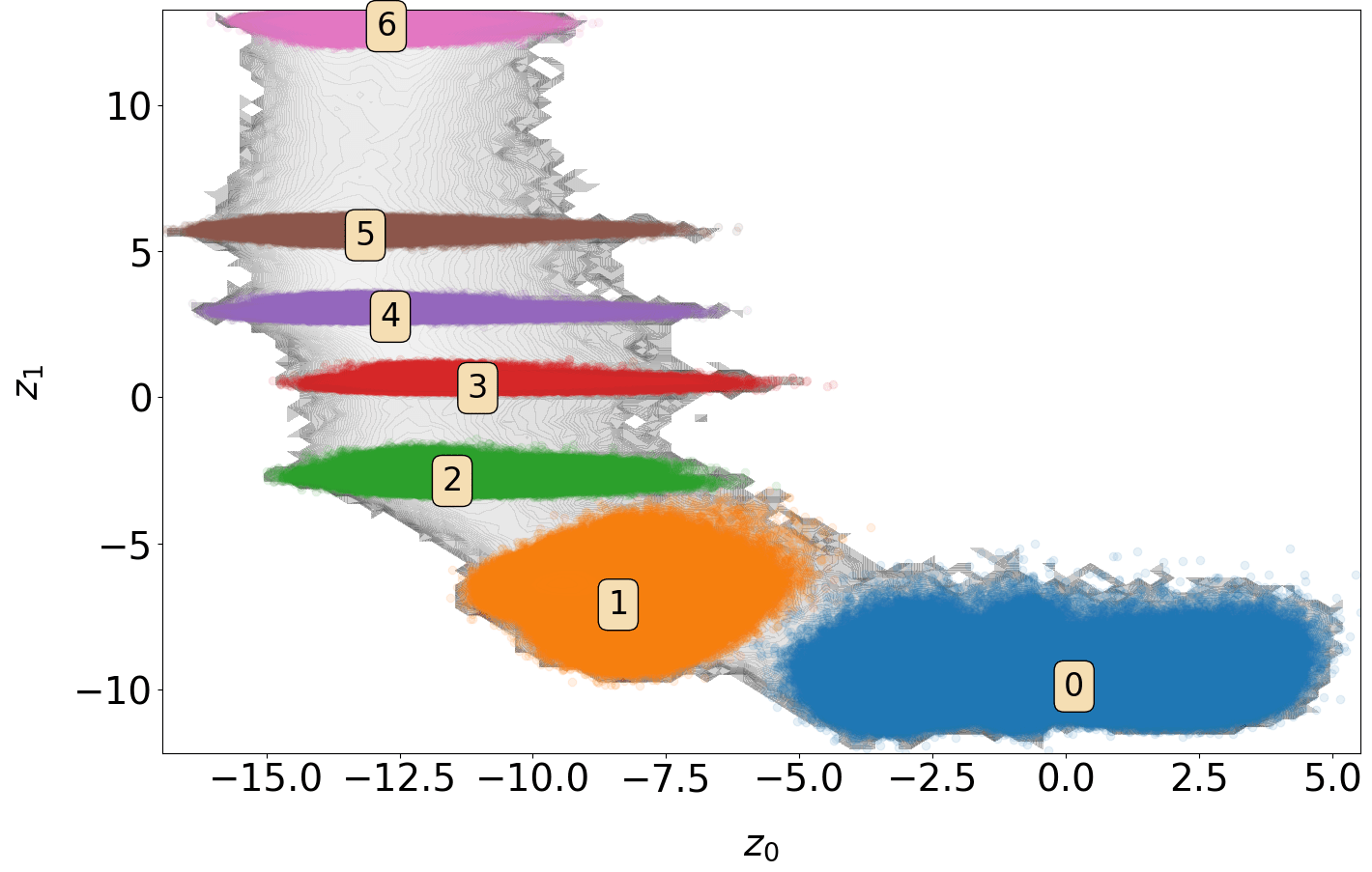}
	\end{subfigure}%
	\\
	\begin{subfigure}[b]{1\textwidth}
	\centering
	\caption{Secondary structures}
	\label{fig:aaqaa-I-structures}	
	\includegraphics[width=0.5\textwidth]{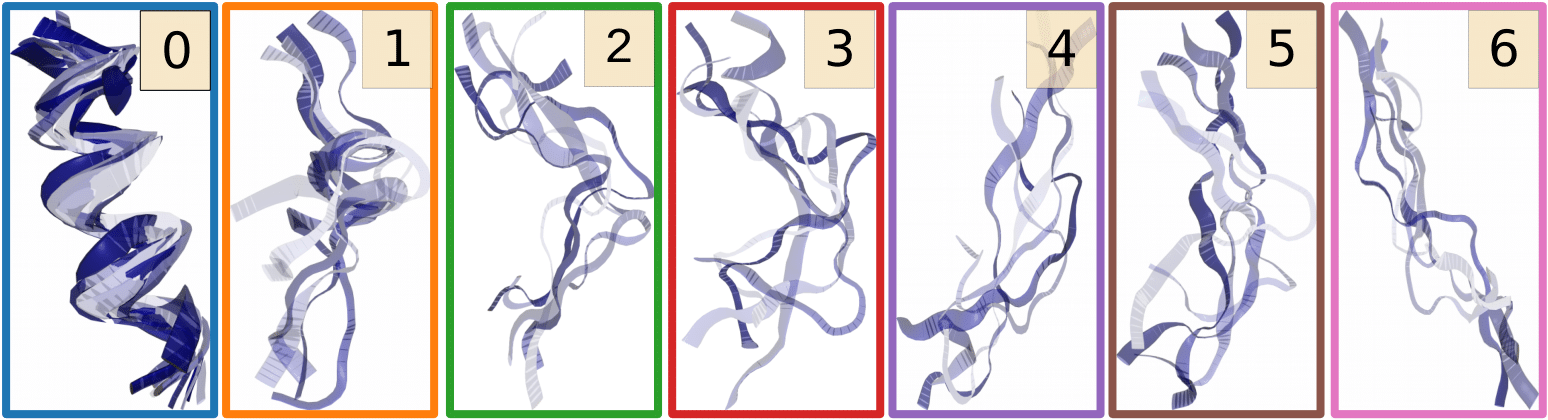}
	\end{subfigure}%
\caption{(a) The clusters obtained for the $\AQA$ peptide - I by the GMVAE after thresholding. (b) The secondary structures closest to the cluster centers.}
\label{fig:aaqaa-I-cluster_strs}
\end{figure}

Panels (a) and (b) of Figure~\ref{fig:aaqaa-I-FELs} show the FELs obtained using the GMVAE and the traditional VAE, respectively. 
As in the model systems, the GMVAE method results in a latent space description with highly separated clusters, while the traditional VAE yields more overlapping states.
The two-dimensional TICA landscape (Figure~\ref{fig:aaqaa-I-tica-landscape}) also separates a number of clearly distinct states, although there are large diffuse regions with relatively low free-energy values.
The clusters obtained via the GMVAE are shown in Figure~\ref{fig:aaqaa-I-clusters}.
Despite employing $k=10$ and obtaining a landscape that appears to have approximately 10 distinct basins, only 7 states (labeled ${0, 1, \dots 6}$) were assigned non-zero membership probabilities (see Figure~\ref{fig:aaqaa-I-qy_prob}).
Since standard metrics for analyzing peptide configurations do not yield a clear organization of the ensemble into a small number of metastable states, the distribution of these quantities are expected to be highly overlapping, even for a good clustering of the input data.
Thus, to more easily visualize the characteristics of the GMVAE clusters, we applied a thresholding scheme, which removes configurations without a membership probability greater than 0.95 (see Section~\ref{sec:threshold} for details and Figure~\ref{fig:aaqaa-I-cluster_pop} for cluster populations).
Figure~\ref{fig:aaqaa-I-structures} shows 5 representative structures closest to the cluster centers.
We stress that these images are intended to give the reader a rough idea of the types of structures contained in each cluster, but do not characterize the variance of structures within the clusters.
This is a disordered ensemble and each cluster necessarily contains a diversity of structures. 
Nevertheless, Figure~\ref{fig:aaqaa-I-structures} indicates that the GMVAE successfully distinguishes between distinct secondary structures within the simulation data.
%

\begin{figure}[htbp] 
\centering
	\begin{subfigure}[b]{0.45\textwidth}
	\centering
	\caption{$\langle f_h \rangle$}
	\label{fig:aaqaa-I-fh}
	\includegraphics[width = 1\textwidth]{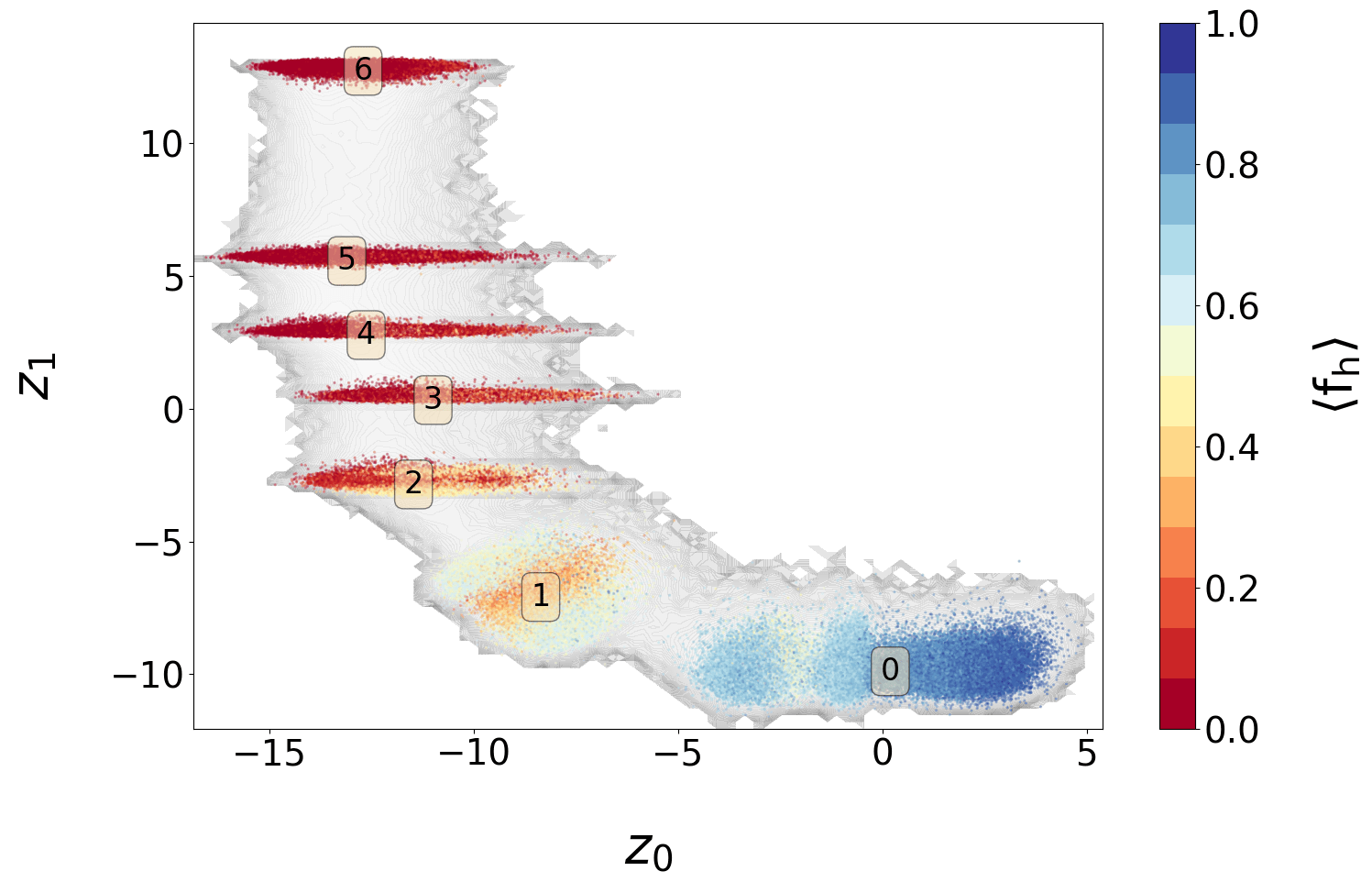}
	\end{subfigure}%
	~
	\begin{subfigure}[b]{0.45\textwidth}
	\centering
	\caption{dRMSD$_\textrm{hel}$}
	\label{fig:aaqaa-I-proj_drmsd_hel}
	\includegraphics[width = 1\textwidth]{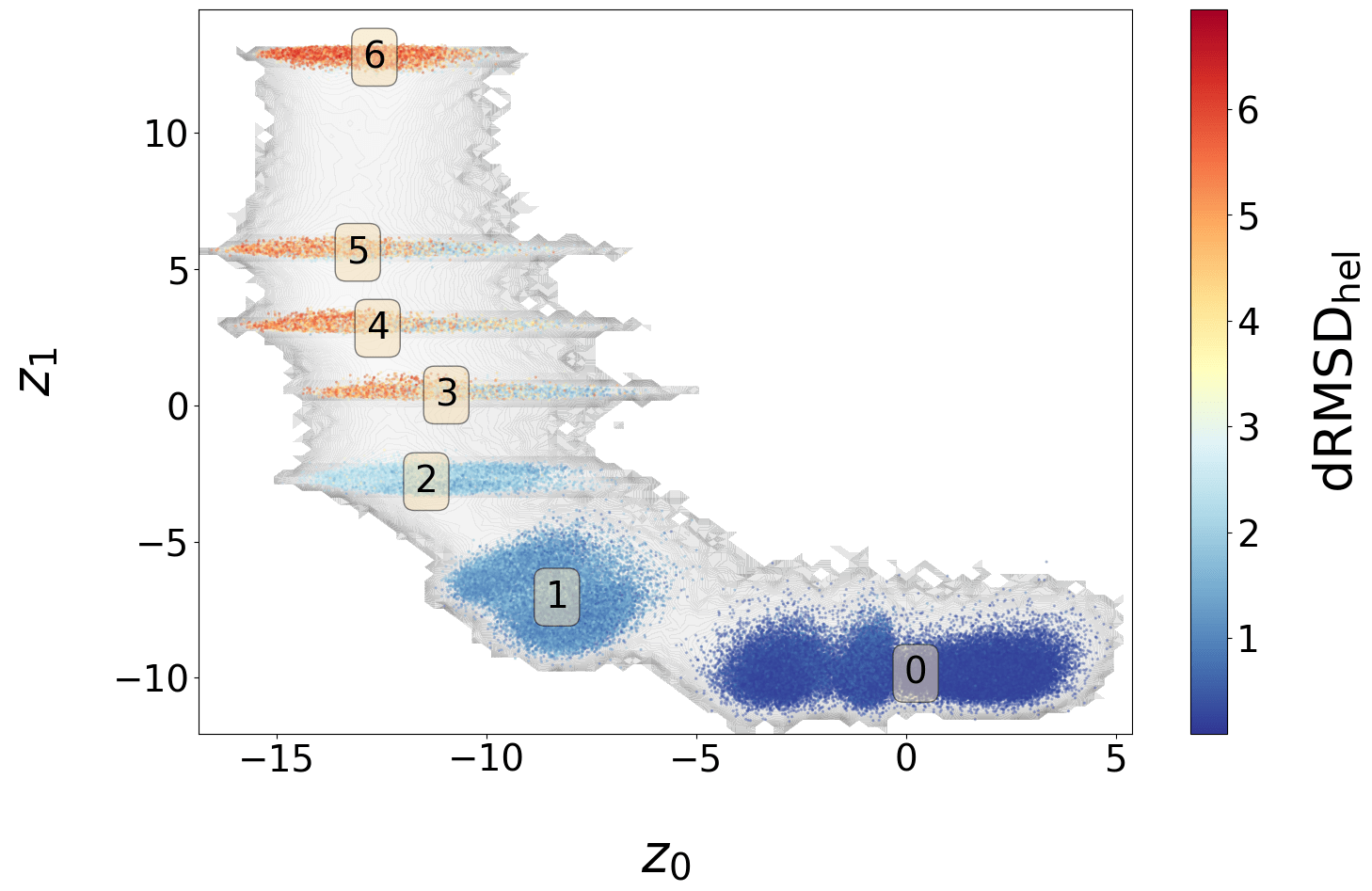}
	\end{subfigure}%
\caption{$\AQA$ - I. (a) Average helical fraction, $\langle f_h \rangle$, analysis.  Colors represents the $\langle f_h \rangle$ values of the corresponding projected data obtained from the GMVAE. (b) dRMSD$_\textrm{hel}$ analysis.}
\label{fig:aaqaa-I-fh-drmsd_hel}
\end{figure}

To characterize the structural properties of the clusters quantitatively, we calculated the distribution of the average fraction of helical segments, $\langle f_h \rangle$. Figure~\ref{fig:aaqaa-I-fh} presents a heat map of $\langle f_h \rangle$ in the latent space. High $\langle f_h \rangle$ values (represented by blue) indicate the presence of helix and helix-like structures, whereas the lower values point to either hairpin- or coil-like secondary structures. 
There is an apparent trend of decreasing average helical content from the lower-right to upper-left regions of the latent space (i.e., from cluster $0$ to $6$).
The VAE and TICA landscapes demonstrate similar trends (Figures~\ref{fig:aaqaa-I-vae-fh} and \ref{fig:aaqaa-I-tica-fh}, respectively), although the VAE does not characterize partially-helical structures as clearly as the GMVAE.
Figure~\ref{fig:aaqaa-I-hist_fh} presents the intra-cluster distributions of $\langle f_h \rangle$, which can be used to assess the quality of the clustering (relative to an alternative clustering).
We expect that an optimal clustering will result in tight, unimodal $\langle f_h \rangle$ distributions.
The GMVAE clustering yields seemingly good distributions for the most and least helical clusters, while the partially-helical clusters appear broader and somewhat bimodal.
For comparison, we consider three alternative clusterings obtained by performing a $k$-means clustering on a given landscape followed by the PCCA+ dynamical coarse-graining method~\cite{roblitz2013fuzzy} to define a set of metastable states (see Section~\ref{sec:msm} for more details): (i) an alternative clustering of the GMVAE landscape (Figure~\ref{fig:aaqaa-I-gmvae_landsc-msm}), (ii) a clustering on the VAE landscape (Figure~\ref{fig:aaqaa-I-vae-msm}), and (iii) a clustering on the TICA landscape (Figure~\ref{fig:aaqaa-I-tica-msm}).
The alternative clustering scheme on the GMVAE landscape, (i), does not improve the intra-cluster distributions of $\langle f_h \rangle$, demonstrating that the GMVAE clustering is reasonable, given the GMVAE embedding.
Similar results were obtained from the VAE clustering, with slightly broader distributions for the most and least helical states.
The TICA clustering resulted in somewhat improved distributions, in the sense that they appear to be mostly unimodal, although some of the distributions appear to be slightly broader.

%

Figure~\ref{fig:aaqaa-I-fh-drmsd_hel}(b) shows the dRMSD$_\textrm{hel}$ values of the projections, where the helicity increases as the dRMSD$_\textrm{hel}$ values decreases. These results are in agreement with the $\langle f_h \rangle$ analysis: as the cluster index increases from $0$ to $6$, the conformations tend to be more extended. 
%
%
The Supporting information (Figures~\ref{fig:aaqaa-I-drmsds} and \ref{fig:aaqaa-I-projs_others}) contains additional characterization of the static properties of the clusters, which further validate the GMVAE embedding and clustering as a reasonable partitioning of the conformational landscape.

\begin{figure}[htbp] 	
\centering
	\includegraphics[width = 0.6\textwidth]{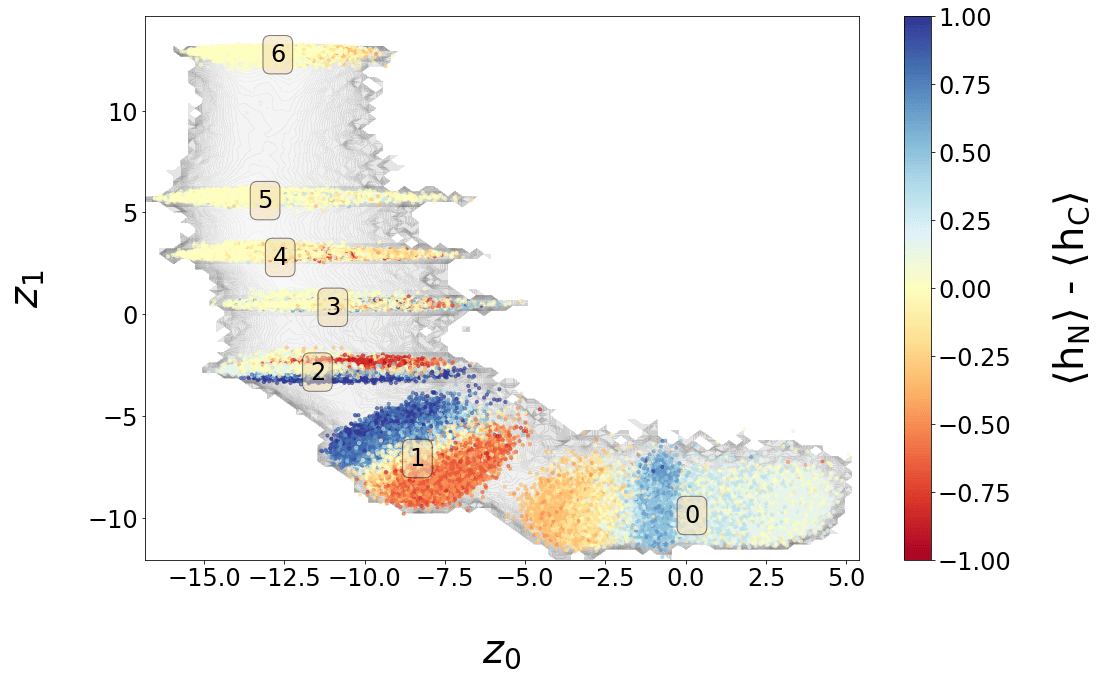}
\caption{Analysis of partially-helical conformations for $\AQA$ - I. Projections are colored according to $\langle h_N \rangle - \langle h_C \rangle$ values.}
\label{fig:aaqaa-I-dhNC}
\end{figure}

We also characterized the average fraction of helical segments on the N- and C-terminus sides of the peptide: $\langle h_N \rangle$ and $\langle h_C \rangle$, repectively (see Section~\ref{sec:peptide_analysis} for more details).
Figure~\ref{fig:aaqaa-I-dhNC} presents the difference of these quantities, $\langle h_N \rangle - \langle h_C \rangle$, plotted along the GMVAE embedding.
Positive values (represented by blue) indicate conformations that contain helical structure on the N-terminus side of the peptide without helical structure on the C-terminus side.
Conversely, negative values (represented by red) indicate conformations that contain helical structure on the C-terminus side of the peptide without helical structure on the N-terminus side.
Values close to zero correspond to either fully helical or non-helical structures. 
Although the GMVAE embedding and clustering separate the most distinct structures in the ensemble (coils and full-helicies), some of the clusters (0, 1, 2) encompass partially-helical conformations on both sides of the peptide (see also Figure~\ref{fig:aaqaa-I-end_fold}).
This is not ideal since kinetic barriers within a cluster will negatively impact the accuracy of a kinetic characterization at the cluster level.
However, it appears that this issue may have more to do with the clustering than the embedding itself, since blue- and red-labeled structures appear to be reasonably separated on the landscape.

Similar to the other examples above, we also constructed an MSM directly from the discretized trajectories of GMVAE cluster indices.
Although thresholding was applied in the results presented here (practically similar to coring methods for constructing kinetic models~\cite{Jain:2012}), we found that this procedure had negligible effect on the accuracy of the resulting MSM.
As shown in Figure~\ref{fig:aaqaa-I-MSM}, the MSM constructed from the GMVAE clustering displayed significant errors in describing, e.g., the decay of probability out of the helix state.
Perhaps this is not so surprising, since coarse-grained MSMs are often only used as a qualitative analysis tool, while higher-resolution kinetic models that characterize configuration space with many microstates are used for quantitative reproduction of simulation kinetics.
Thus, to more carefully assess the GMVAE embedding and to more easily compare to the VAE and TICA results, we constructed a higher-resolution MSM by performing $k$-means to define microstates on the landscape (Figure~\ref{fig:aaqaa-I-gmvae_landsc-msm}).
Although the resulting model demonstrates improved accuracy according to the Chapman-Kolmogorov test, the probability decay out of the metastable states occurs on a fast time scale relative to the chosen lag time.
This may be indicative of poorly defined dividing surfaces between metastable states.
The kinetic models constructed from the VAE and TICA landscapes (Figures~\ref{fig:aaqaa-I-vae-msm} and \ref{fig:aaqaa-I-tica-msm}, respectively) demonstrate similar quickly decaying probabilities.
Although coring procedures could be applied to attempt to fix this problem, it indicates that there are fundamental limitations of all of these landscapes in terms of characterizing the long timescale simulation kinetics. 
There are several possible reasons for these difficulties, including (i) the limitation of our embeddings to two dimensions, (ii) the limitation of the chosen input features in characterizing kinetically-distinct structures, (iii) the presence of many low-lying barriers along the potential energy landscape of this disordered ensemble, and (iv) the poor sampling of relatively rare transitions to the full helix conformation.
Further investigation of this issue is beyond the scope of this initial study of the performance of the GMVAE, and is left for future work.

\subsection{$\AQA$ Peptide - II} \label{sec:aaqaa-II}
\begin{figure}[htbp] 	
\centering
	\begin{subfigure}[b]{0.45\textwidth}
	\centering
	\caption{FEL via the GMVAE}
	\label{fig:aaqaa-II-fel}
    \includegraphics[width = 1\textwidth]{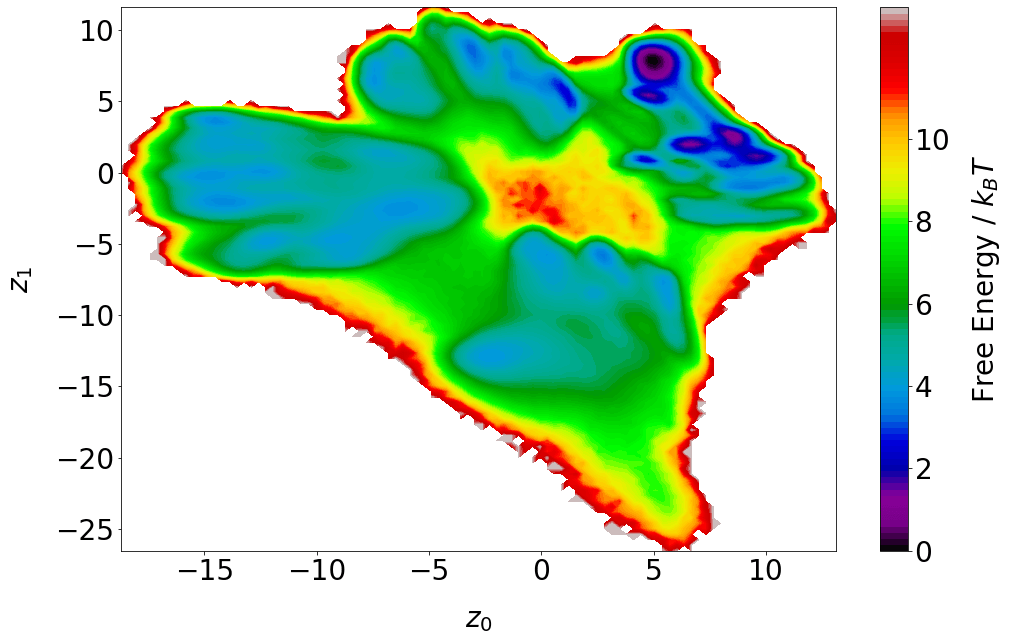}
	\end{subfigure}
	~
	\begin{subfigure}[b]{0.45\textwidth}
	\centering
	\caption{Clusters}
	\label{fig:aaqaa-II-clusters}	
	\includegraphics[width = 1\textwidth]{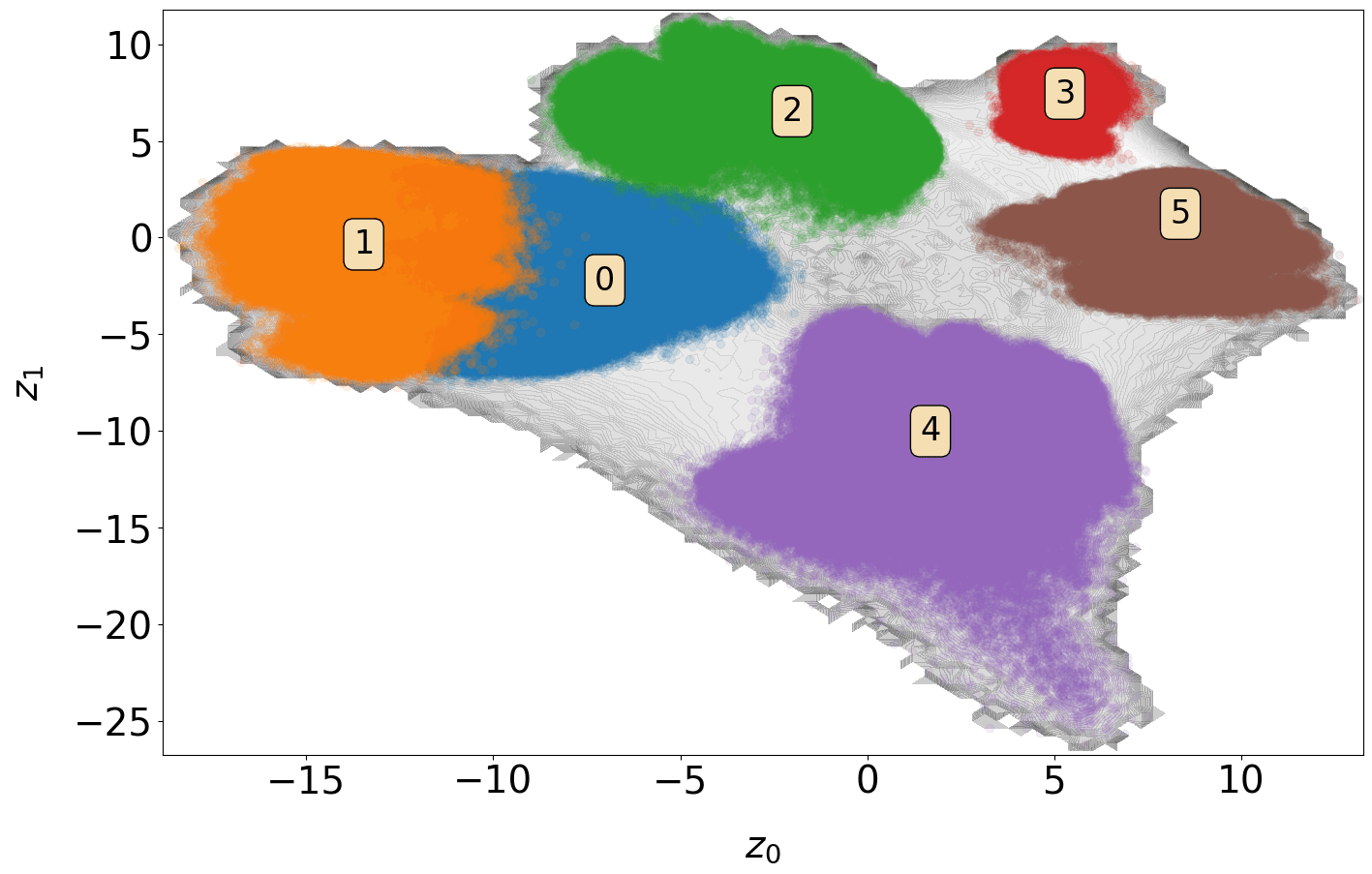}
	\end{subfigure}%
	\\
	\begin{subfigure}[b]{1\textwidth}
	\centering
	\caption{Secondary structures}
	\label{fig:aaqaa-II-structures}	
	\includegraphics[width=0.5\textwidth]{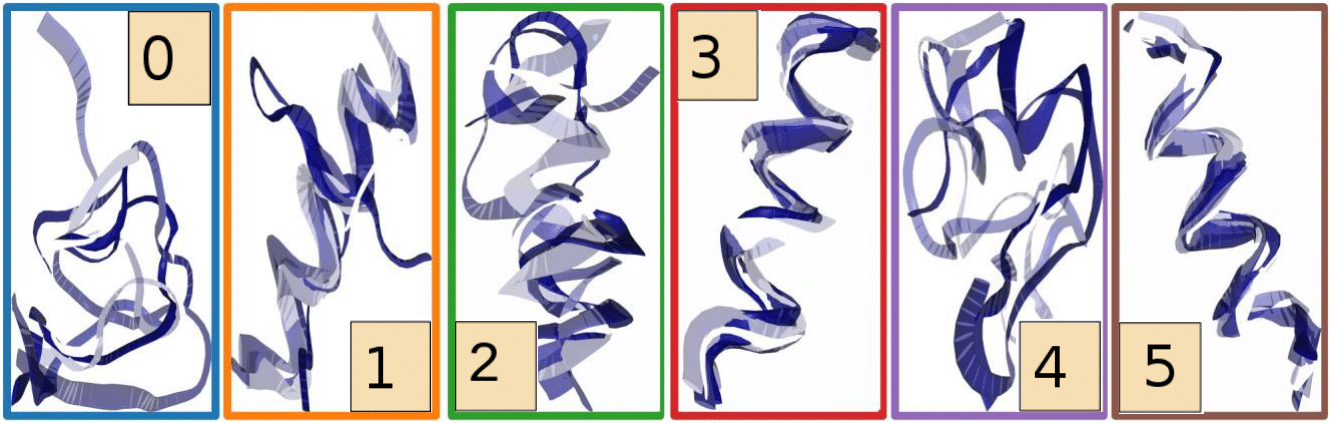}
	\end{subfigure}%
\caption{The GMVAE results for $\AQA$ peptide - II. (a) Free-energy landscape. (b) The clusters obtained after thresholding. (c) The secondary structures closest to the cluster centers.}
\label{fig:aaqaa-II-fel-c-str}
\end{figure}

\begin{figure}[htbp] 	
\centering
	\begin{subfigure}[b]{0.45\textwidth}
	\centering
	\caption{$\langle f_h \rangle$}
	\label{fig:aaqaa-II-proj_fh}
	\includegraphics[width = 1\textwidth]{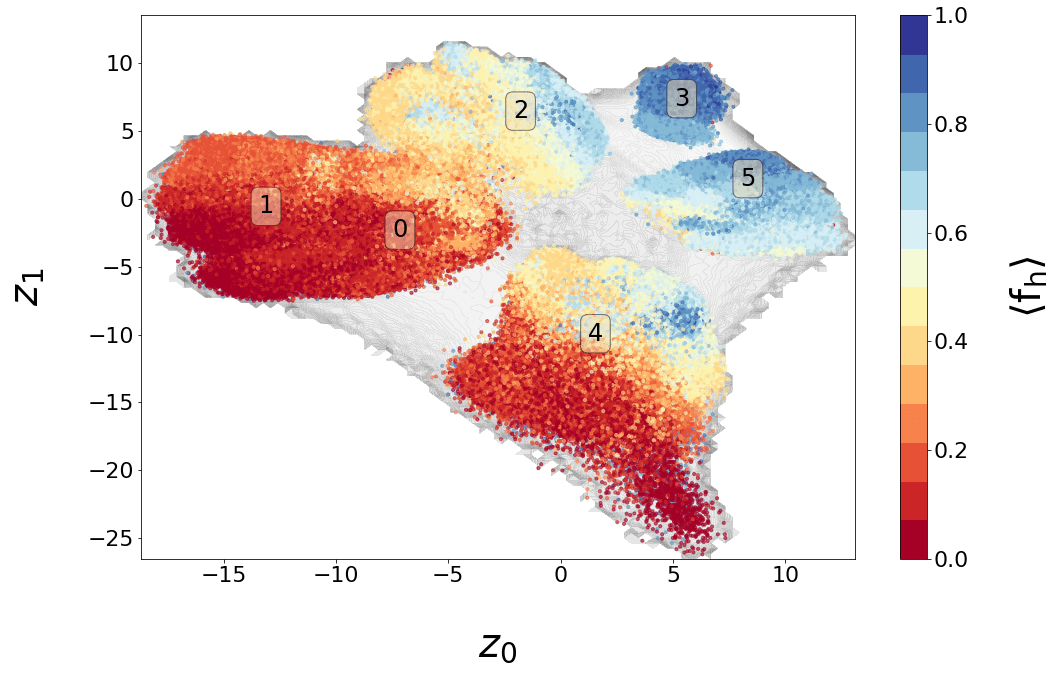}
	\end{subfigure} 
	~
	\begin{subfigure}[b]{0.45\textwidth}
	\centering
	\caption{dRMSD$_\textrm{hel}$}
	\label{fig:aaqaa-II-proj_drmsd}
	\includegraphics[width = 1\textwidth]{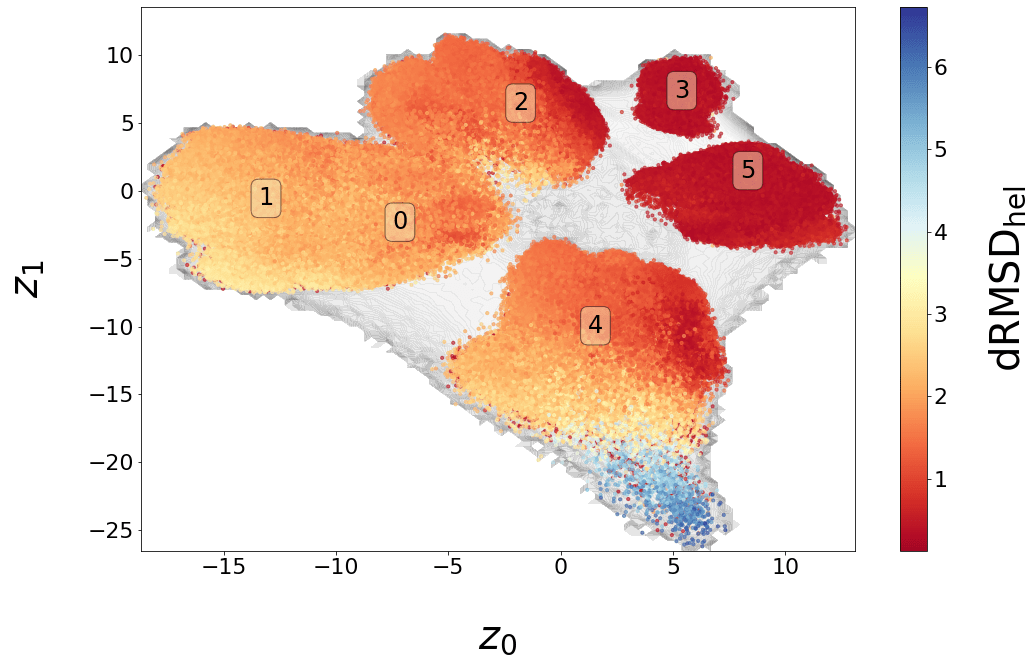}
	\end{subfigure} 
\caption{Projections for the $\AQA$ peptide - II. (a) $\langle f_h \rangle$, (b) dRMSD$_\textrm{hel}$.}
\label{fig:aaqaa_II-fh_drmsd}
\end{figure}

To investigate the impact of the low sampling of helical structures on the GMVAE embedding, as in the $\AQA$ - I simulations presented above, we also considered a second set of simulations which primarily samples helical- and hairpin-like structures, while only rarely sampling fully-coil structures.
(Please see the Supporting Information for more details about the differences between the two sets of simulations). In addition to the dihedral angles, normalized pairwise distances between residues that are more than $3$ residues apart were included as input features. Figure~\ref{fig:aaqaa-II-fel-c-str} presents the obtained GMVAE FEL (panel (a)), the corresponding clustering of 6 metastable states (panel (b)), and overlays of five structures that are closest to the cluster centers (panel (c)). 
The GMVAE embedding demonstrates significant separation of metastable states, relative to the landscape obtained with a standard VAE (Figure~\ref{fig:aaqaa-II-vae-fel}).
Similar to the previous ensemble ($\AQA$ - I), Figure~\ref{fig:aaqaa_II-fh_drmsd} shows the separation of structures according to $\langle f_h \rangle$ (panel (a)), and dRMSD$_\textrm{hel}$ (panel (b)). 
The VAE and TICA landscapes demonstrate similar trends (Figures~\ref{fig:aaqaa-II-vae-landscape} and \ref{fig:aaqaa-II-tica-landscape}, respectively).
The intra-cluster $\langle f_h \rangle$ distributions are shown in~Figure~\ref{fig:aaqaa-II-hist_fh}. 
The majority of the fully-helical structures are in cluster 3 and 5, while clusters 0, 1, 2 and 4 contain hairpin-like structures as well as partial helicies.
The coil structures are gathered in the bottom-most part of the landscape (in cluster 4), though not separated as a distinct cluster by the GMVAE.
The distributions are broader and less unimodal than those determined from the previous set of simulations, although these can be somewhat improved with the alternative clustering scheme on the GMVAE landscape (Figure~\ref{fig:aaqaa-II-gmvae_landsc-hist_fh}).
Similar results are also obtained from the VAE and TICA landscapes (Figures~\ref{fig:aaqaa-II-vae-hist_fh} and \ref{fig:aaqaa-II-tica-hist_fh}, respectively).

\begin{figure}[htbp] 	
\centering
	\begin{subfigure}[b]{0.4\textwidth}
	\centering
	\label{fig:aaqaa-II-dhNC}
	\includegraphics[height = 0.25\textheight, width=\textwidth]{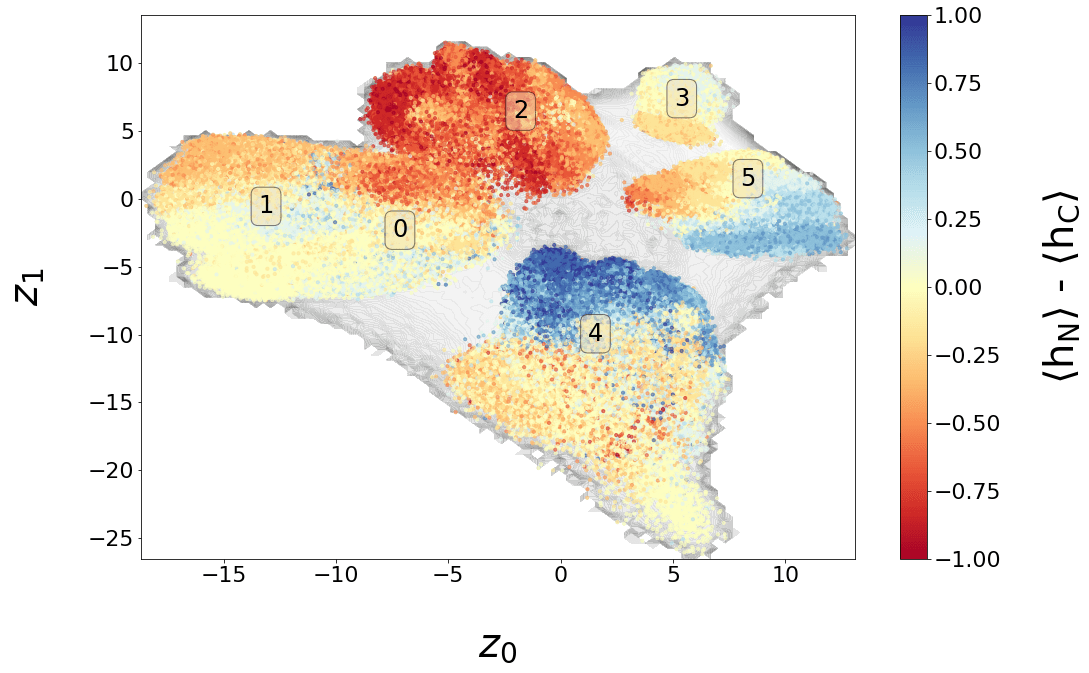}
	\end{subfigure} 
	~
	\begin{subfigure}[b]{0.56\textwidth}
	\centering
	\label{fig:aaqaa-II-end_fold}
	\includegraphics[height = 0.27\textheight, width=\textwidth]{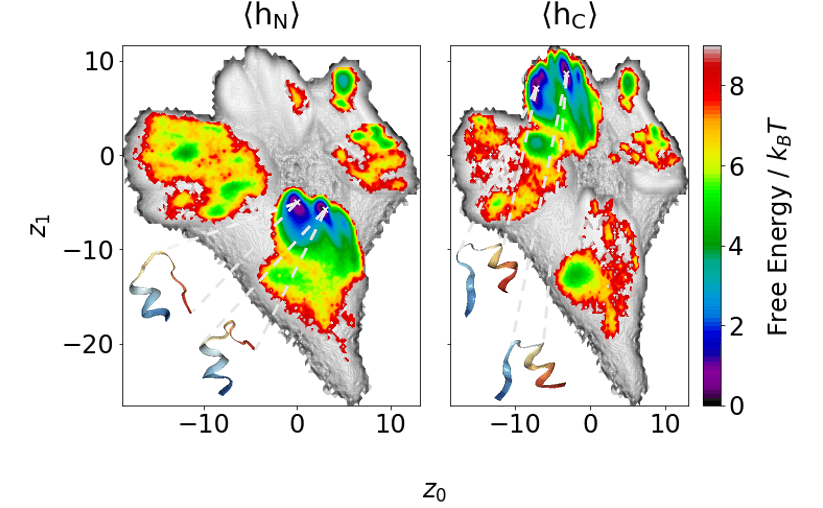}
	\end{subfigure}
\caption{The N- and C-terminus end folding analysis for the $\AQA$ peptide - II.
(Left) The difference in the average values of the two-end foldings, $\langle h_N \rangle - \langle h_C \rangle$. (Right) Distribution of the N- (on the left, $\langle h_N \rangle \geq 0.8$) and C-end (on the right, $\langle h_N \rangle \leq - 0.8$.)}
\label{fig:aaqaa-II-foldings}
\end{figure}

Figure~\ref{fig:aaqaa-II-foldings} presents the characterization of the N- and C-terminus, partially-helical conformations.
In contrast to the $\AQA$ - I embedding, the GMVAE embedding and clustering for $\AQA$ - II more clearly separates the distinct types of structures.
It appears that this difference may be due to the increased sampling of helical structures in $\AQA$ - II, although the inclusion of pairwise distances as additional input features may also have played a role.
N- and C-terminus partially-helical structures are mostly located in clusters 4 and 2, respectively, while both types of structures can be found to a lesser extent in cluster 5.
Although the VAE and TICA landscapes also appear to largely distinguish between distinct partially-helical structures (Figures~\ref{fig:aaqaa-II-vae-landscape} and \ref{fig:aaqaa-II-tica-landscape}, resepectively), the GMVAE landscape provides a significantly better clustering of these two distinct sets of conformations.
%

Despite the improved description of partially-helical structures, the MSM constructed directly from the GMVAE clustering for $\AQA$ - II displayed similar discrepancies to the model built for $\AQA$ - I (Figure~\ref{fig:aaqaa-II-MSM}).
Moreover, the high-resolution MSMs constructed from the GMVAE, VAE, and TICA landscapes (Figures~\ref{fig:aaqaa-II-gmvae_landsc-msm}, \ref{fig:aaqaa-II-vae-msm}, and \ref{fig:aaqaa-II-tica-msm}, respectively) displayed very fast decay of probability out of the identified metastable states, as in the $\AQA$ - I example.

\section{Discussion and Conclusions} \label{sec:conclusion}

Variational autoencoders are quickly making an impact in
the field of molecular simulations due to the inherent focus of
the architecture on retaining the essential features of the system.
Control over the \emph{topology} of the latent space can increase the
performance and interpretability of these methods by making a direct
connection to the physics of the system through our physical intuition: 
an ideal free-energy landscape characterizes basins that are well-separated
by the largest barriers along the higher-dimensional potential energy 
landscape.
To explicitly enforce such features, we propose a Gaussian mixture model 
as the prior distribution in the latent space.  

The performance of the Gaussian mixture variational autoencoder
(GMVAE) was illustrated on two standard toy-model systems and on the standard benchmark alanine dipeptide,  as well as on a
challenging 15-residue-long disordered peptide. For each example, the
GMVAE circumvents the aggregation of points in the latent space
characteristic of traditional variational autoencoders. Instead, samples that are
structurally distinct are clearly separated, leading to a latent space
that displays apparent metastable basins and barriers.
The GMVAE introduces a categorical variable that probabilistically
assigns samples to a set of underlying clusters, each of which is
Gaussian distributed. Thus, the approach combines the commonly
distinct tasks of dimensionality reduction and clustering into a
unified framework. In the absence of dimensionality reduction, the
GMVAE retains the characteristics of the system within the latent
space, while providing an accurate assignment between clusters.
Remarkably, in the case of limited dimensionality reduction, the GMVAE
identifies the inherent clustering of the input data, insensitive to
the cluster-number hyperparameter. 

Beyond statics, there have been several recent autoencoder
architectures aiming at the characterization of molecular kinetics. Several of these methods
directly incorporate kinetic information in the loss function, either
by reconstructing time-lagged samples or by approximating the
dynamical propagator~\cite{mardt2018vampnets, lusch2018deep,
wehmeyer2018time, hernandez2018variational, wang2019past,
chen2019capabilities}. The interpretability of the latent space is
becoming a feature of increasing interest: Hern{\'a}ndez \emph{et al.}
recently proposed an approach for identifying the most important input
features for determining the one-dimensional latent-space representation
within a time-lagged VAE framework~\cite{hernandez2018variational},
while Wang \emph{et al.} relied on a linear encoder to interpret the
relevant coordinates of interest~\cite{wang2019past}. 
Here, we argue that incorporating physical constraints into the architecture
helps to construct an interpretable model for the kinetics, even when kinetic information is not
used for learning the representation. 
The GMVAE architecture attempts to better mimic the shape of an ideal free-energy landscape within the 
latent space.
In particular, the presence of barriers that separate metastable clusters determines the relevant kinetic 
properties through the separation of timescales between
intra- and inter-basin transitions.

We report extremely encouraging results for constructing kinetic models
from representations learned from static information alone. 
For the two toy models and for alanine dipeptide, the resulting Markov state models demonstrate 
excellent properties, as monitored by the implied timescale and Chapman-Kolmogorov (CK) tests. 
The disordered ensemble of the $\AQA$ peptide proves more challenging:
the CK test shows discrepancies for the decay of probability out of the longest-lived metastable states.
Although higher-resolution MSMs constructed directly from the GMVAE landscape demonstrated an improved description of the simulation kinetics, these models were unable to resolve the longest timescale processes.
Comparisons of two distinct peptide ensembles clarified the role that sampling can play in distinguishing distinct partially-helical structures on the GMVAE landscape.
It remains unclear to what extent the restriction of our embeddings to two dimensions or the choice of input features prevented the GMVAE (as well as the more standard methods considered) from better describing the simulation kinetics.
Moreover, the presence of many low-lying barriers along the potential energy landscape of this disordered ensemble may cause fundamental challenges in obtaining a clear few-metastable-state characterization of the conformational landscape.
Thus, we propose that, in conjunction with simpler test systems that clearly assess a method's performance, such examples are important for significant advancements in data-driven characterizations of molecular simulation trajectories.


While we defer a more detailed investigation of these issues for future work, we highlight the promising performance of the GMVAE demonstrated through our results.
First, in the context of static equilibrium properties, the incorporation of the Gaussian mixture model as a prior distribution on the latent space closely links our physical intuition about ideal free-energy landscapes, resulting in an inherently more interpretable latent space.
Secondly, our results show encouraging performance when constructing kinetic models from the learned representations---an aspect that is entirely absent in the loss function, representing an independent validation of the procedure.

\section{Acknowledgments}  \label{sec:ack}
The authors thank Kiran H. Kanekal and Omar Valsson for critical reading of the manuscript. 
JFR is grateful to the BiGmax consortium and participants of the {\it BiGmax Big Data Summer School} for insightful discussions.
YBV acknowledges foreign collaborative research study support by The Scientific and Technological Research Council of Turkey, T{{\"U}}B{{\.{I}}}TAK- B{{\.{I}}}DEB, under the 2214-A programme. TB acknowledges financial support by the Emmy Noether program of the
Deutsche Forschungsgemeinschaft (DFG) and the long program Machine
Learning for Physics and the Physics of Learning at the Institute for
Pure and Applied Mathematics (IPAM).

\bibliographystyle{unsrt}
\bibliography{gmvae_paper,references_MPIP,references_PSU}
\clearpage


\setcounter{equation}{0}
\setcounter{figure}{0}
\setcounter{table}{0}
\setcounter{page}{1}
\setcounter{section}{0}

\newcommand{\Go}{{G\={o}}}
\newcommand{\Ca}{{\rm C}_{\alpha}}
\newcommand{\Cb}{{\rm C}_{\beta}}
\newcommand{\tausb}{\mathcal{T}^{\rm S}}
\newcommand{\epsNC}{\epsilon_{\rm nc}}
\newcommand{\epsDB}{\epsilon_{\rm db}}
\newcommand{\epsHP}{\epsilon_{\rm hp}}

\renewcommand{\thefigure}{S\arabic{figure}}
\renewcommand\thesection{S.\Roman{section}}
\renewcommand\thesubsection{\thesection.\arabic{subsection}}
\renewcommand{\thepage}{S.\arabic{page}}

\begin{center}
  \textbf{\large Supporting Information for: \\
Interpretable embeddings from molecular simulations using Gaussian mixture variational autoencoders}\\[.2cm]
  Yasemin Bozkurt Varolg{{\"u}}ne{{\c{s}}},$^{1,2,*}$ Tristan Bereau,$^{1}$ and Joseph F. Rudzinski$^1$\\[.1cm]
  {\itshape ${}^1$Max Planck Institute for Polymer Research, Mainz 55128, Germany\\
  ${}^2$Department of Electrical \& Electronics Engineering, Koc University, Sariyer, Istanbul 34450, Turkey}\\
  ${}^*$ bozkurty@mpip-mainz.mpg.de\\
\end{center}

%

%
\section{Overview}
This document presents additional results to the main text. X$' \in \mathbb{R}^n$ denotes the reconstructions. The sampling operation in the reconstructions (shown in the decoding part of Figure~\ref{fig:encoding-decoding}), corresponds to taking the means of the Gaussians for simplicity. 

\clearpage
\section{One dimensional 4-well potential} \label{sec:app-1d4w}
The trajectory data is obtained as suggested in~\cite{chen2019nonlinear}, and using the code provided in~\cite{1d4wdata}. $100 \times 100$ transition probability matrix is obtained among the equally-spaced 100 bins in the interval [-1, 1] as follows 
\begin{equation} \label{eq:1d4w_sim}
  P_{ij}=\begin{cases}
    C_i \exp{(-(V_i - V_j))}, & \text{if $|i - j| \leq 1$}\\
    0, & \text{otherwise} \,\, ,
  \end{cases}
\end{equation}
where $V_i$ and $V_j$ are the potential energies at the centers
of bins i and j, which are defined according to the potential of the form: $V (X) = 2( X^8 + 0.8 e^{-80X^2} + 0.2 e^{-80(X-0.5)^2} + 0.5 e^{-40(X+0.5)^2})$, and $C_i$ is the normalization factor. The system is initialized randomly, and propagated according to $P_{ij}$ $5 \times 10^6$ steps in time. 

\begin{figure}[htbp]
	\centering
	\includegraphics[width = 0.33\textwidth]{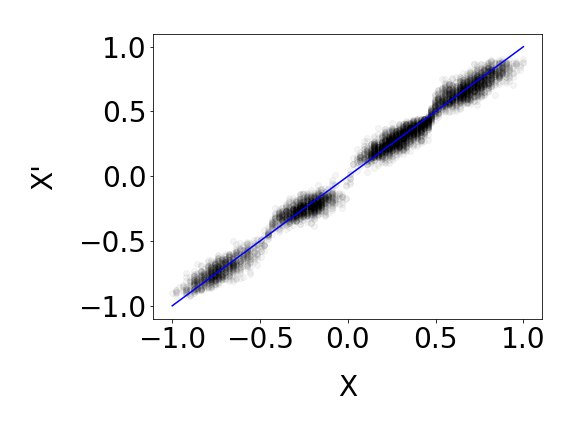}
	\caption{X vs X$'$ for one-dimensional 4-well potential. Reconstructions are obtained via the GMVAE.}
	\label{fig:app-1d4w-reconst}
\end{figure}

Figure~\ref{fig:app-1d4w-reconst} shows the reconstructions in a scatter plot. The $X=X'$ line shows the lossless reconstructions.  
\clearpage
\section{M{{\"u}}ller-Brown potential} \label{sec:app-mb}
The trajectory data is obtained as suggested in~\cite{hernandez2018variational}, and using the code provided in~\cite{mbdata}. Two dimensional potential energy is defined as:
\begin{equation}
V(X_0, X_1) = \sum_{j=0}^{3} A_j \exp[a_j (X_0 - x_{0, j})^2 + b_j (X_0 - x_{0, j})(X_1 - x_{1, j}) + c_j (X_1 - x_{1, j})^2 ] \,\, ,
\end{equation}
where $x = (X_0, X_1)$ is the two-dimensional coordinate, and $A, a, b, c, x_0$ and $y_0$ are the standard parameters~\cite{muller1979location} such that $A = (-200, -100, -170, 15)$, $a = (-1 , -1, 6.5, 0.7)$, $b = (0, 0, 11, 0.6)$, $c = (-10, -10, -6,5, -0.7)$, $x_0 = (1, 0, -0.5, -1)$, $x_1 = (0, 0.5, 1.5, 1)$.  The trajectory data is generated using 30 trajectories of 10000 steps simulated with Brownian dynamics: 
\begin{equation}
\frac{d x}{d t} = - \frac{\Delta V(x)}{k T} + \sqrt{2D}{R(t)} \,\, ,
\end{equation}
where $k T = 1.5 \times 10^4$ joules, and $D = 10^{-2}$ meters-squared per second, and $R(t)$ is a delta-correlated Gaussian process with zero
mean. 

\begin{figure}[htbp]
\centering
	\begin{subfigure}[b]{0.32\textwidth}
	\centering
	\caption{True labels}
	\label{fig:mb-labels}
	\includegraphics[width = 1\textwidth]{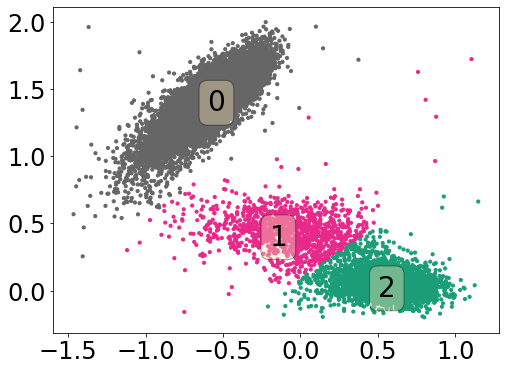}
	\end{subfigure}
	~ 
	\begin{subfigure}[b]{0.32\textwidth}
	\centering
	\caption{X0 vs X0$'$}
	\label{fig:mb-reconstX1}	
	\includegraphics[width = 1\textwidth]{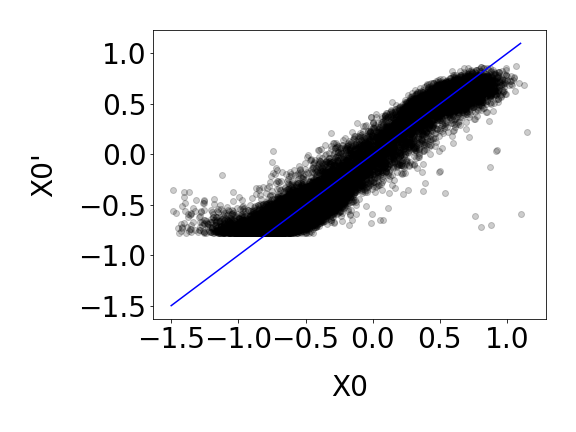}
	\end{subfigure}
	~	
	\begin{subfigure}[b]{0.32\textwidth}
	\centering
	\caption{X1 vs X1$'$}
	\label{fig:mb-reconstX2}
	\includegraphics[width = 1\textwidth]{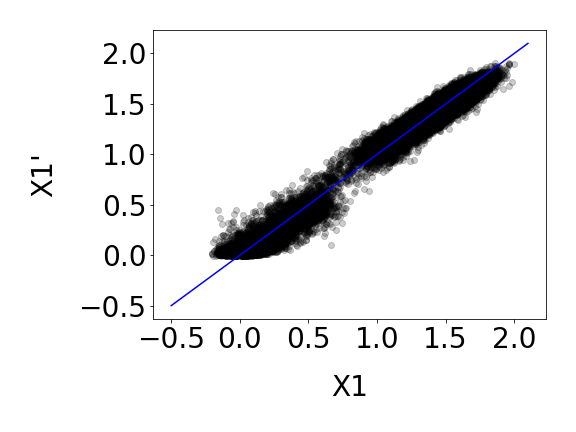}
	\end{subfigure}
\caption{True label definitions and X vs X$'$ for M{{\"u}}ller-Brown potential. Reconstructions  are obtained via the GMVAE. (a) True labels. (b) Reconstructions in the first dimension. (c) Reconstructions in the second dimension. }
\label{fig:app-mb}
\end{figure}

The true labels are defined as shown in Figure~\ref{fig:mb-labels}. Figures ~\ref{fig:mb-reconstX1} and ~\ref{fig:mb-reconstX2} show the reconstructions. 

\begin{figure}[htbp]
\centering
        \includegraphics[height = 0.25\textheight]{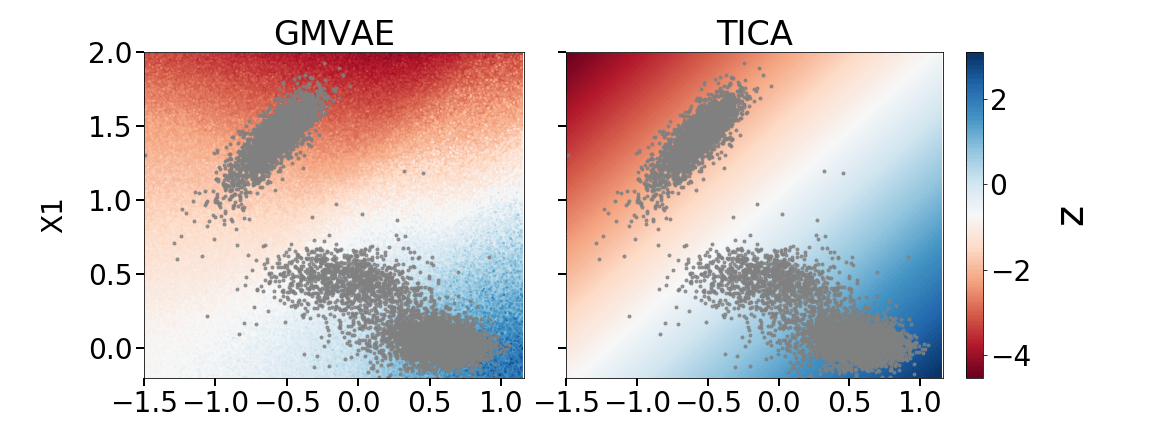}
        \caption{Projections via the GMVAE and TICA. The GMVAE learns the nonlinear dividing surface in the low-dimensional space. }
        \label{fig:mb-manifold}
\end{figure}

Figure~\ref{fig:mb-manifold} further demonstrates the ability of the GMVAE to learn a nonlinear manifold that separates the three distinct free-energy basins, compared with time-lagged independent component analysis (TICA), which can only find a linear separatrix for the basins. Figure~\ref{fig:mb-manifold}, showing the projections obtained with the GMVAE and TICA, was constructed following~\cite{hernandez2018variational}, with the colors indicating values of the latent variable while the gray dots correspond to trajectory data.

\clearpage
\section{Alanine dipeptide} \label{sec:app-ala2}
As the input features, dihedral angles and pairwise distances for heavy atoms that are provided in~\cite{Markovmodel} for three simulations of length 250 ns each are used. Dihedral angles are transformed to their $\sin / \cos$ representaions, and the pairwise distances whose variance are low are removed from the feature set (using kurtosis function from scipy.stats library~\cite{2019arXiv190710121V}, with threshold value of 0.03). 
\begin{figure}[htbp]
\centering
	\begin{subfigure}[b]{0.48\textwidth}
	\centering
	\caption{Dihedral angles}
	\label{fig:ala2-features-dih}
	\includegraphics[width = 1\textwidth]{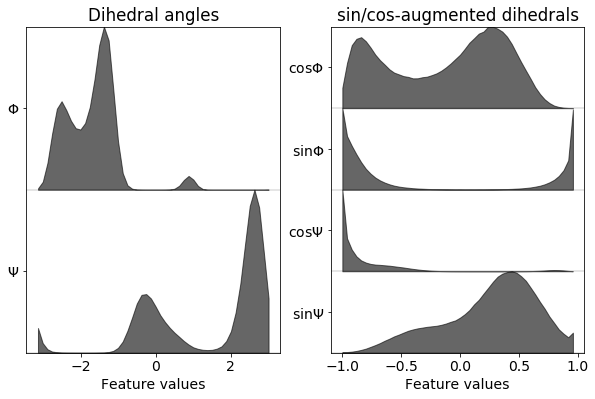}
	\end{subfigure}
	~ 
	\begin{subfigure}[b]{0.48\textwidth}
	\centering
	\caption{Pairwise distances}
	\label{fig:ala2-features-pw}	
	\includegraphics[width = 1\textwidth]{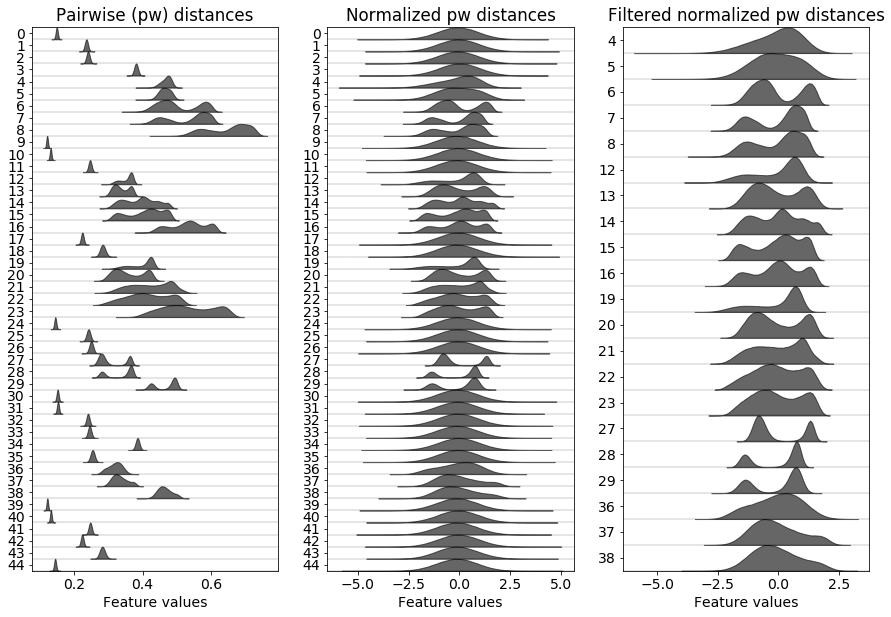}
	\end{subfigure}
\caption{Processing of the features: (a) the dihedral angles along the backbone, and (b) the pairwise distance between heavy atoms. }
\label{fig:ala2-features}
\end{figure}

\begin{figure}[h!]
\centering
	\begin{subfigure}[b]{0.32\textwidth}
	\centering
	\caption{Metastable states}
	\label{fig:ala2-true-clusters}
	\includegraphics[height = 0.15\textheight]{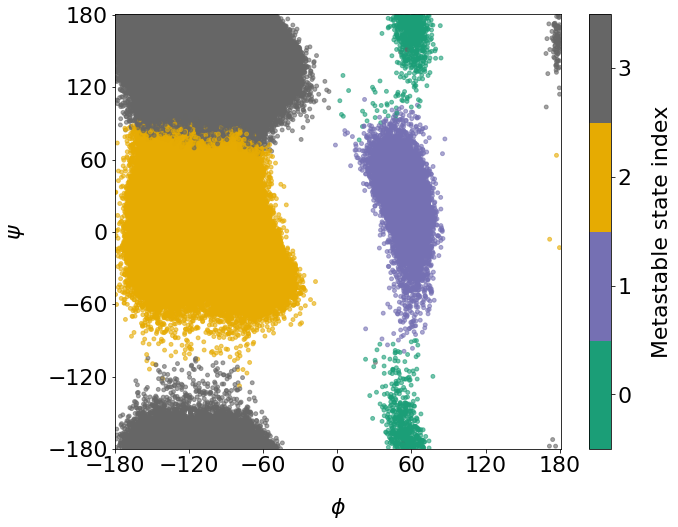}
	\end{subfigure}
	~
	\begin{subfigure}[b]{0.32\textwidth}
	\centering
	\caption{The true metastable states}
	\label{fig:ala2-meta-hist}
	\includegraphics[height = 0.15\textheight]{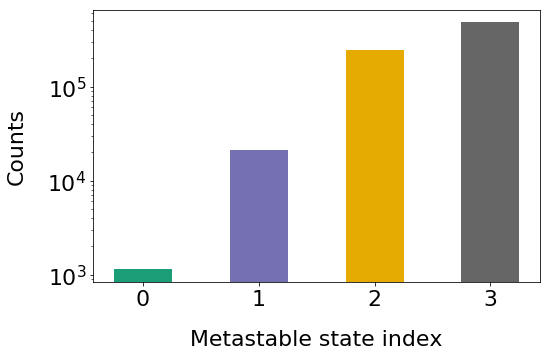}
	\end{subfigure}
	~
	\begin{subfigure}[b]{0.32\textwidth}
	\centering
	\caption{The GMVAE clusters}
	\label{fig:ala2-gmvae-cluster-hist}	
	\includegraphics[height = 0.15\textheight]{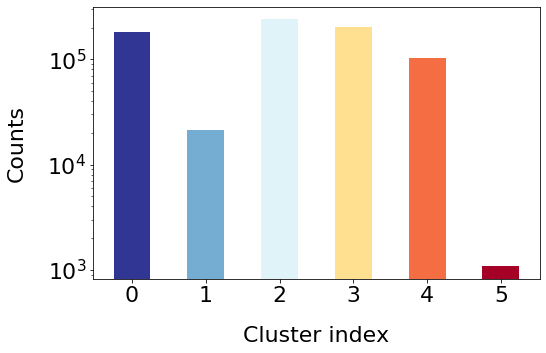}
	\end{subfigure}
\caption{Metastable states from TICA, and PCCA+ on the Ramachandran plot. The histograms for the (a) true metastable states, (b) GMVAE clusters.}
\label{fig:ala2-hist}
\end{figure}

We applied TICA to the set of pairwise distances only, followed by a kinetic coarse-graining with the PCCA+ method into 4 metastable states.
Figure~\ref{fig:ala2-true-clusters} presents the resulting clusters plotted on the Ramachandran plot.
Figures~\ref{fig:ala2-meta-hist} and ~\ref{fig:ala2-gmvae-cluster-hist} show the histograms of these metastable states, and the GMVAE clusters, respectively.

\begin{figure}[htbp] 	
\centering
	\begin{subfigure}[b]{0.48\textwidth}
	\centering
	\caption{FEL via the GMVAE}
	\label{fig:ala2-gmvae-fel_np}
	\includegraphics[height = 0.22\textheight]{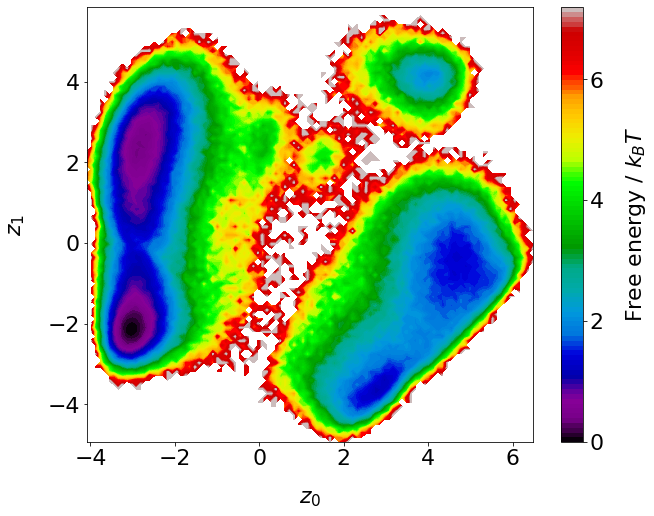}
	\end{subfigure}%
	~
	\begin{subfigure}[b]{0.48\textwidth}
	\centering
	\caption{Clusters}
	\label{fig:ala2-gmvae-clusters_np}
	\includegraphics[height = 0.22\textheight]{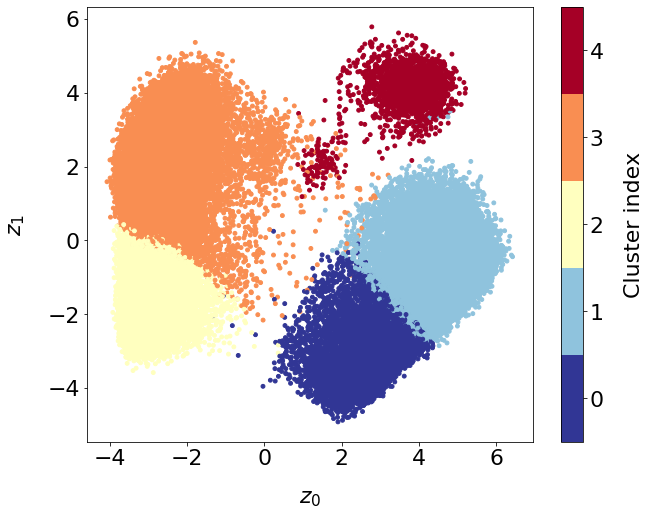}
	\end{subfigure}
	\\
	\begin{subfigure}[b]{0.48\textwidth}
	\centering
	\caption{Clusters on the Ramachandran plot}
	\label{fig:ala2-gmvae-rama_np}
	\includegraphics[height = 0.22\textheight]{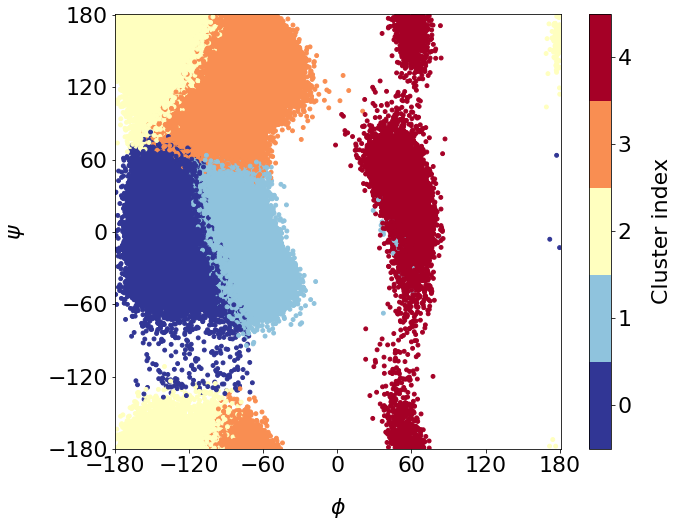}
	\end{subfigure}%
	~
	\begin{subfigure}[b]{0.48\textwidth}
	\centering
	\caption{Cluster counts}
	\label{fig:ala2-gmvae-cluster-hist_np}
	\includegraphics[height = 0.22\textheight, width = 0.8\textwidth]{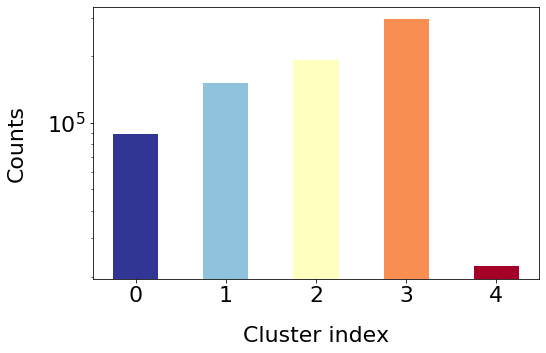}
	\end{subfigure}%
\caption{(a) FEL obtained for the alanine dipeptide by the GMVAE in a separate fully-converged training. The GMVAE clusters on the (b) GMVAE landscape, (c) Ramachandran plot. (d) Cluster counts.}
\label{fig:ala2_gmvae_np}
\end{figure}
\begin{figure}[htbp] 	
\centering
	\begin{subfigure}[b]{0.48\textwidth}
	\centering
	\caption{Implied timescales}
	\label{fig:ala2-its_np}
	\includegraphics[height = 0.22\textheight]{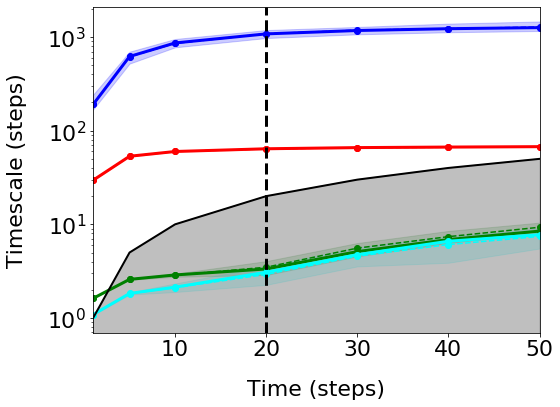}
	\end{subfigure}%
	~
	\begin{subfigure}[b]{0.48\textwidth}
	\centering
	\caption{Chapman-Kolmogorov test}
	\label{fig:ala2-ck_np}
	\includegraphics[height = 0.22\textheight]{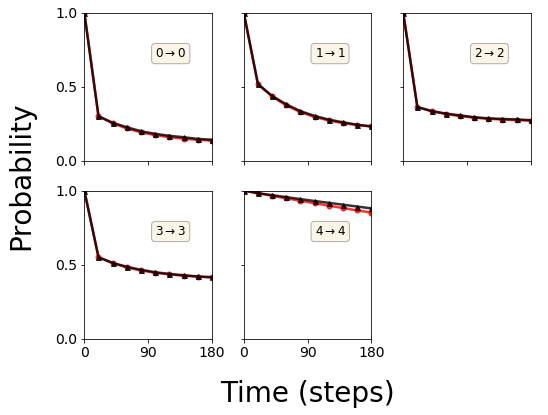}
	\end{subfigure}
\caption{Markovianity check of the MSM built for alanine dipeptide via the GMVAE. (a) Implied timescales. (b) Chapman-Kolmogorov test (at lag=20 steps).}
\label{fig:ala2_MSM_np}
\end{figure}
\begin{figure}[htbp]
\centering
	\begin{subfigure}[b]{0.4\textwidth}
	\centering
	\caption{FEL via the VAE}
	\label{fig:ala2-vae-fel}
	\includegraphics[height = 0.20\textheight]{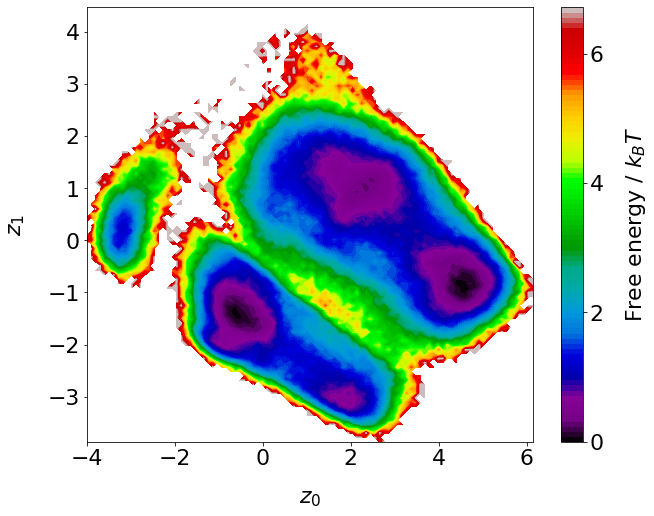}
	\end{subfigure}
	~ 
	\begin{subfigure}[b]{0.4\textwidth}
	\centering
	\caption{The true metastable state partitions}
	\label{fig:ala2-vae-true-states}	
	\includegraphics[height = 0.20\textheight]{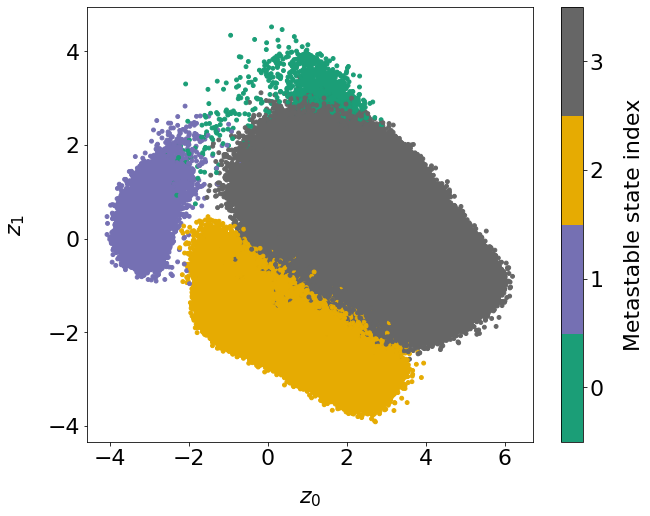}
	\end{subfigure}
\caption{VAE results for alanine dipeptide. (a) The FEL obtained by the VAE, (b) the true metastable state partitions on this landscape.}
\label{fig:ala2-vae}
\end{figure}

\clearpage
\section{AAQAA$_\textrm{3}$ peptide - I} \label{sec:app-aaqaa-I}

\subsection{Coarse-grained Peptide Model}
We employ a simple physics-based peptide model that was previously used to investigate structural-kinetic relationships in helix-coil transitions~\cite{Rudzinski:2018b,Rudzinski:2018}.
The model employs three attractive interactions, following standard \Go-type models~\cite{Go:1975,Cheung:2002,Clementi:2004,Chan:2011}:
{\it{(i)}} a native contact (nc) attraction, $U_{\rm nc}$, employed between pairs of $\Ca$ atoms which lie within a certain distance in the native structure, i.e., the $\alpha$-helix, of the peptide,
{\it{(ii)}} a desolvation barrier (db) interaction, $U_{\rm db}$, also employed between native contacts, and
{\it{(iii)}} a hydrophobic (hp) attraction, $U_{\rm hp}$, employed between all pairs of $\Cb$ atoms of the amino acid side chains.
We employed the same functional forms as in many previous studies~\cite{Chan:2011}, with a tunable prefactor, $\epsilon_{i}$, for each of the interactions.
The model considered here employed the prefactors $\epsNC = 12.5$, $\epsDB = 0.4\epsNC$, and $\epsHP = 0.2\epsNC$, while performing simulations at a temperature of 280~K.
In addition to these simple coarse-grained interactions, a standard AA force field, AMBER99sb~\cite{Hornak:2006}, is also partially incorporated to model both the steric interactions between all non-hydrogen atoms and also the specific local conformational preferences along the chain.

Molecular dynamics simulations of $\AQA$ were performed with the Gromacs 4.5.3 simulation suite~\cite{Hess:2008} in the constant NVT ensemble, while employing the stochastic dynamics algorithm with a friction coefficient $\gamma = (2.0~\tausb)^{-1}$ and a time step of $1 \times 10^{-3}~\tausb$.
For each model, 100 independent simulations were performed with starting conformations varying from full helix to full coil.
Each simulation was performed for $100,000~\tausb$, recording the system every $0.5~\tausb$.
The CG unit of time, $\tausb$, can be determined from the fundamental units of length, mass, and energy of the simulation model, but does not provide any meaningful description of the dynamical processes generated by the model.
In this case, $\tausb = 1$~ps.
\clearpage
\subsection{GMVAE Landscape and the Cluster Assignments} \label{sec:aaqaa-I-gmvae-all}
\begin{figure}[htbp]
	\centering
	\includegraphics[height = 0.3\textheight]{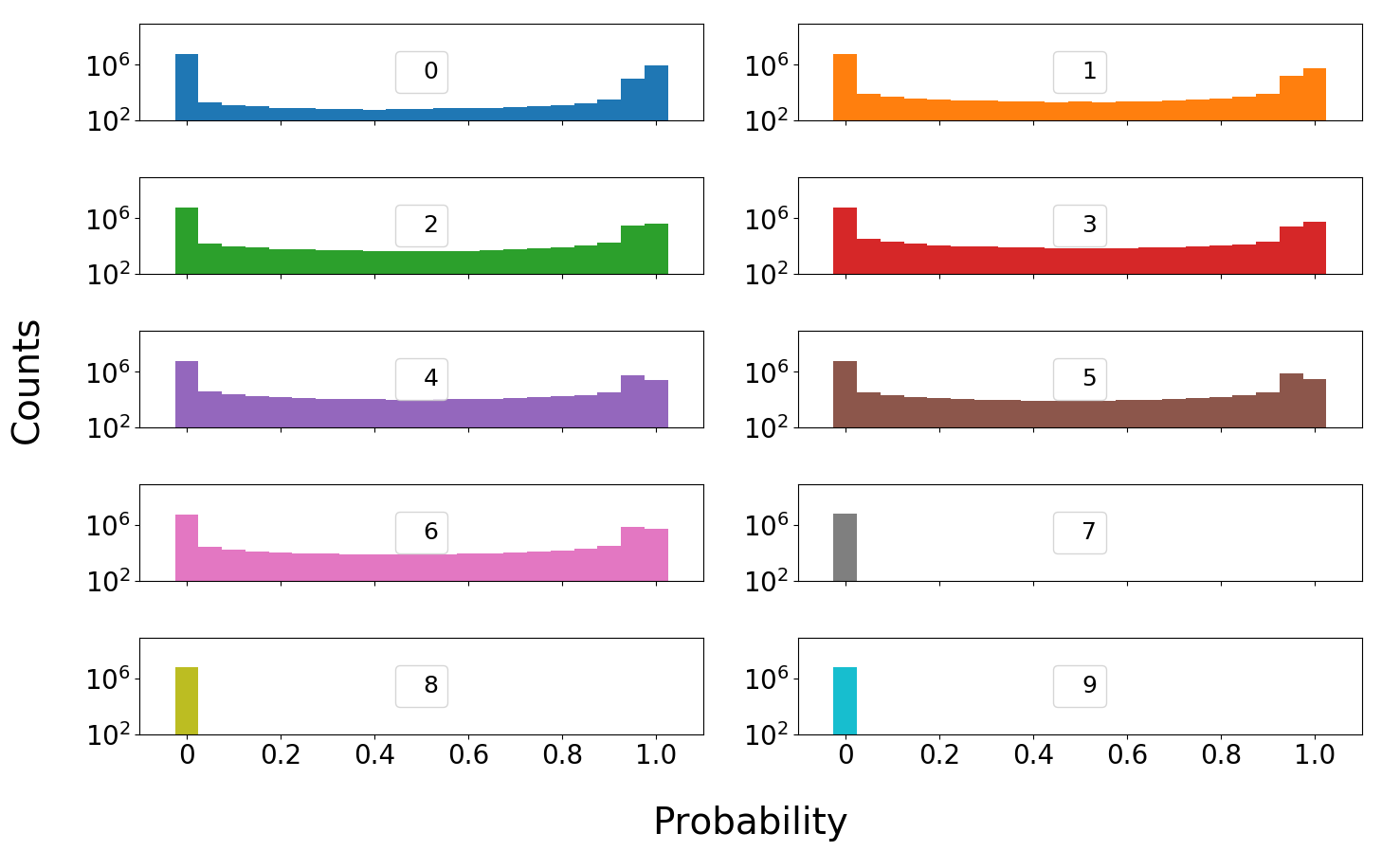}
	\caption{The population distribution as a function of probability of belonging to each of the clusters after the training. None of the data points is assigned to clusters $7-8-9$.}
	\label{fig:aaqaa-I-qy_prob}
\end{figure}

\begin{figure}[htbp]
	\centering
	\includegraphics[height = 0.2\textheight]{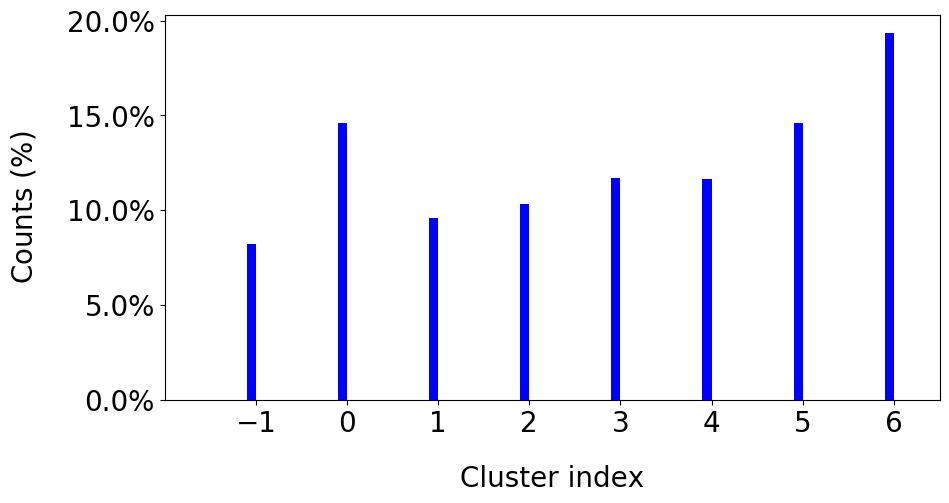}
	\caption{Cluster populations}
	\label{fig:aaqaa-I-cluster_pop}
\end{figure}

\begin{figure}[htbp] 	
\centering
	\includegraphics[width = 1\textwidth]{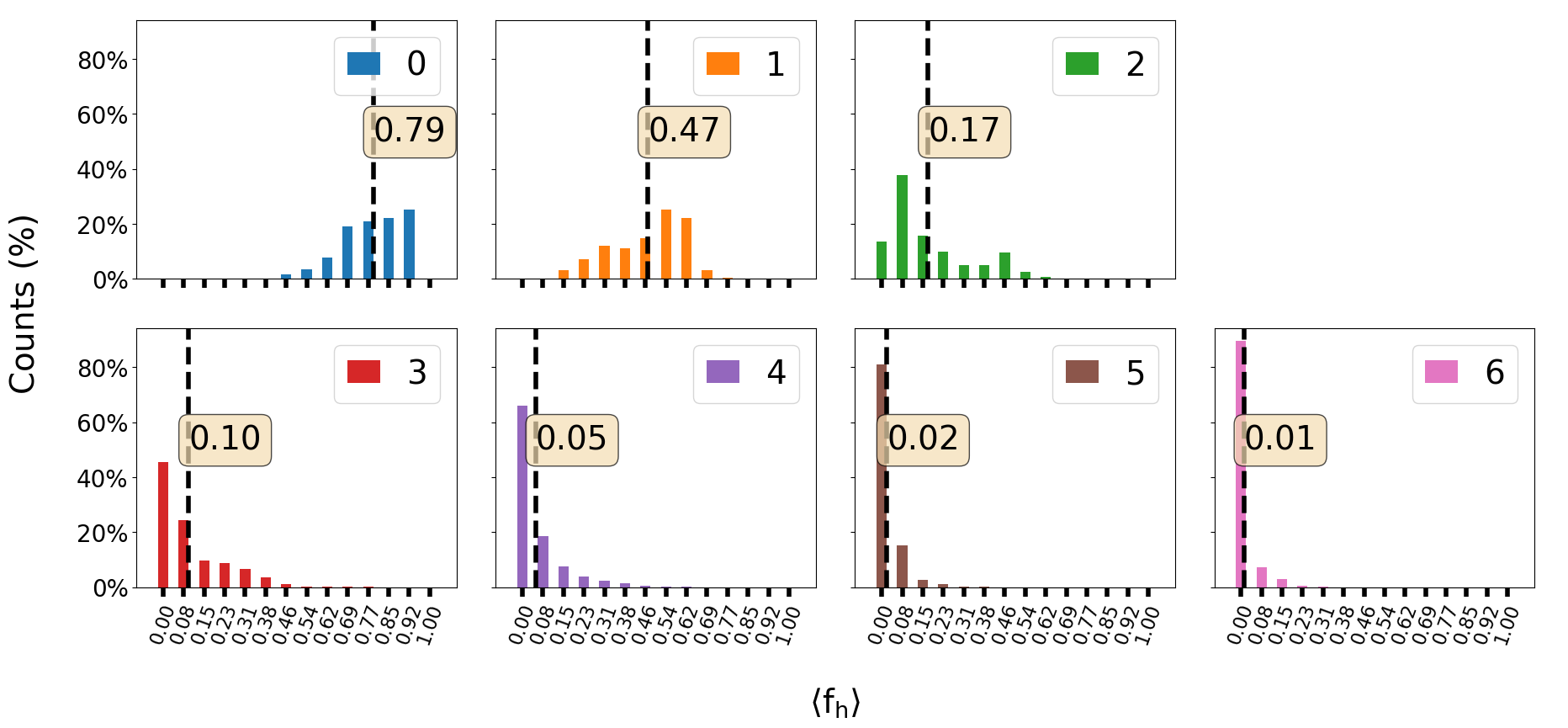}
	\caption{Intra-cluster distributions of average helical fraction, $\langle f_h \rangle$, ($\AQA$ - I). The dashed lines indicate the average values, which are also written in the text boxes.}
	\label{fig:aaqaa-I-hist_fh}
\end{figure}

\begin{figure}[htbp] 	
\centering
	\begin{subfigure}[b]{0.6\textwidth}
	\centering
	\caption{Histogram plots of dRMSD values}
	\label{fig:aaqaa-I-drmsd_histograms}	
	\includegraphics[height= 0.22\textheight]{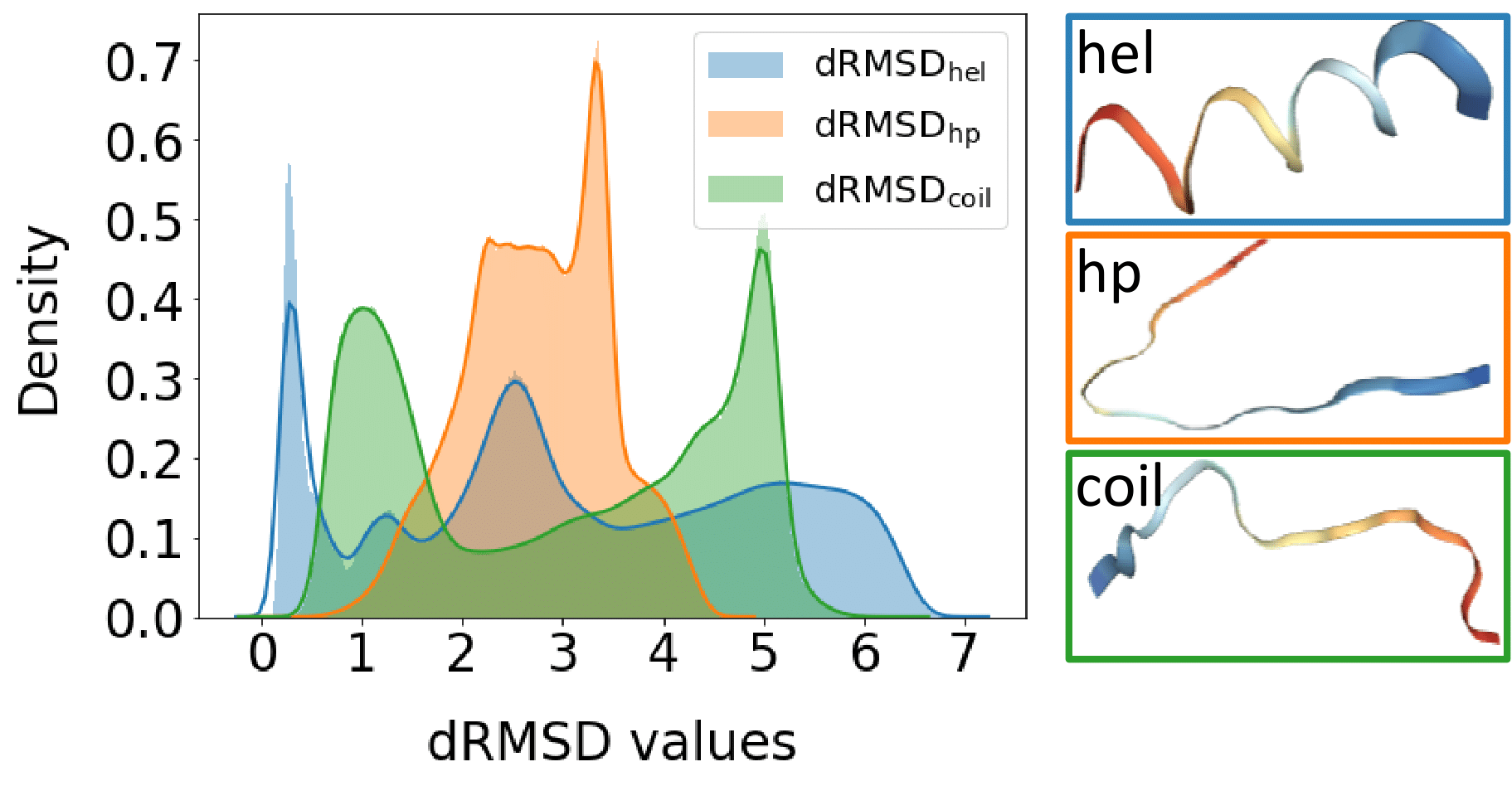}
	\end{subfigure} 
	~
	\begin{subfigure}[b]{0.35\textwidth}
	\centering
	\caption{Scatter plot of dRMSD$_\textrm{hel}$ vs. dRMSD$_\textrm{hp}$}	
	\label{fig:aaqaa-I-scatter_hel_hp}	
	\includegraphics[height= 0.22\textheight]{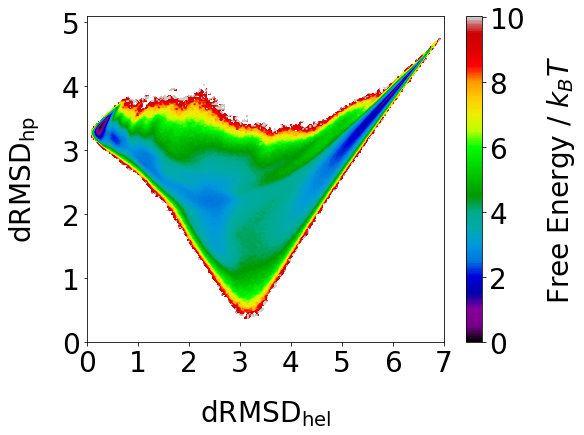}
	\end{subfigure}%
	\\ 
	\begin{subfigure}[b]{1\textwidth}
	\centering
	\caption{Sampled regions of dRMSD$_\textrm{hel}$ vs. dRMSD$_\textrm{hp}$ for each cluster}
	\label{fig:aaqaa-I-scatter_hel_hp_interc}
	\includegraphics[height = 0.4\textheight]{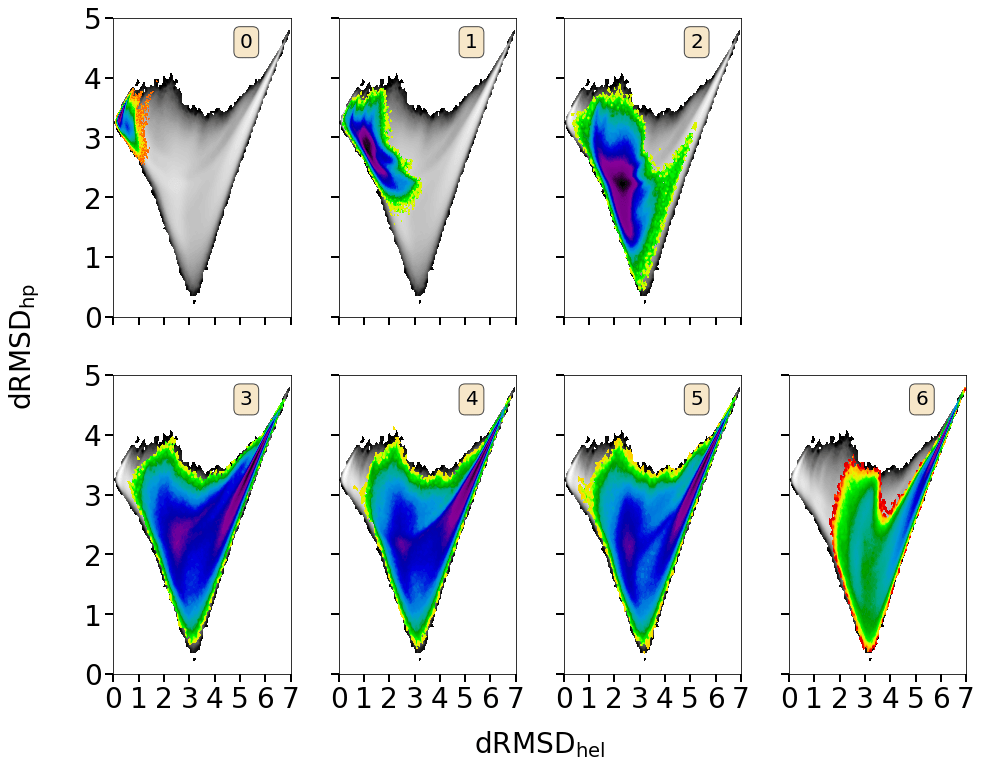} 
	\end{subfigure}
\caption{dRMSD analysis. (a) The density histograms of dRMSD values for helix, hairpin, and coil structures (with the visualized reference structures). (b) Scatter plot of dRMSD$_\textrm{hel}$ vs. dRMSD$_\textrm{hp}$ with densities. (c) Sampled regions of dRMSD$_\textrm{hel}$ vs. dRMSD$_\textrm{hp}$ in each of the cluster. The same colormap is used in (b) and (c). }
\label{fig:aaqaa-I-drmsds}
\end{figure}

Since the GMVAE method is a probabilistic clustering method, each data point has a probability of assignment to each of the $k$ clusters. For a data point $d_i$, the probability of assignment to each of the clusters has probability values $p_{d_i, 0}, p_{d_i, 1}, \dots p_{d_i, k-1}$ for $k$ number of clusters. In the ideal case, all of the probability values except the true cluster is equal to $0$, and the true cluster has a value of $1$. Figure~\ref{fig:aaqaa-I-qy_prob} separately shows the histogram of probability distributions of all of the data points for a cluster. For instance, for cluster $0$, the probability distributions are accumulated at probability values $0$ and $1$. In other words, with high certainty, cluster $0$ is differentiated from the others. None of the data points is assigned to clusters $7-8-9$. Note that although the network is initially trained for $10$ clusters, it is not possible to separate more than $7$ clusters under the specified loss function. This suggests a way to find the inherent number of clusters, i.e., metastable states, provided that $k$ is chosen larger than that true value. 
Cluster ID's are obtained after a thresholding step as explained in Section~\ref{sec:threshold}. Figure~\ref{fig:aaqaa-I-cluster_pop} shows the cluster populations. Cluster -1 indicates the datapoints that are not assigned to any of the clusters. 
Figure~\ref{fig:aaqaa-I-hist_fh} shows the inter-cluster $\langle f_h \rangle$ distributions. 
To further characterize the clustering of secondary structures, we separately calculated dRMSDs with respect to three reference structures: helix (hel), hairpin-like (hp), and extended (coil). 
Figure~\ref{fig:aaqaa-I-drmsd_histograms} presents both the reference structures (right) and corresponding dRMSD distributions (left). 
The first and the second small peaks in the dRMSD$_\textrm{hel}$ distribution represent helical conformations, while the peak corresponding to dRMSD$_\textrm{hel}$ values between ~$2-3.5$ hints at the presence of the hairpin-like structures. 
Note that there is an offset in dRMSD$_\textrm{hp}$ values due to (i) the scarcity of the well-defined hairpins in the trajectory data, and (ii) the subjectivity involved in choosing the reference structures. 
By plotting the two-dimensional free-energy surface along dRMSD$_\textrm{hel}$ and dRMSD$_\textrm{hp}$, shown in Figure~\ref{fig:aaqaa-I-scatter_hel_hp}, the distinct secondary structures can be separated. 
The conformations with dRMSD$_\textrm{hel}$ values below 1.8 are helical, whereas the minimum in the upper right with dRMSD$_\textrm{hel}$ greater than 4 is comprised of extended structures.
The energy minimum in the middle (enclosed in the region with dRMSD$_\textrm{hel}$ values between $1.8$ and $3$, and dRMSD$_\textrm{hp}$ values between $0$ and $3$) contains hairpin-like structures. 
Figure~\ref{fig:aaqaa-I-scatter_hel_hp_interc} presents inter-cluster free-energy surfaces along dRMSD$_\textrm{hel}$ and dRMSD$_\textrm{hp}$, generated by considering only conformations within a single cluster. 
%
%
\begin{figure}[htbp] 	
\centering
	\begin{subfigure}[b]{0.48\textwidth}
	\centering
	\caption{dRMSD$_\textrm{hp}$}
	\label{fig:aaqaa-I-proj_drmsd_hp}
	\includegraphics[width = 1\textwidth]{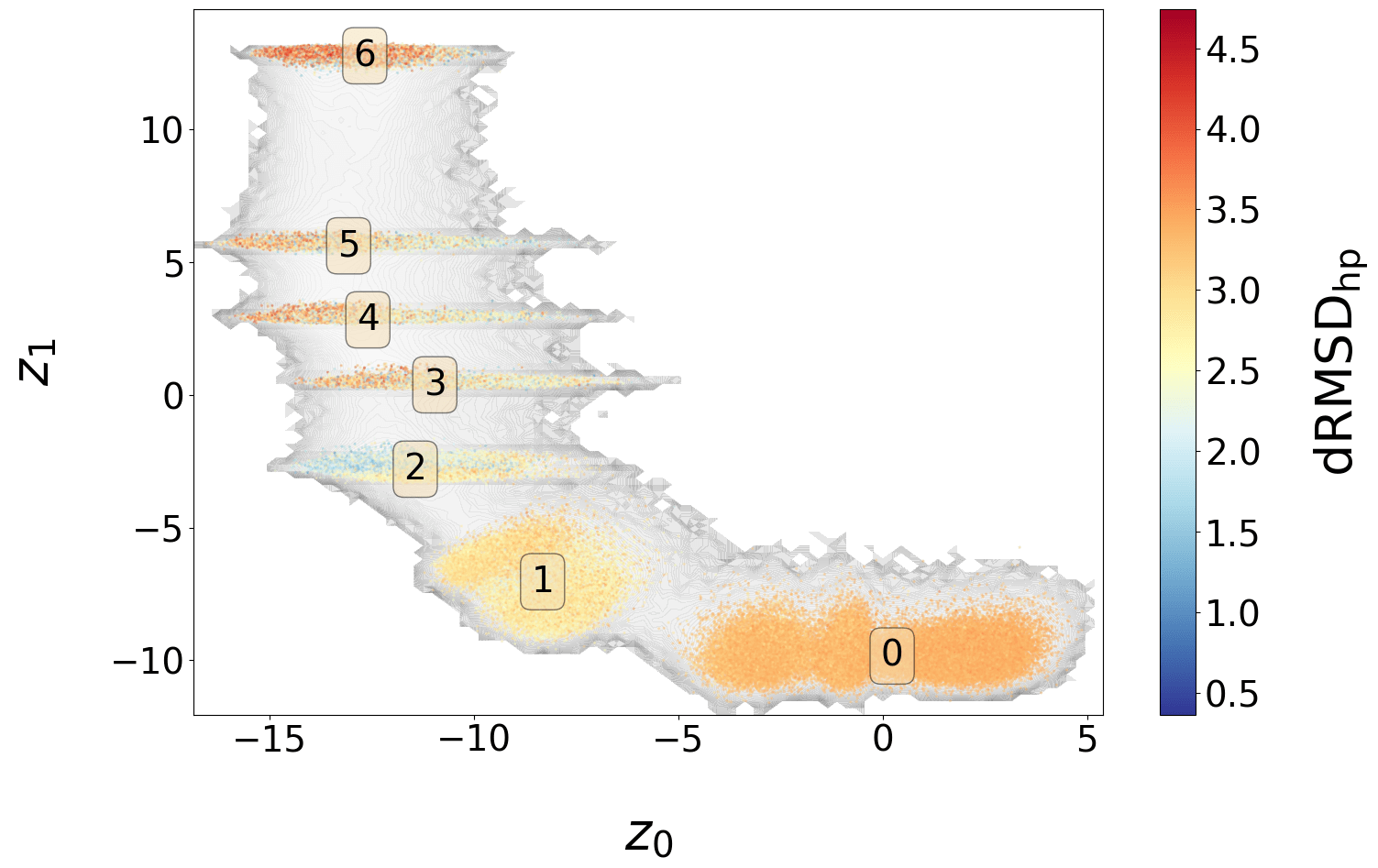}
	\end{subfigure}  
	~
	\begin{subfigure}[b]{0.48\textwidth}
	\centering
	\caption{dRMSD$_\textrm{coil}$}
	\label{fig:aaqaa-I-proj_drmsd_coil}
	\includegraphics[width = 1\textwidth]{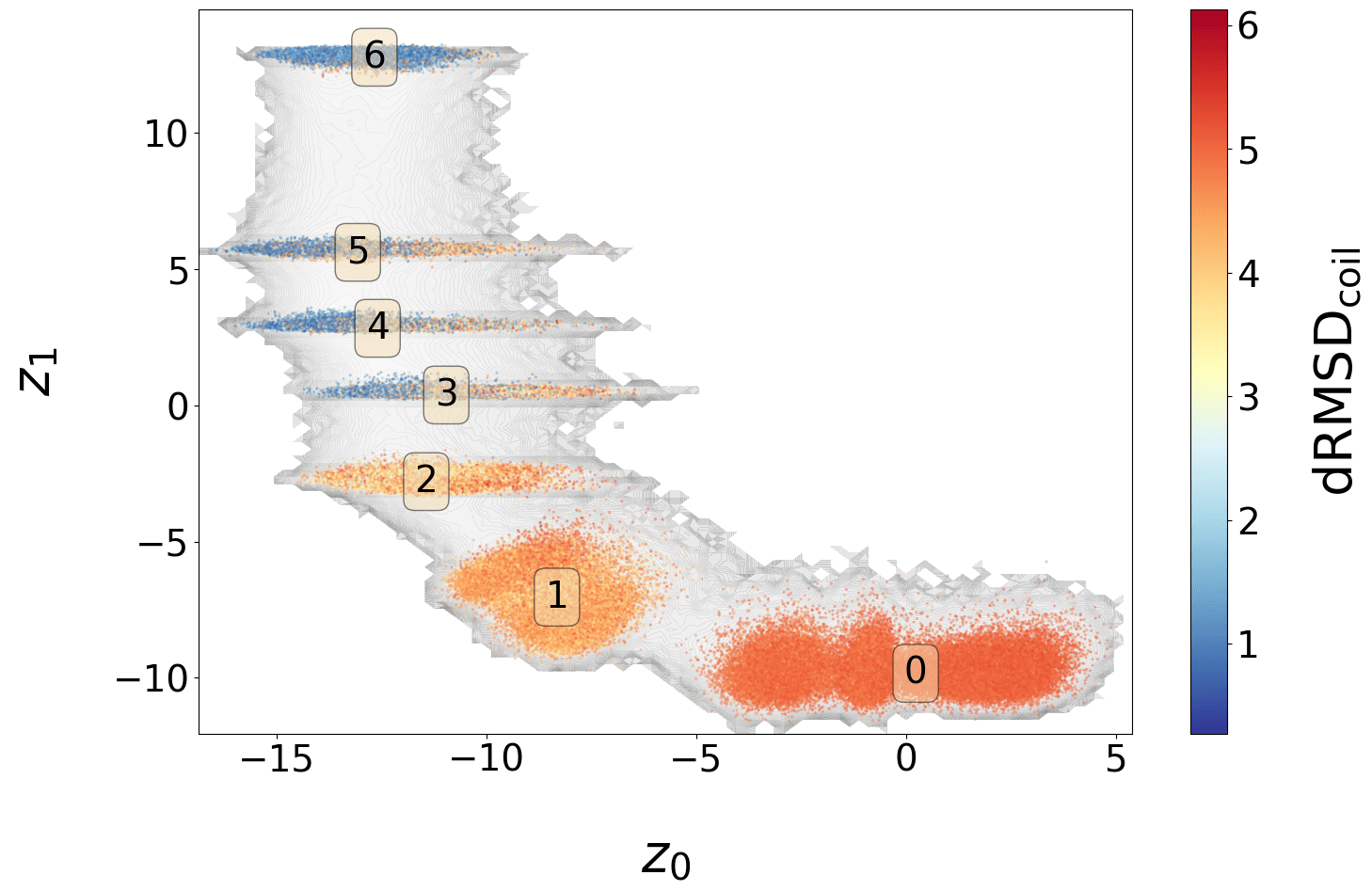}
	\end{subfigure}%
	\\
	\begin{subfigure}[b]{0.48\textwidth}
	\centering
	\caption{$\textrm{R}_\textrm{g}$}
	\label{fig:aaqaa-I-proj_Rg}
	\includegraphics[width = 1\textwidth]{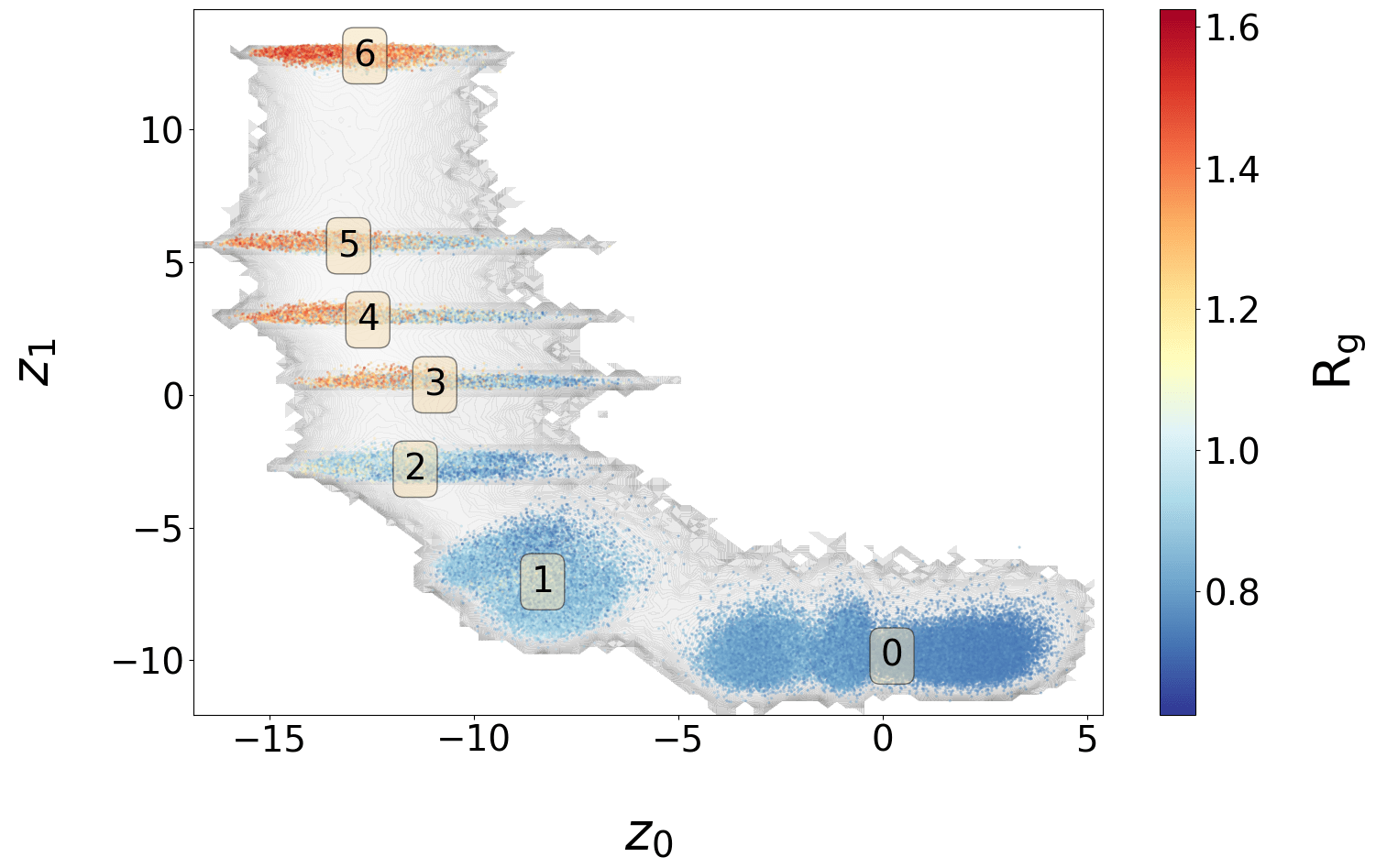}
	\end{subfigure}%
\caption{Projections colored according to (a) dRMSD$_\textrm{hel}$, (b) dRMSD$_\textrm{hp}$, (c) dRMSD$_\textrm{coil}$, and (d) $\textrm{R}_\textrm{g}$}
\label{fig:aaqaa-I-projs_others}
\end{figure}

In addition to $\langle f_h \rangle$ and dRMSD$_\textrm{hel}$, the \emph{radius of gyration} $\textrm{R}_\textrm{g}$ distribution is also analyzed. $\textrm{R}_\textrm{g}$ measures the mass-weighted deviations from center of mass, and gives an idea on the overall spread and compactness of the molecule, and is calculated as
\begin{equation} \label{eq:Rg}
R_g = \sqrt{\dfrac{\sum_{i} m_i ||\textbf{r}_i - \textbf{r}_c||^2}{\sum_{i} m_i}} \,\, ,
\end{equation}
where $m_i$ is the mass of atom $i$, $\textbf{r}_i$ the coordinates of atom $i$, and $\textbf{r}_c$ is the coordinates of the center of mass. Figures~\ref{fig:aaqaa-I-proj_drmsd_hel}, ~\ref{fig:aaqaa-I-proj_drmsd_hp}, ~\ref{fig:aaqaa-I-proj_drmsd_coil}, and ~\ref{fig:aaqaa-I-proj_Rg} show the heat map of dRMSD$_\textrm{hel}$, dRMSD$_\textrm{hp}$, dRMSD$_\textrm{coil}$, and R$_\textrm{g}$ on the FEL obtained via the GMVAE, respectively.  

\begin{figure}[htbp] 	
\centering
	\begin{subfigure}[b]{0.48\textwidth}
	\centering
	\caption{Implied timescales}
	\label{fig:aaqaa-I-its}
	\includegraphics[height = 0.22\textheight]{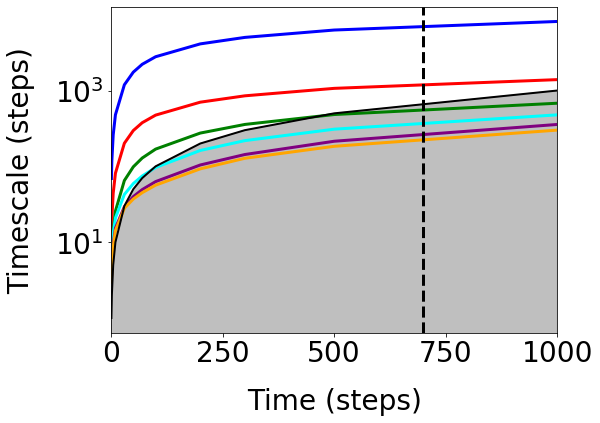}
	\end{subfigure}%
	~
	\begin{subfigure}[b]{0.48\textwidth}
	\centering
	\caption{Chapman-Kolmogorov Test}
	\label{fig:aaqaa-I-ck}
	\includegraphics[height = 0.22\textheight]{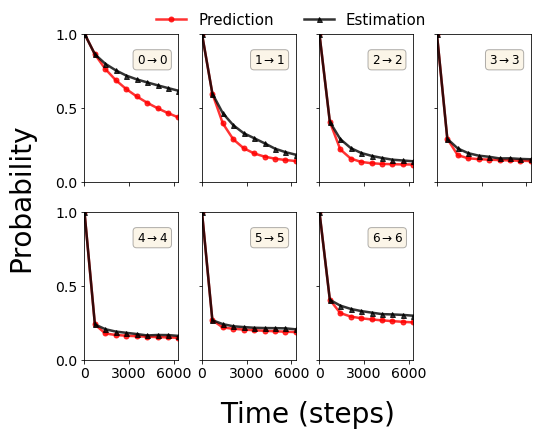}
	\end{subfigure}
\caption{Markovianity check of the MSM built for the $\AQA$ peptide - I via using the cluster labels from the GMVAE. (a) Implied timescales. (b) Chapman-Kolmogorov test (at lag=700 steps)}
\label{fig:aaqaa-I-MSM}
\end{figure}

\begin{figure}[htbp] 	
\centering
	\begin{subfigure}[b]{0.48\textwidth}
	\centering
	\caption{Cluster centers}
	\label{fig:aaqaa-I-gmvae_landsc-ccenters}
	\includegraphics[height = 0.22\textheight]{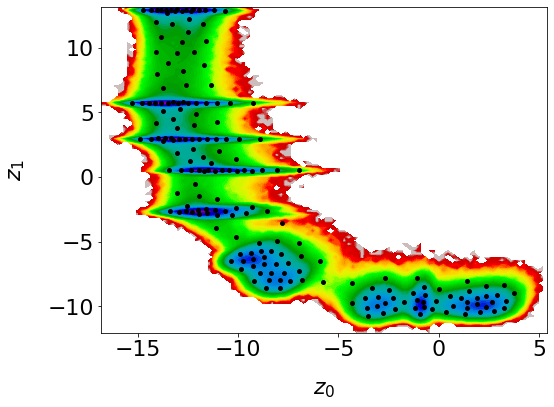}
	\end{subfigure}  
	~
	\begin{subfigure}[b]{0.48\textwidth}
	\centering
	\caption{Implied timescales}
	\label{fig:aaqaa-I-gmvae_landsc-its}
	\includegraphics[height = 0.22\textheight]{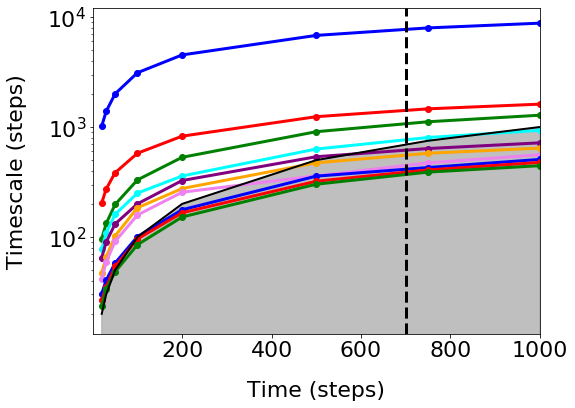}
	\end{subfigure}  
	\\
	\begin{subfigure}[b]{0.48\textwidth}
	\centering
	\caption{Chapman Kolmogorov test}
	\label{fig:aaqaa-I-gmvae_landsc-ck}
	\includegraphics[height = 0.22\textheight]{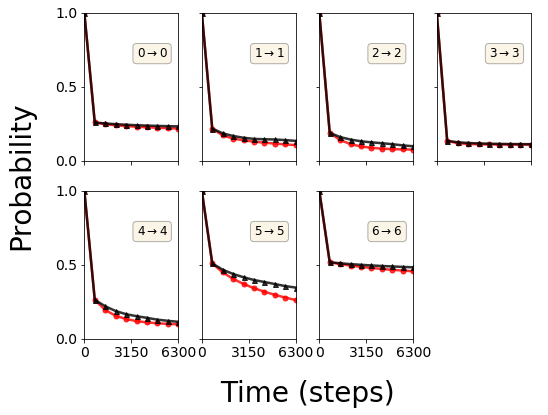}
	\end{subfigure}%
	~
	\begin{subfigure}[b]{0.48\textwidth}
	\centering
	\caption{Clusters}
	\label{fig:aaqaa-I-gmvae_landsc-states}
	\includegraphics[height = 0.22\textheight]{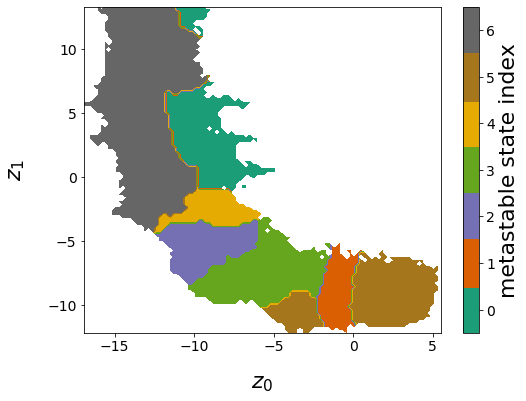}
	\end{subfigure}%
\caption{Kinetic analysis on the GMVAE landscape for $\AQA$ - I}
\label{fig:aaqaa-I-gmvae_landsc-msm}
\end{figure}

As a final characterization of the clustering, we constructed an MSM directly from the discretized trajectories of GMVAE cluster indices.
Although thresholding was applied in the results presented here (practically similar to coring methods for constructing kinetic models~\cite{Jain:2012}), we found that this procedure had negligible effect on the accuracy of the resulting MSM.
Figure~\ref{fig:aaqaa-I-its} presents implied timescale test.
The kinetic model can resolve two of longest characteristic processes and demonstrates reasonable convergence, although there is a small increase in the longest timescale with increasing lag time.
This subtle discrepancy already indicates that there may be some issues with the accuracy of the kinetic model.
An MSM is constructed at lag time $700$ to balance between the convergence of the timescales and the resolution of shorter timescale processes.
Figure~\ref{fig:aaqaa-I-ck} presents the CK test from this model.
There are significant errors in the description of probability decay from each of the metastable states (i.e., clusters), especially states 0 and 1.
First, we note that coarse-grained MSMs (i.e., MSMs built on a small number of metastable states) are often not expected to be quantitatively accurate due to difficulties in accurately defining the dividing surfaces between states~\cite{Bowman:2014}.
However, we anticipate that it should be possible to make a more accurate coarse-grained MSM for this particular simulation trajectory.
The discrepancies in the model can then originate from two coupled problems: (i) the GMVAE latent space definition places structures close together that are kinetically distinct (i.e., there are hidden barriers) or (ii) the GMVAE clustering fails to identify/separate distinct metastable states.
The FEL within the latent space (Figure~\ref{fig:aaqaa-I-FEL_gmvae}) contains clearly separated basins that are not identified as unique clusters by the GMVAE.
In particular, within clusters 0 and 1, there seems to be $2$ and $3$ separate states, respectively.
Figure~\ref{fig:aaqaa-I-hist_fh} shows that cluster 0 (1) contains structures with a range of helicities ranging from 0.46-1.0 (0.15-0.69).
According to the conventional picture of the helix-coil transition, the overarching kinetics can be described by two timescales: (i) the rate at which a single helical segment is formed and (ii) the elongation rate of helical segments along the chain.
By grouping together conformations with a single helical segment and several helical segments, the GMVAE has convoluted these two timescales, resulting in non-Markovianity in the kinetics described on these clusters.
To further clarify the source of these errors, we constructed an MSM in the conventional way, directly from the latent space distribution.
More specifically, we applied k-means algorithm with $1000$ cluster centers, and then applied PCCA+~\cite{roblitz2013fuzzy}.
In order to enable comparison, we continued with the previous number of metastable states ($7$).
The CK test for the resulting model with lag time $\tau = 700$ (obtained from the implied timescale test, Figure~\ref{fig:aaqaa-I-gmvae_landsc-its}) is presented in Figure~\ref{fig:aaqaa-I-gmvae_landsc-ck}, and demonstrates slightly improved accuracy.
\begin{figure}[htbp] 	
\centering
	\includegraphics[width = 0.9\textwidth]{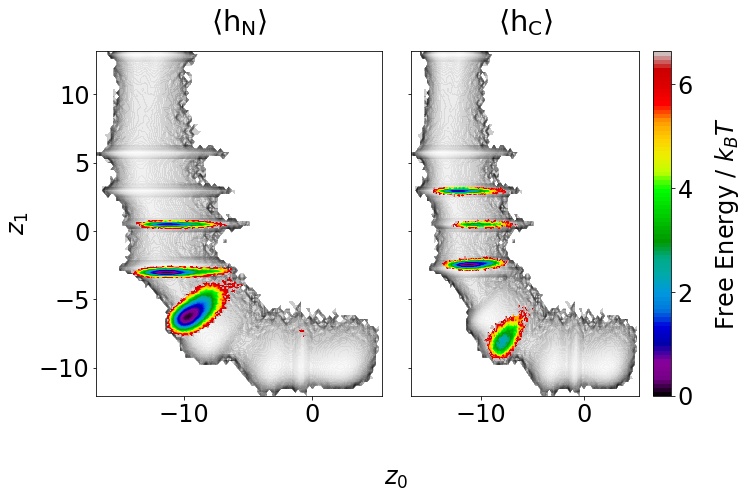}
\caption{Distributions for $\langle h_N \rangle \geq 0.8$ (on the left), $\langle h_N \rangle \leq - 0.8$ (on the right).}
\label{fig:aaqaa-I-end_fold}
\end{figure}

\clearpage
\subsection{GMVAE Landscape Only (Without Using the Cluster Assignments)} \label{sec:aaqaa-I-gmvae-landsc-only}
\begin{figure}[htbp]
	\centering
	\includegraphics[height = 0.2\textheight]{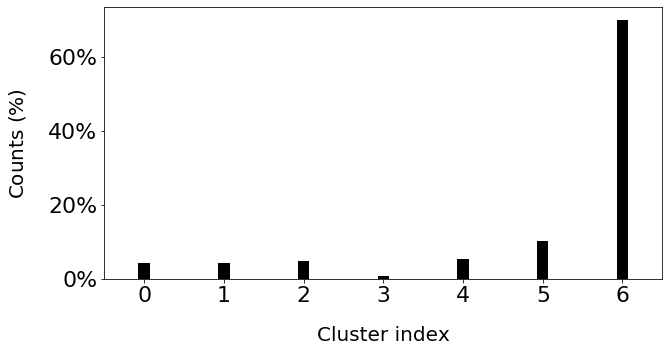}
	\caption{Cluster populations for $\AQA$ - I from PCCA+}
	\label{fig:aaqaa-I-gmvae_landsc-cluster_pop}
\end{figure}
\begin{figure}[htbp]
\centering
	\includegraphics[width = 1\textwidth]{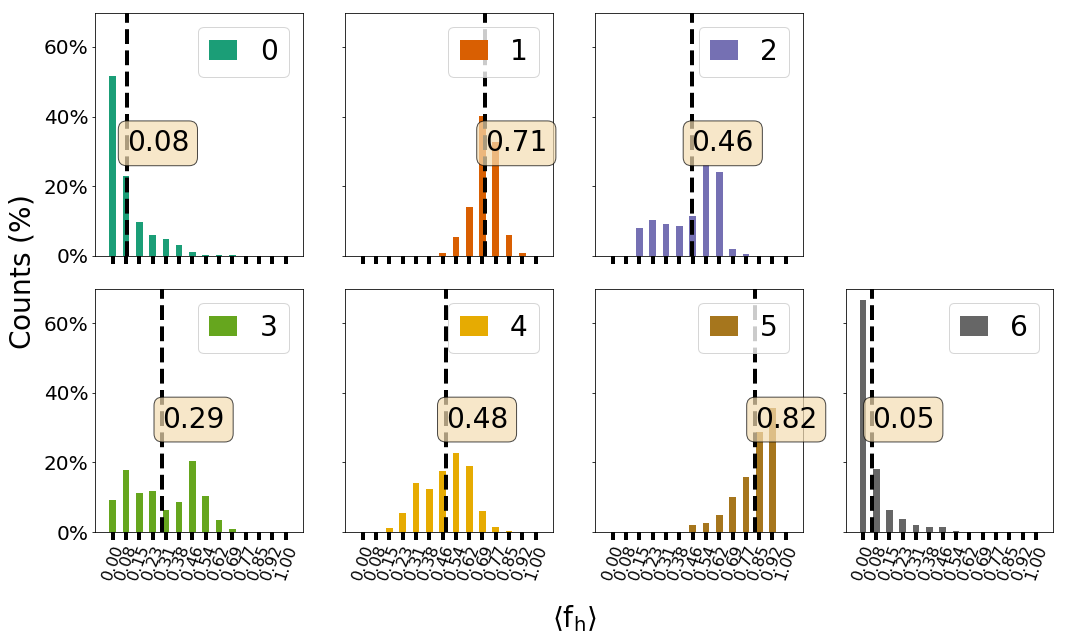}
	\caption{Intra-cluster distributions of average helical fraction, $\langle f_h \rangle$, ($\AQA$ - I) for the clusters obtained with PCCA+). The dashed lines indicate the average values, which are also written in the text boxes.}
	\label{fig:aaqaa-I-gmvae_landsc-hist_fh}
\end{figure}
%
%

\clearpage
\subsection{TICA Results} \label{sec:aaqaa-I-tica}
2D TICA projections are obtained at lag time $\tau = 20$ steps.
\begin{figure}[htbp] 	
\centering
	\begin{subfigure}[b]{0.48\textwidth}
	\centering
	\caption{FEL}
	\label{fig:aaqaa-I-tica-fel}
	\includegraphics[height = 0.22\textheight]{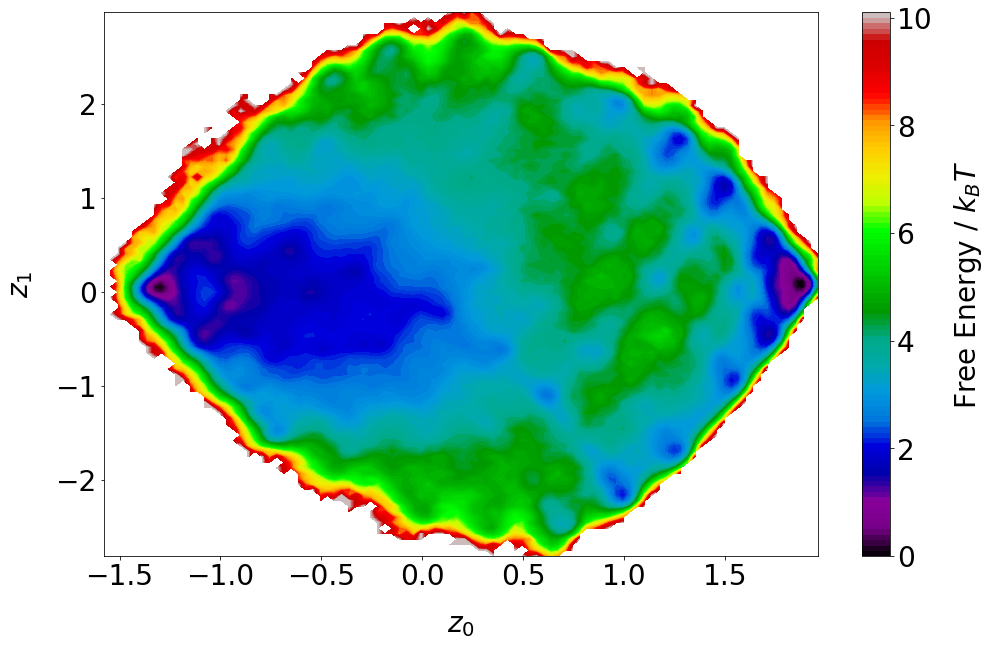}
	\end{subfigure}  
	~
	\begin{subfigure}[b]{0.48\textwidth}
	\centering
	\caption{$\langle f_h \rangle$}
	\label{fig:aaqaa-I-tica-fh}
	\includegraphics[height = 0.22\textheight]{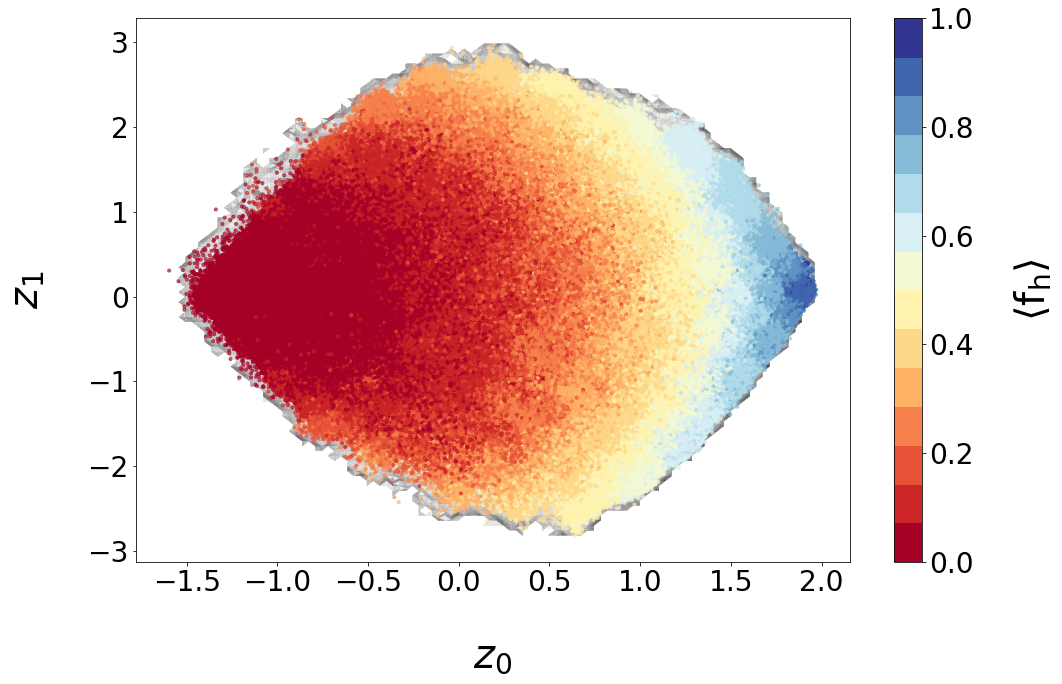}
	\end{subfigure}%
	\\
	\begin{subfigure}[b]{0.4\textwidth}
	\centering
	\caption{$\langle h_N \rangle - \langle h_C \rangle$}
	\label{fig:aaqaa-I-tica-fold1}
	\includegraphics[height = 0.22\textheight]{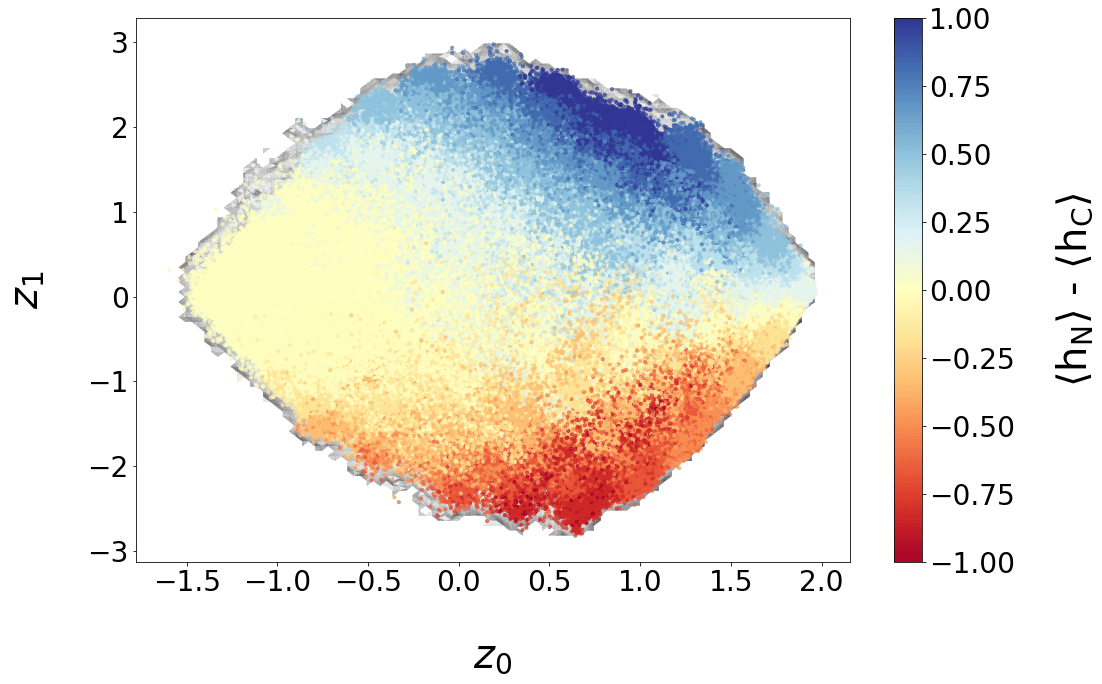}
	\end{subfigure}  
\caption{TICA results for $\AQA$ - I}
\label{fig:aaqaa-I-tica-landscape}
\end{figure}
\begin{figure}[htbp] 	
\centering
	\begin{subfigure}[b]{0.48\textwidth}
	\centering
	\caption{Cluster centers}
	\label{fig:aaqaa-I-tica-ccenters}
	\includegraphics[height = 0.22\textheight]{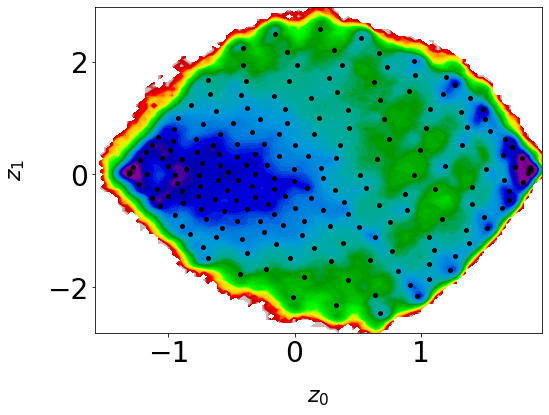}
	\end{subfigure}  
	~
	\begin{subfigure}[b]{0.48\textwidth}
	\centering
	\caption{Implied timescales}
	\label{fig:aaqaa-I-tica-its}
	\includegraphics[height = 0.22\textheight]{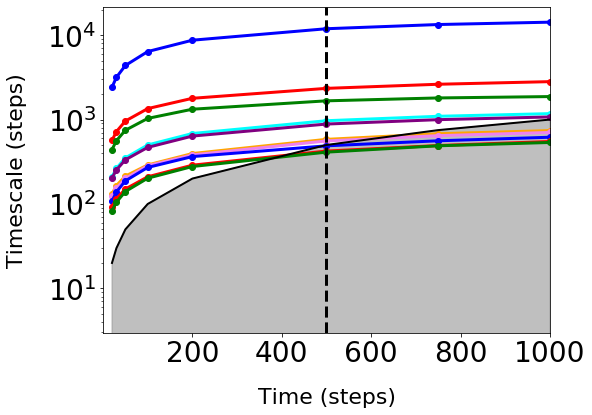}
	\end{subfigure}  
	\\
	\begin{subfigure}[b]{0.48\textwidth}
	\centering
	\caption{Chapman Kolmogorov test}
	\label{fig:aaqaa-I-tica-ck}
	\includegraphics[height = 0.22\textheight]{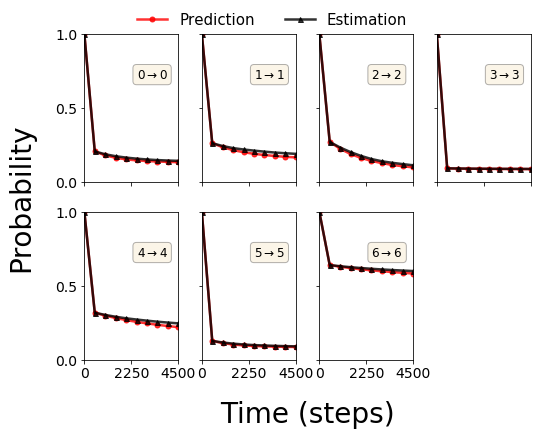}
	\end{subfigure}%
	~
	\begin{subfigure}[b]{0.48\textwidth}
	\centering
	\caption{Clusters}
	\label{fig:aaqaa-I-tica-states}
	\includegraphics[height = 0.22\textheight]{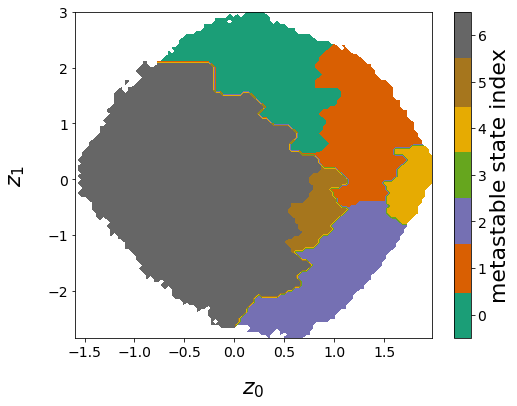}
	\end{subfigure}%
\caption{Kinetic analysis on TICA landscape for $\AQA$ - I}
\label{fig:aaqaa-I-tica-msm}
\end{figure}
\begin{figure}[htbp]
	\centering
	\includegraphics[height = 0.2\textheight]{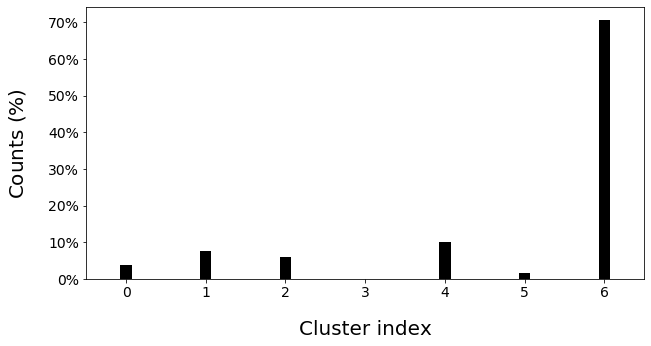}
	\caption{Cluster populations for $\AQA$ - I from TICA + PCCA+}
	\label{fig:aaqaa-I-tica-cluster_pop}
\end{figure}
\begin{figure}[htbp]
\centering
	\includegraphics[width = 1\textwidth]{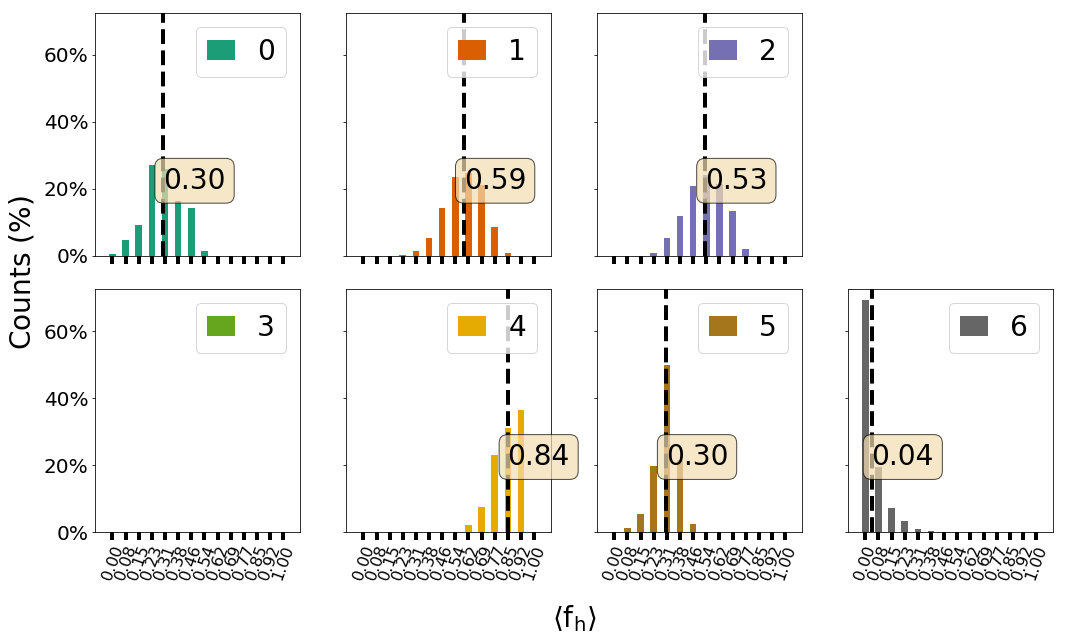}
	\caption{Intra-cluster distributions of average helical fraction, $\langle f_h \rangle$, ($\AQA$ - I) for the clusters obtained with PCCA+). The dashed lines indicate the average values, which are also written in the text boxes. Note that cluster $3$ is an empty cluster.}
	\label{fig:aaqaa-I-tica-hist_fh}
\end{figure}
\clearpage
\subsection{VAE Results} \label{sec:aaqaa-I-vae}
\begin{figure}[htbp] 	
\centering
	\begin{subfigure}[b]{0.48\textwidth}
	\centering
	\caption{FEL}
	\label{fig:aaqaa-I-vae-fel}
	\includegraphics[height = 0.22\textheight]{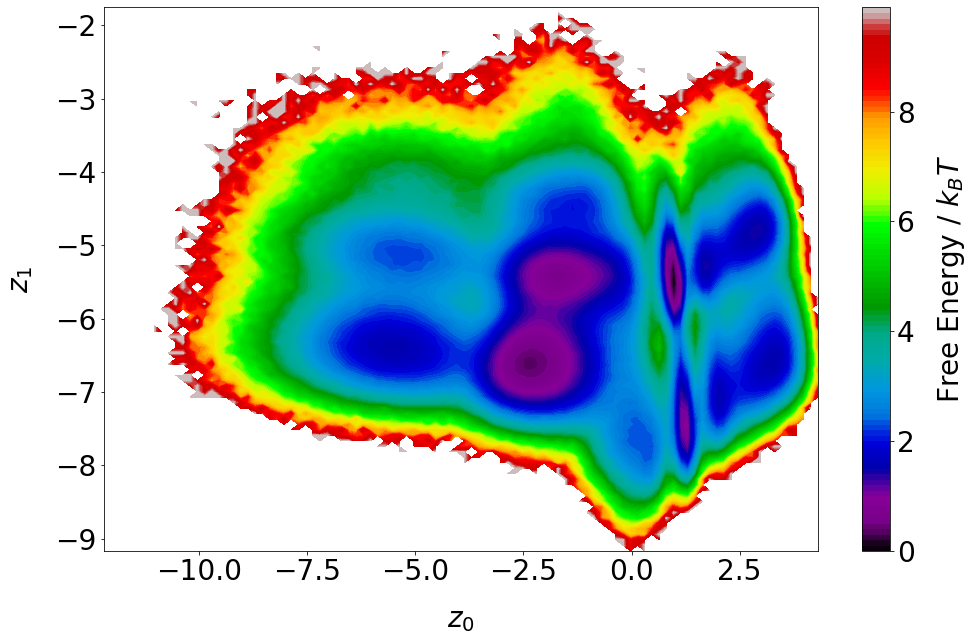}
	\end{subfigure}  
	~
	\begin{subfigure}[b]{0.48\textwidth}
	\centering
	\caption{$\langle f_h \rangle$}
	\label{fig:aaqaa-I-vae-fh}
	\includegraphics[height = 0.22\textheight]{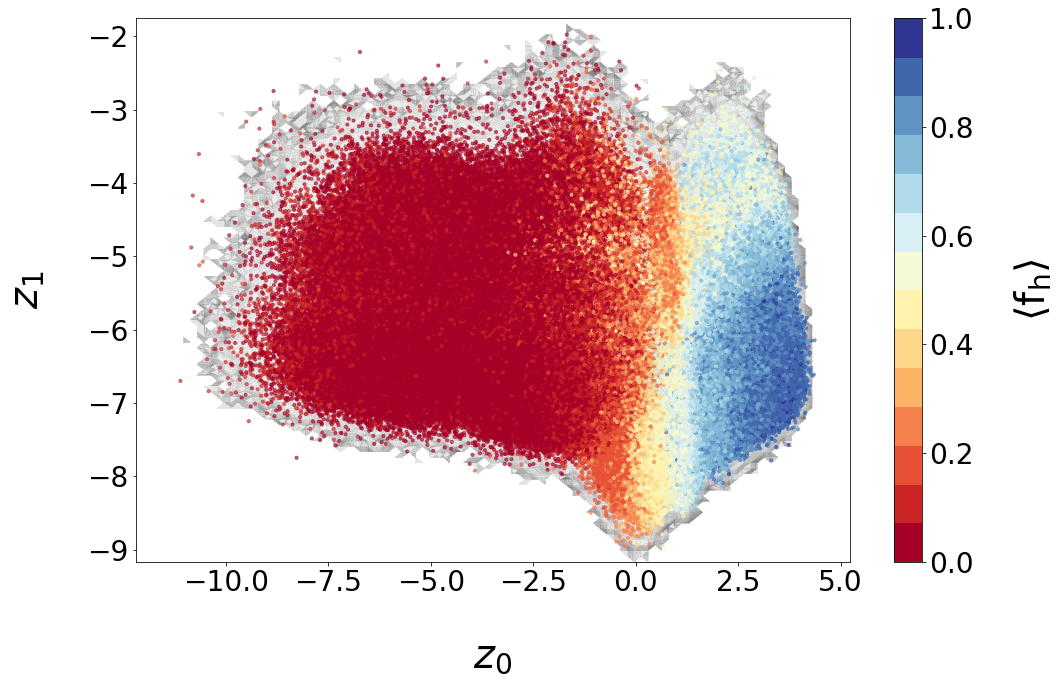}
	\end{subfigure}%
		\\
	\begin{subfigure}[b]{0.4\textwidth}
	\centering
	\caption{$\langle h_N \rangle - \langle h_C \rangle$}
	\label{fig:aaqaa-I-vae-fold1}
	\includegraphics[height = 0.22\textheight]{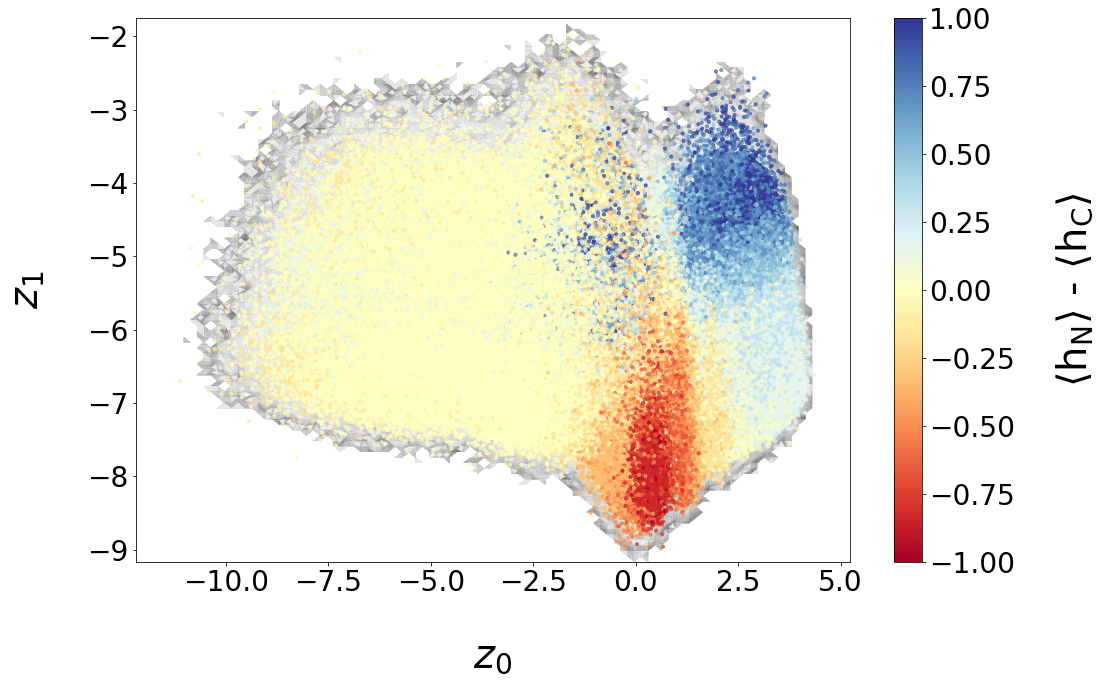}
	\end{subfigure}  
\caption{VAE results for $\AQA$ - I}
\label{fig:aaqaa-I-vae-landscape}
\end{figure}
\begin{figure}[htbp] 	
\centering
	\begin{subfigure}[b]{0.48\textwidth}
	\centering
	\caption{Cluster centers}
	\label{fig:aaqaa-I-vae-ccenters}
	\includegraphics[height = 0.22\textheight]{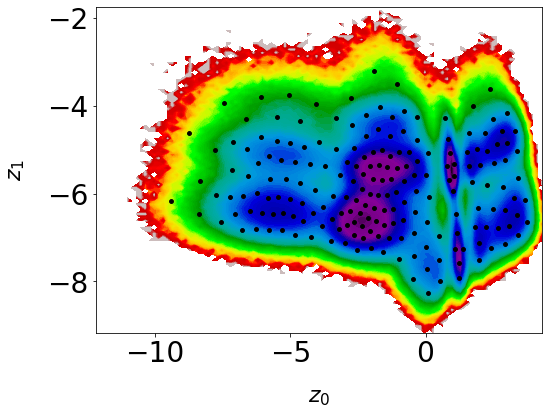}
	\end{subfigure}  
	~
	\begin{subfigure}[b]{0.48\textwidth}
	\centering
	\caption{Implied timescales}
	\label{fig:aaqaa-I-vae-its}
	\includegraphics[height = 0.22\textheight]{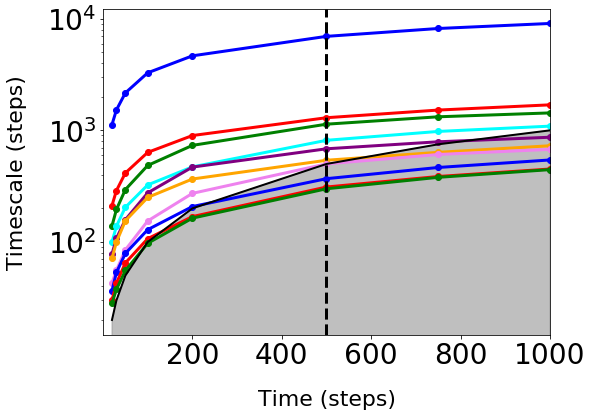}
	\end{subfigure}  
	\\
	\begin{subfigure}[b]{0.48\textwidth}
	\centering
	\caption{Chapman Kolmogorov test}
	\label{fig:aaqaa-I-vae-ck}
	\includegraphics[height = 0.22\textheight]{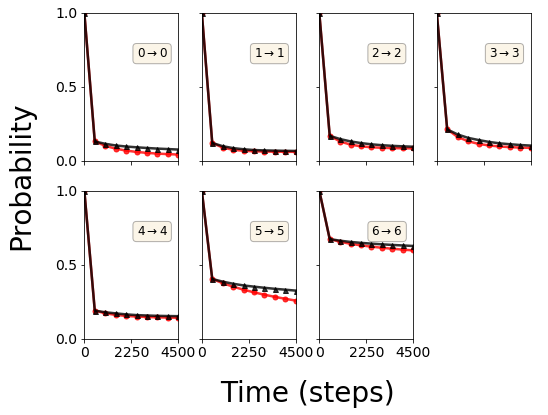}
	\end{subfigure}%
	~
	\begin{subfigure}[b]{0.48\textwidth}
	\centering
	\caption{Clusters}
	\label{fig:aaqaa-I-vae-states}
	\includegraphics[height = 0.22\textheight]{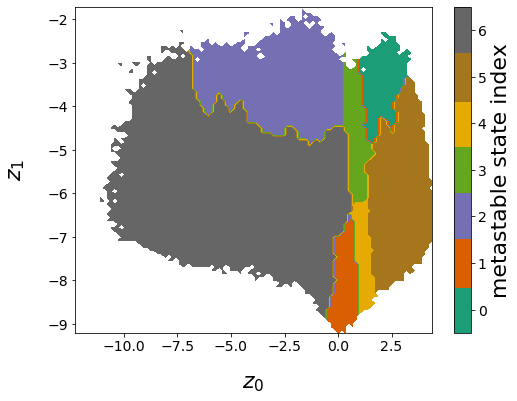}
	\end{subfigure}%
\caption{Kinetic analysis on the VAE landscape for $\AQA$ - I}
\label{fig:aaqaa-I-vae-msm}
\end{figure}
\begin{figure}[htbp]
	\centering
	\includegraphics[height = 0.2\textheight]{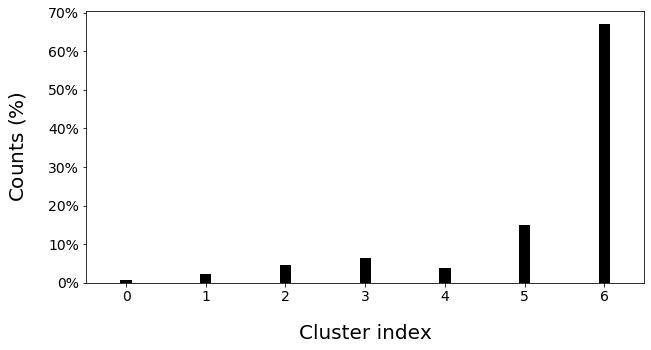}
	\caption{Cluster populations for $\AQA$ - I from VAE + PCCA+}
	\label{fig:aaqaa-I-vae-cluster_pop}
\end{figure}
\begin{figure}[htbp]
\centering
	\includegraphics[width = 1\textwidth]{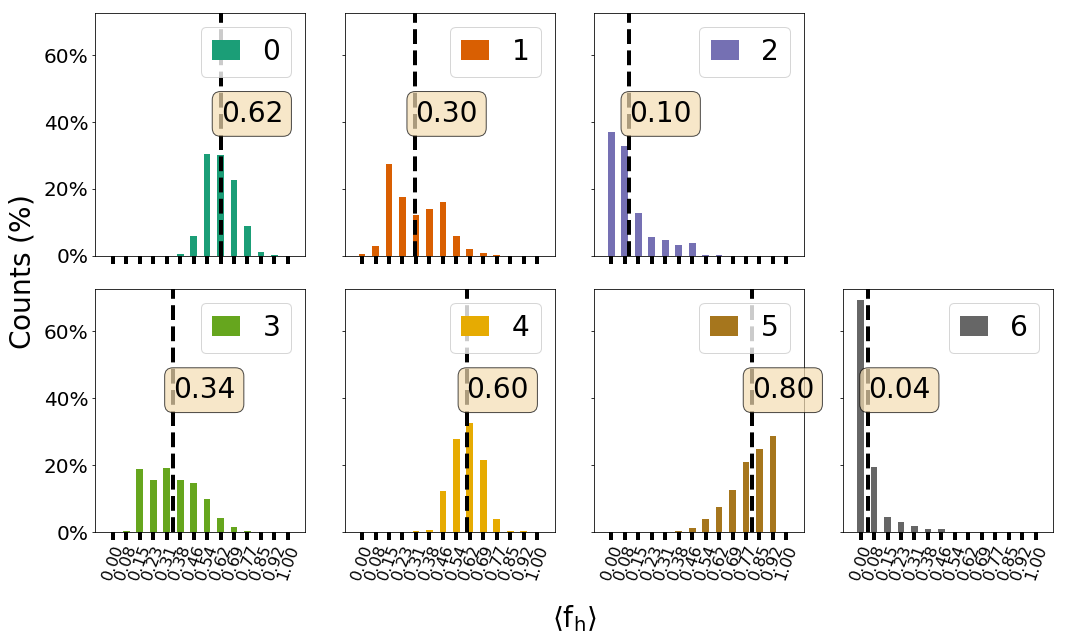}
	\caption{Intra-cluster distributions of average helical fraction, $\langle f_h \rangle$, ($\AQA$ - I) for the clusters obtained with PCCA+). The dashed lines indicate the average values, which are also written in the text boxes.}
	\label{fig:aaqaa-I-vae-hist_fh}
\end{figure}

%
\clearpage{}

\section{AAQAA$_\textrm{3}$ peptide - II} \label{sec:app-aaqaa-II}

We also considered an alternative coarse-grained model, with energetic prefactors $\epsNC = 10.92$, $\epsDB = 0.2\epsNC$, and $\epsHP = 0.5\epsNC$, while performing simulations at a temperature of 300~K.

\clearpage
\subsection{GMVAE Landscape and the Cluster Assignments} \label{sec:aaqaa-II-gmvae-all}
\begin{figure}[htbp]
	\centering
	\includegraphics[height = 0.2\textheight]{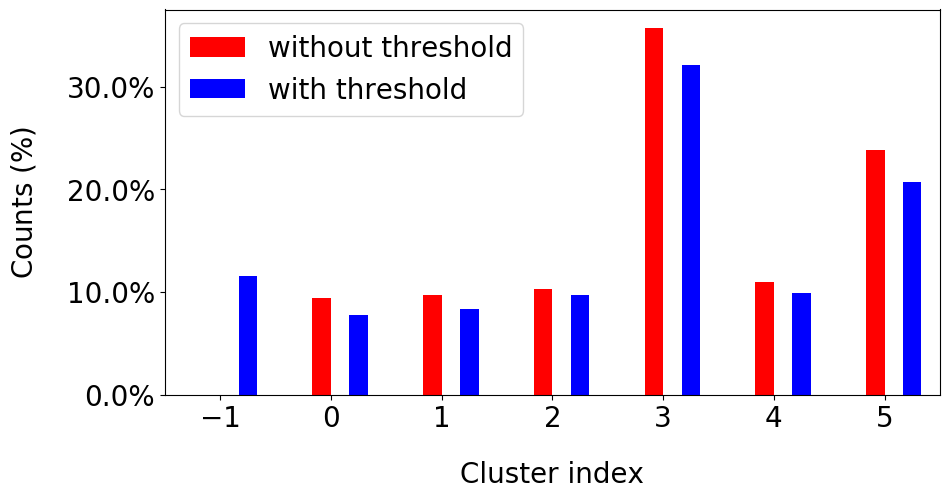}
	\caption{Cluster populations from the GMVAE.}
	\label{fig:aaqaa-II-cluster_pop}
\end{figure}
\begin{figure}[htbp]
\centering
	\includegraphics[width = 1\textwidth]{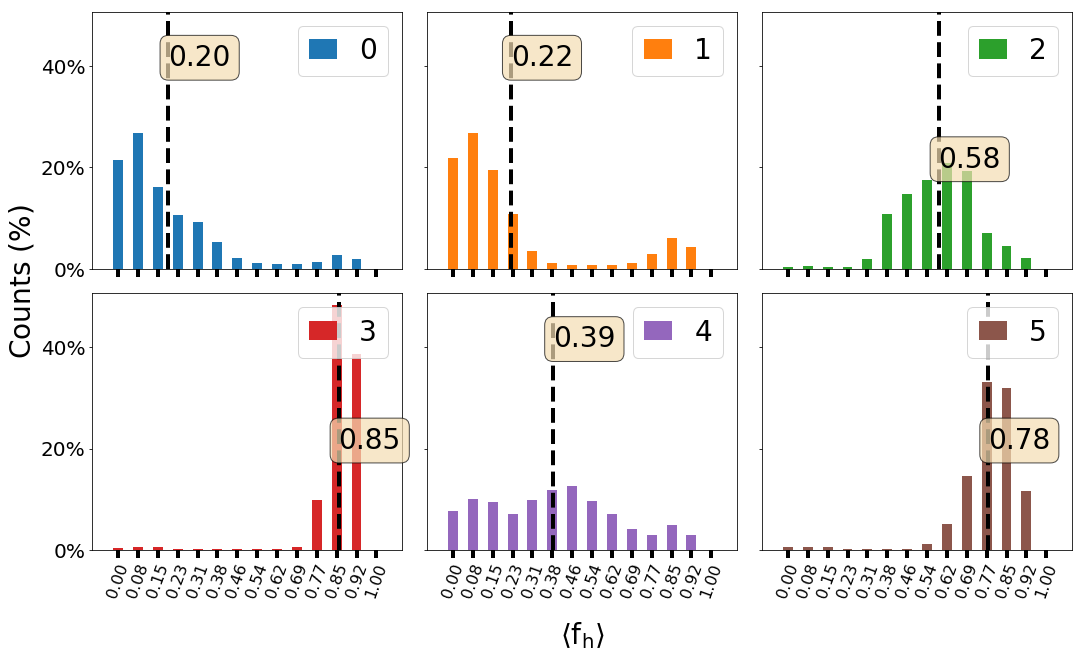}
	\caption{Intra-cluster distributions of average helical fraction, $\langle f_h \rangle$, ($\AQA$ - II). The dashed lines indicate the average values, which are also written in the text boxes.}
	\label{fig:aaqaa-II-hist_fh}
\end{figure}
\begin{figure}[htbp] 	
\centering
	\begin{subfigure}[b]{0.48\textwidth}
	\centering
	\caption{Implied timescales}
	\label{fig:aaqaa-II-its}
	\includegraphics[height = 0.22\textheight]{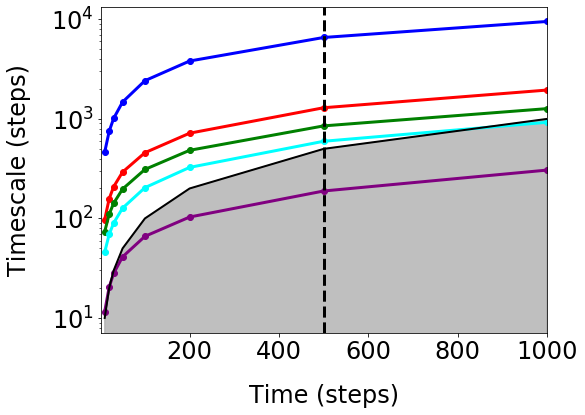}
	\end{subfigure}
	~
	\begin{subfigure}[b]{0.48\textwidth}
	\centering
	\caption{Chapman-Kolmogorov Test}
	\label{fig:aaqaa-II-ck}
	\includegraphics[height = 0.22\textheight]{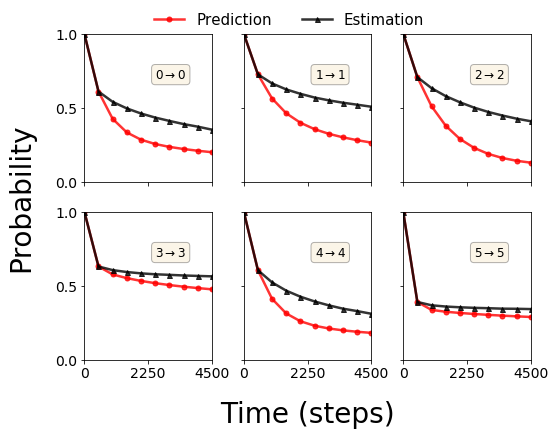}
	\end{subfigure}
\caption{Markovianity check of the MSM built for the $\AQA$ - II via using the cluster labels from the GMVAE. (a) Implied timescales. (b) Chapman-Kolmogorov test (at lag=500 steps)}
\label{fig:aaqaa-II-MSM}
\end{figure}

\clearpage{}
\subsection{GMVAE Landscape Only (Without Using the Cluster Assignments)} \label{sec:aaqaa-II-gmvae-landsc-only}
\begin{figure}[htbp] 	
\centering
	\begin{subfigure}[b]{0.48\textwidth}
	\centering
	\caption{Cluster centers}
	\label{fig:aaqaa-II-gmvae_landsc-ccenters}
	\includegraphics[height = 0.22\textheight]{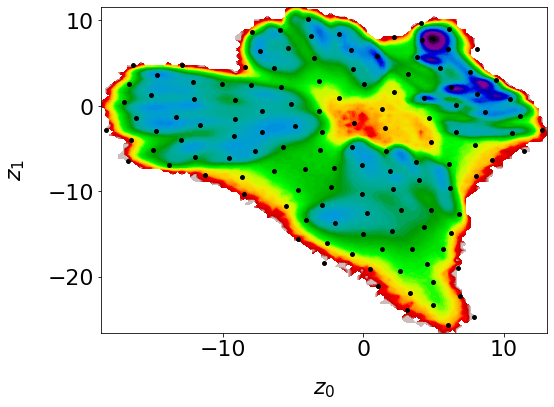}
	\end{subfigure}  
	~
	\begin{subfigure}[b]{0.48\textwidth}
	\centering
	\caption{Implied timescales}
	\label{fig:aaqaa-II-gmvae_landsc-its}
	\includegraphics[height = 0.22\textheight]{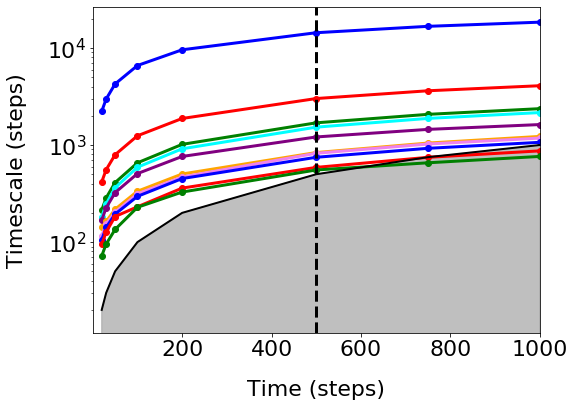}
	\end{subfigure}  
	\\
	\begin{subfigure}[b]{0.48\textwidth}
	\centering
	\caption{Chapman Kolmogorov test}
	\label{fig:aaqaa-II-gmvae_landsc-ck}
	\includegraphics[height = 0.22\textheight]{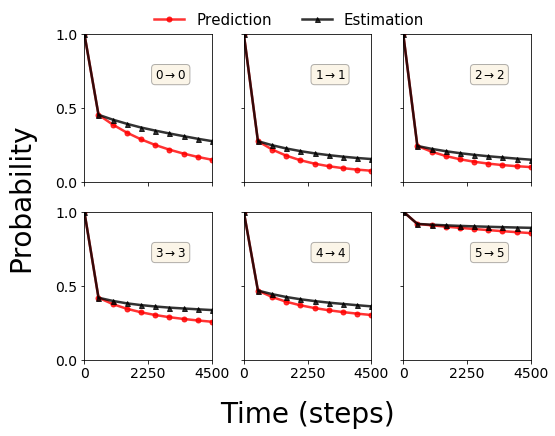}
	\end{subfigure}%
	~
	\begin{subfigure}[b]{0.48\textwidth}
	\centering
	\caption{Clusters}
	\label{fig:aaqaa-II-gmvae_landsc-states}
	\includegraphics[height = 0.22\textheight]{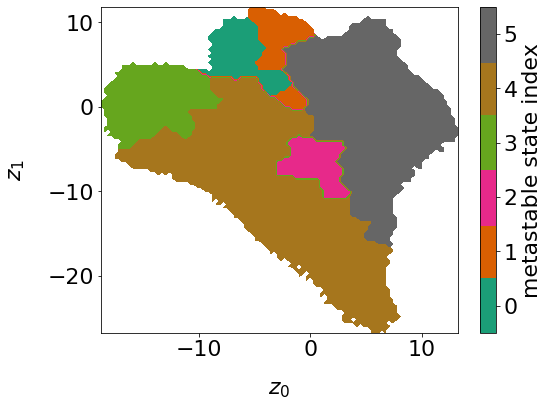}
	\end{subfigure}%
\caption{Kinetic analysis on the GMVAE landscape for $\AQA$ - II}
\label{fig:aaqaa-II-gmvae_landsc-msm}
\end{figure}
\begin{figure}[htbp]
	\centering
	\includegraphics[height = 0.2\textheight]{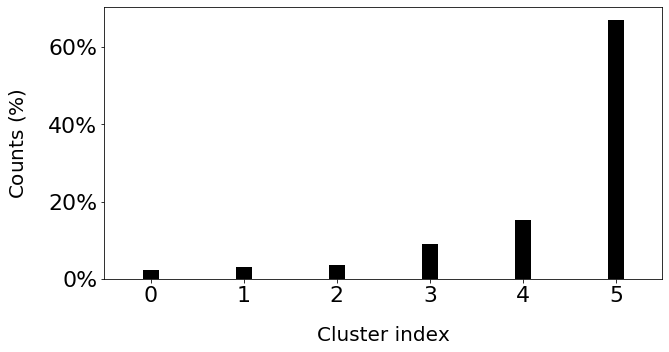}
	\caption{Cluster populations for $\AQA$ - II from PCCA+}
	\label{fig:aaqaa-II-gmvae_landsc-cluster_pop}
\end{figure}
\begin{figure}[htbp]
\centering
	\includegraphics[width = 1\textwidth]{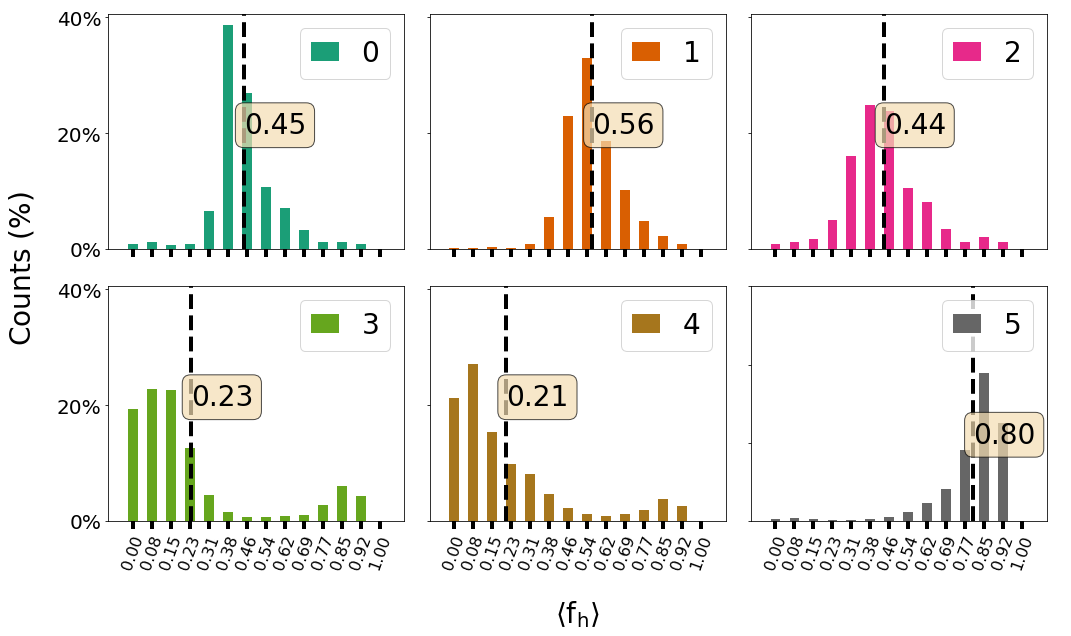}
	\caption{Intra-cluster distributions of average helical fraction, $\langle f_h \rangle$, ($\AQA$ - II) for the clusters obtained with PCCA+). The dashed lines indicate the average values, which are also written in the text boxes.}
	\label{fig:aaqaa-II-gmvae_landsc-hist_fh}
\end{figure}

\clearpage
\subsection{TICA Results} \label{sec:aaqaa-II-tica}
2D TICA projections are obtained at lag time $\tau = 20$ steps.
\begin{figure}[htbp] 	
\centering
	\begin{subfigure}[b]{0.48\textwidth}
	\centering
	\caption{FEL}
	\label{fig:aaqaa-II-tica-fel}
	\includegraphics[height = 0.22\textheight]{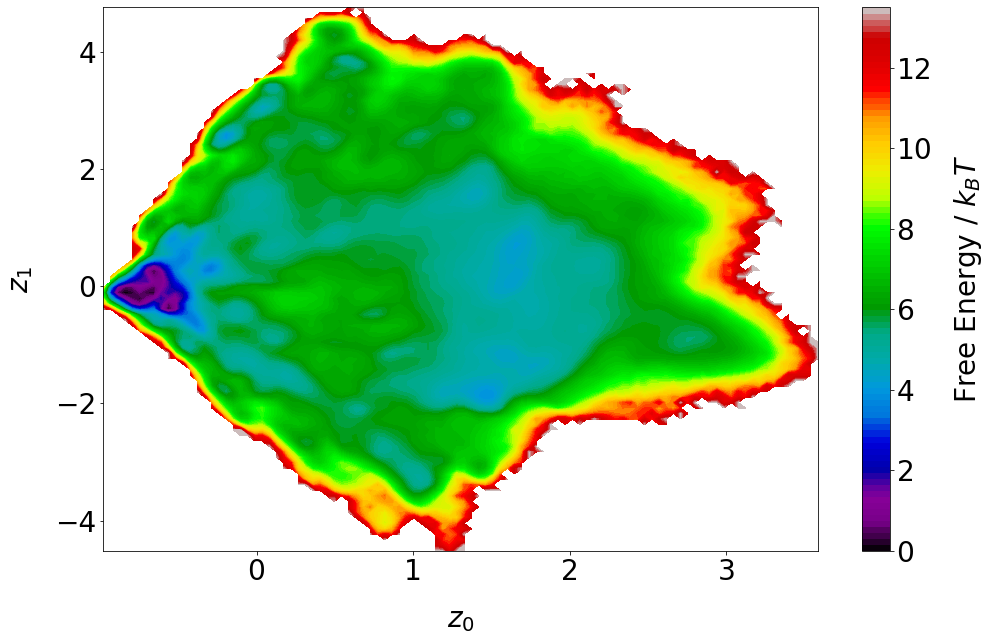}
	\end{subfigure}  
	~
	\begin{subfigure}[b]{0.48\textwidth}
	\centering
	\caption{$\langle f_h \rangle$}
	\label{fig:aaqaa-II-tica-fh}
	\includegraphics[height = 0.22\textheight]{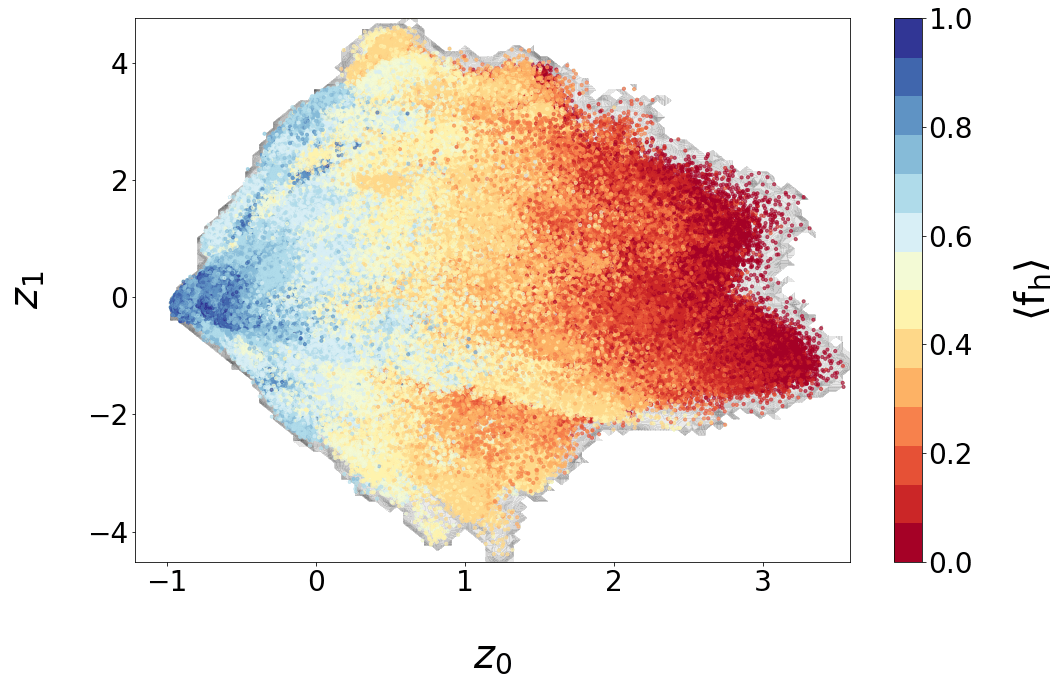}
	\end{subfigure}%
	\\
	\begin{subfigure}[b]{0.4\textwidth}
	\centering
	\caption{$\langle h_N \rangle - \langle h_C \rangle$}
	\label{fig:aaqaa-II-tica-fold1}
	\includegraphics[height = 0.22\textheight]{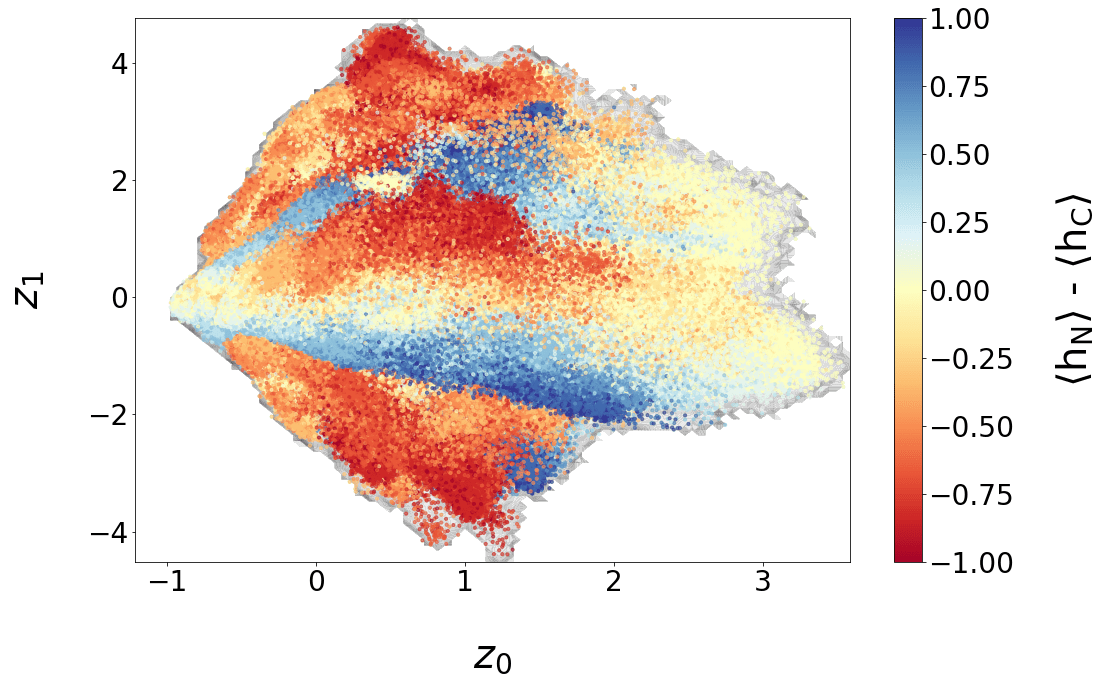}
	\end{subfigure}  
\caption{TICA results for $\AQA$ - II}
\label{fig:aaqaa-II-tica-landscape}
\end{figure}
\begin{figure}[htbp] 	
\centering
	\begin{subfigure}[b]{0.48\textwidth}
	\centering
	\caption{Cluster centers}
	\label{fig:aaqaa-II-tica-ccenters}
	\includegraphics[height = 0.22\textheight]{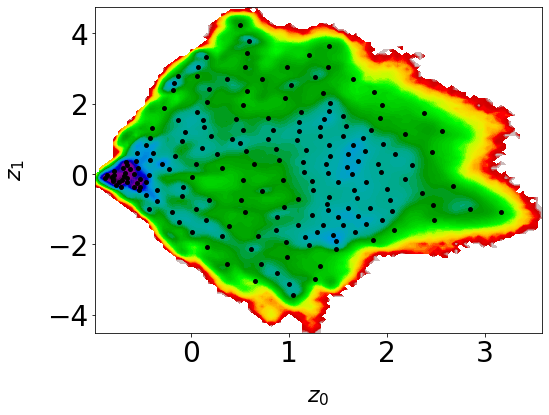}
	\end{subfigure}  
	~
	\begin{subfigure}[b]{0.48\textwidth}
	\centering
	\caption{Implied timescales}
	\label{fig:aaqaa-II-tica-its}
	\includegraphics[height = 0.22\textheight]{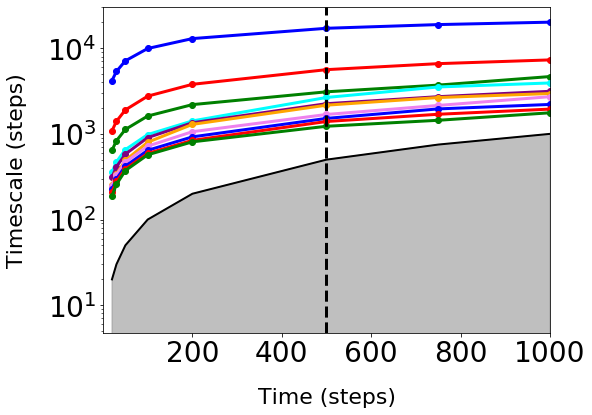}
	\end{subfigure}  
	\\
	\begin{subfigure}[b]{0.48\textwidth}
	\centering
	\caption{Chapman Kolmogorov test}
	\label{fig:aaqaa-II-tica-ck}
	\includegraphics[height = 0.22\textheight]{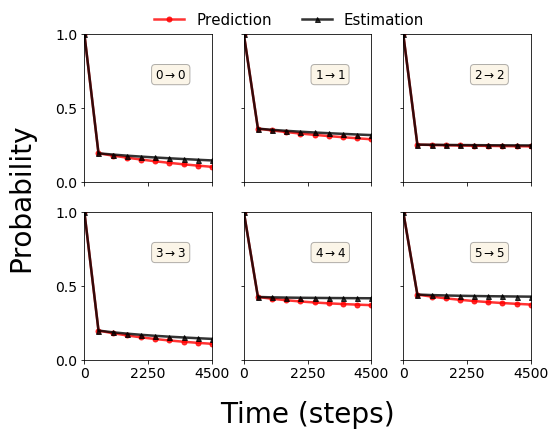}
	\end{subfigure}%
	~
	\begin{subfigure}[b]{0.48\textwidth}
	\centering
	\caption{Clusters}
	\label{fig:aaqaa-II-tica-states}
	\includegraphics[height = 0.22\textheight]{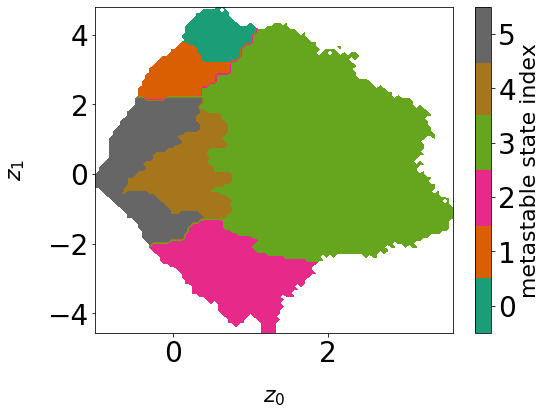}
	\end{subfigure}%
\caption{Kinetic analysis on TICA landscape for $\AQA$ - II}
\label{fig:aaqaa-II-tica-msm}
\end{figure}
\begin{figure}[htbp]
	\centering
	\includegraphics[height = 0.2\textheight]{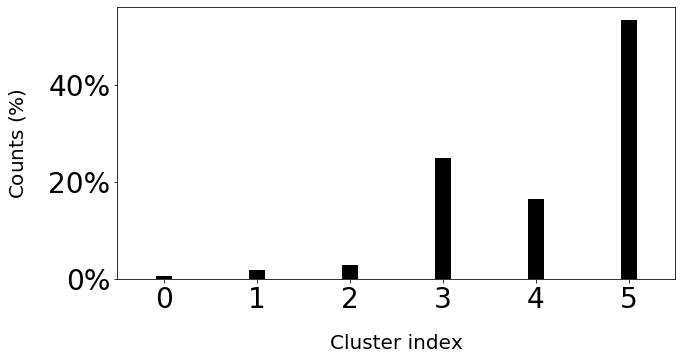}
	\caption{Cluster populations for $\AQA$ - II from TICA + PCCA+}
	\label{fig:aaqaa-II-tica-cluster_pop}
\end{figure}
\begin{figure}[htbp]
\centering
	\includegraphics[width = 1\textwidth]{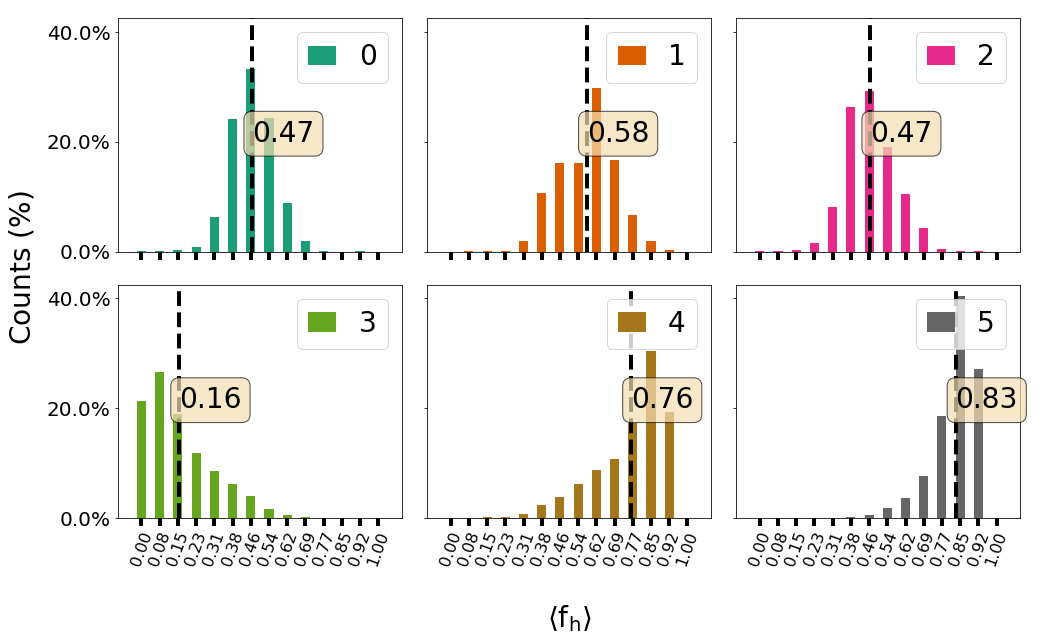}
	\caption{Intra-cluster distributions of average helical fraction, $\langle f_h \rangle$, ($\AQA$ - II) for the clusters obtained with PCCA+). The dashed lines indicate the average values, which are also written in the text boxes.}
	\label{fig:aaqaa-II-tica-hist_fh}
\end{figure}
\clearpage
\subsection{VAE Results} \label{sec:aaqaa-II-vae}
\begin{figure}[htbp] 	
\centering
	\begin{subfigure}[b]{0.48\textwidth}
	\centering
	\caption{FEL}
	\label{fig:aaqaa-II-vae-fel}
	\includegraphics[height = 0.22\textheight]{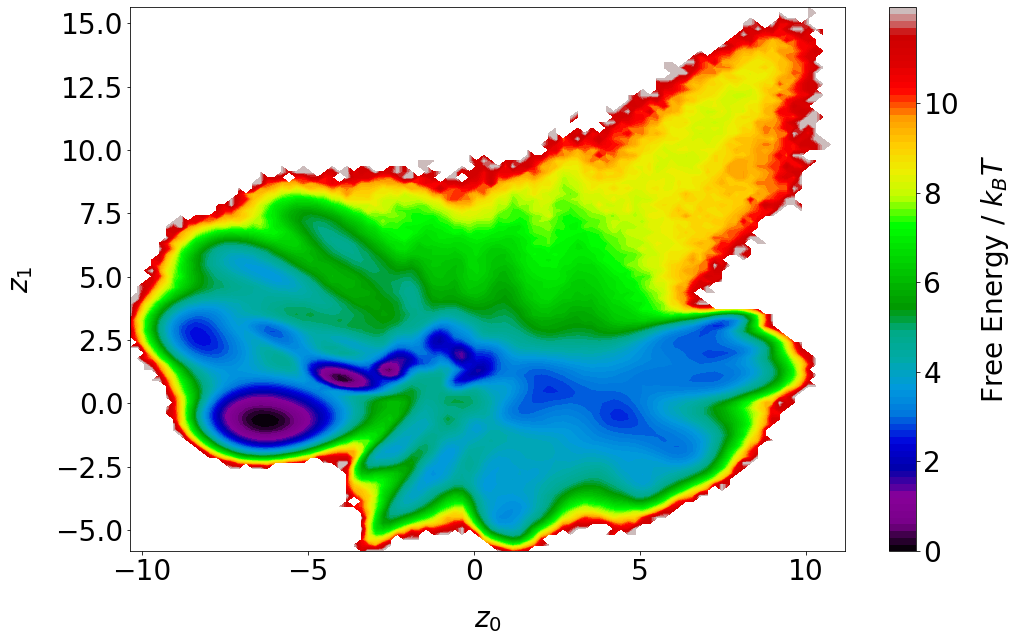}
	\end{subfigure}  
	~
	\begin{subfigure}[b]{0.48\textwidth}
	\centering
	\caption{$\langle f_h \rangle$}
	\label{fig:aaqaa-II-vae-fh}
	\includegraphics[height = 0.22\textheight]{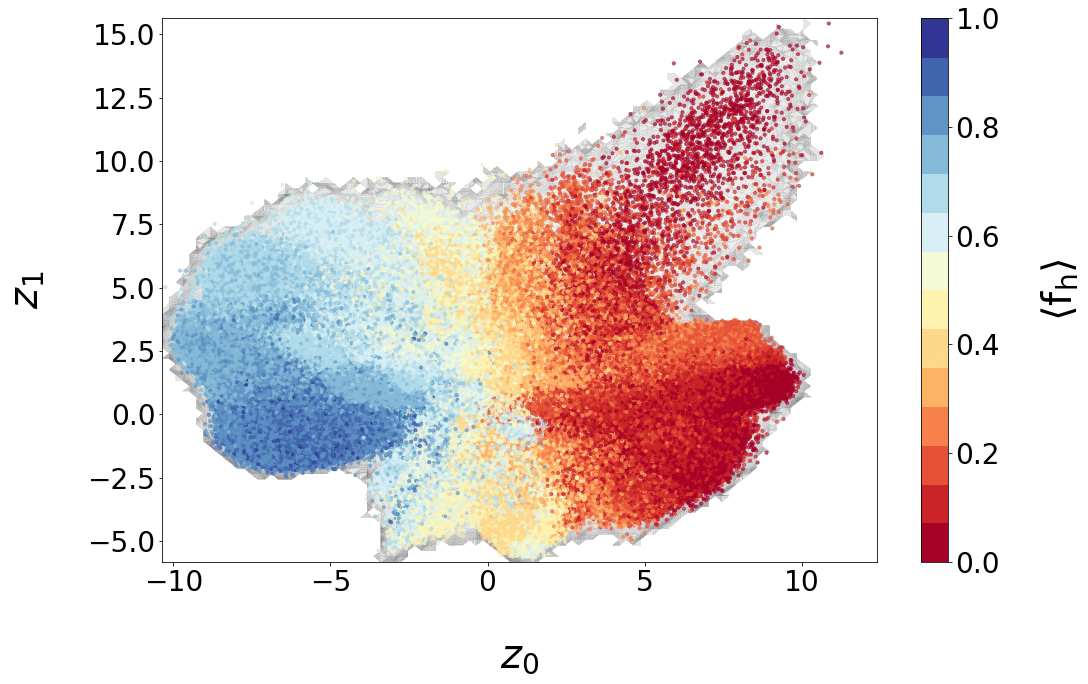}
	\end{subfigure}%
	\\
	\begin{subfigure}[b]{0.4\textwidth}
	\centering
	\caption{$\langle h_N \rangle - \langle h_C \rangle$}
	\label{fig:aaqaa-II-vae-fold1}
	\includegraphics[height = 0.22\textheight]{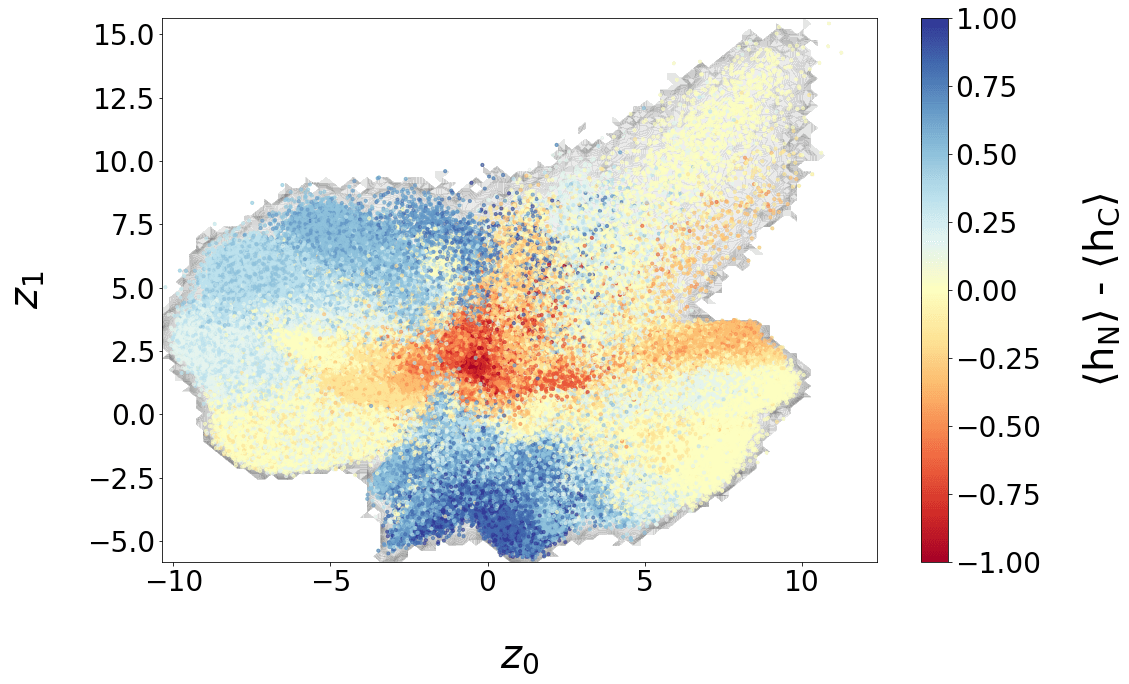}
	\end{subfigure}  
\caption{VAE results for $\AQA$ - II}
\label{fig:aaqaa-II-vae-landscape}
\end{figure}
\begin{figure}[htbp] 	
\centering
	\begin{subfigure}[b]{0.48\textwidth}
	\centering
	\caption{Cluster centers}
	\label{fig:aaqaa-II-vae-ccenters}
	\includegraphics[height = 0.22\textheight]{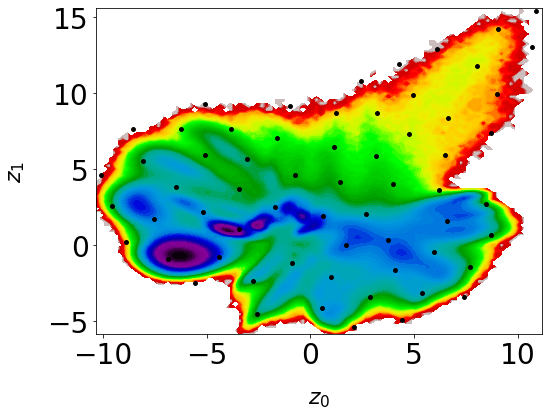}
	\end{subfigure}  
	~
	\begin{subfigure}[b]{0.48\textwidth}
	\centering
	\caption{Implied timescales}
	\label{fig:aaqaa-II-vae-its}
	\includegraphics[height = 0.22\textheight]{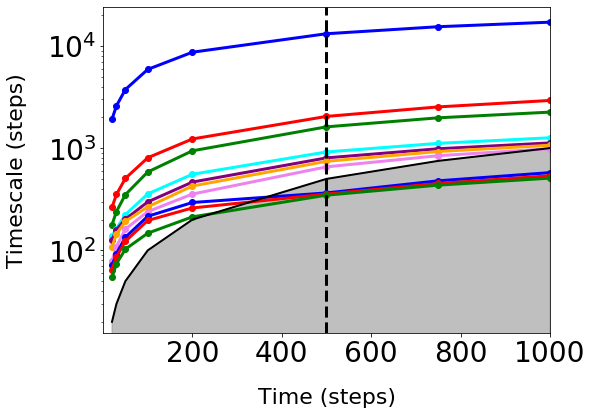}
	\end{subfigure}  
	\\
	\begin{subfigure}[b]{0.48\textwidth}
	\centering
	\caption{Chapman Kolmogorov test}
	\label{fig:aaqaa-II-vae-ck}
	\includegraphics[height = 0.22\textheight]{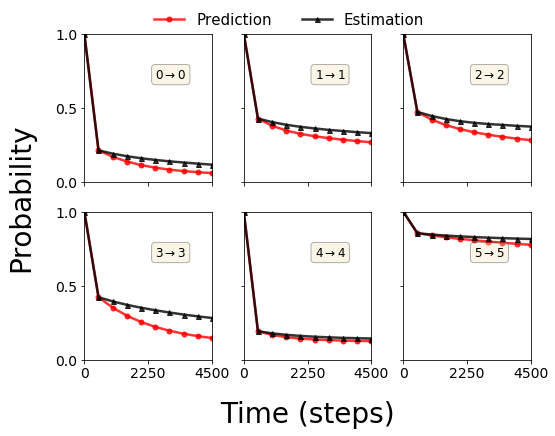}
	\end{subfigure}%
	~
	\begin{subfigure}[b]{0.48\textwidth}
	\centering
	\caption{Clusters}
	\label{fig:aaqaa-II-vae-states}
	\includegraphics[height = 0.22\textheight]{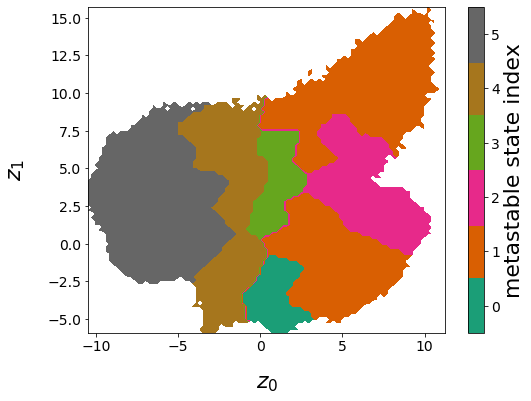}
	\end{subfigure}%
\caption{Kinetic analysis on the VAE landscape for $\AQA$ - II}
\label{fig:aaqaa-II-vae-msm}
\end{figure}
\begin{figure}[htbp]
	\centering
	\includegraphics[height = 0.2\textheight]{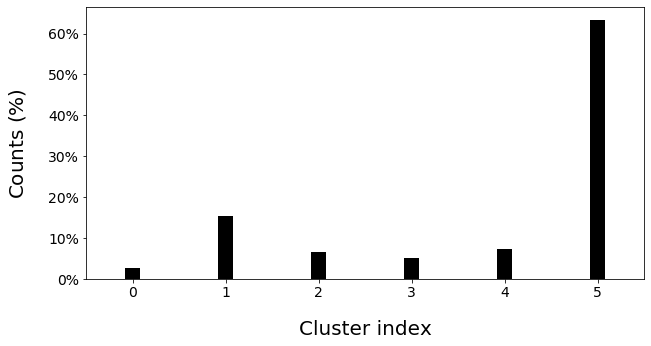}
	\caption{Cluster populations for $\AQA$ - II from VAE + PCCA+}
	\label{fig:aaqaa-II-vae-cluster_pop}
\end{figure}
\begin{figure}[htbp]
\centering
	\includegraphics[width = 1\textwidth]{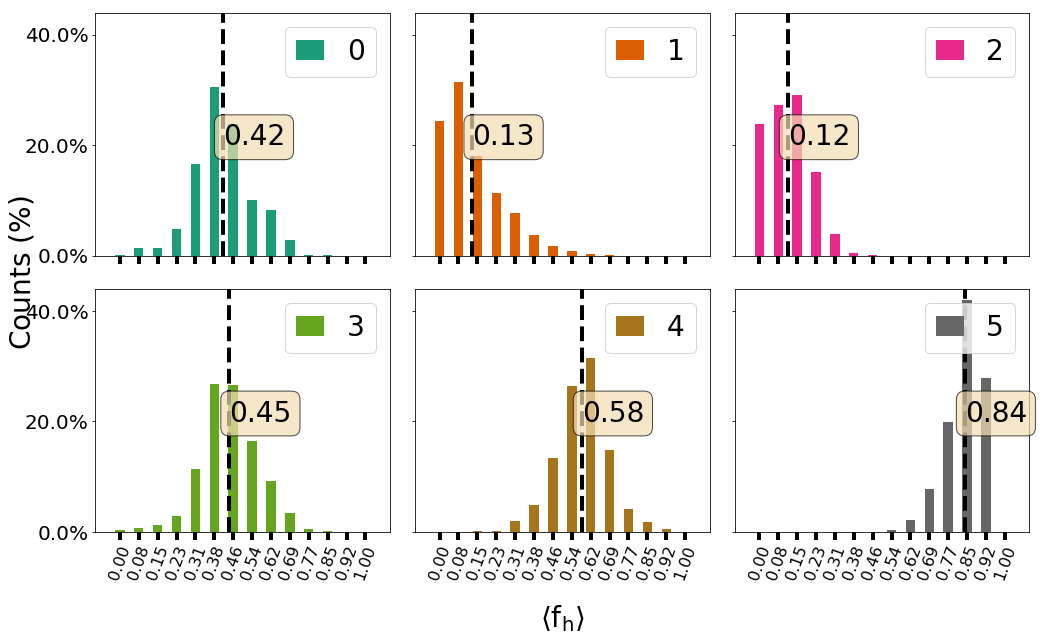}
	\caption{Intra-cluster distributions of average helical fraction, $\langle f_h \rangle$, ($\AQA$ - II) for the clusters obtained with PCCA+). The dashed lines indicate the average values, which are also written in the text boxes.}
	\label{fig:aaqaa-II-vae-hist_fh}
\end{figure}
%
%


\end{document}